\documentclass[12pt]{article}
\usepackage{amsfonts,amssymb,epsfig,amsmath}
\usepackage{color}

\usepackage[enableskew]{youngtab}

\renewcommand{\baselinestretch}{1.2}

\def\det{{\rm det}}

\newcommand{\be}{\begin{eqnarray}}
\newcommand{\ee}{\end{eqnarray}}
\newcommand{\nn}{\nonumber}
\newcommand{\bn}{\begin{enumerate}}
\newcommand{\en}{\end{enumerate}}

\newcommand{\sk}[1]{\textbf{sk: {#1}}}

\begin{document}

\makeatletter \@addtoreset{equation}{section} \makeatother
\renewcommand{\theequation}{\thesection.\arabic{equation}}
\renewcommand{\thefootnote}{\alph{footnote}}

\begin{titlepage}

\begin{center}
\hfill {\tt SNUTP14-006}\\

\vspace{2cm}

{\Large\bf General instanton counting and 5d SCFT}

\vspace{2cm}

\renewcommand{\thefootnote}{\alph{footnote}}

{\large Chiung Hwang$^1$, Joonho Kim$^2$, Seok Kim$^2$ and Jaemo Park$^1$}

\vspace{1cm}

\textit{$^1$Department of Physics \& Postech Center for Theoretical Physics (PCTP),\\
Postech, Pohang 790-784, Korea.}\\

\vspace{0.2cm}

\textit{$^2$Department of Physics and Astronomy \& Center for
Theoretical Physics,\\
Seoul National University, Seoul 151-747, Korea.}

\vspace{0.7cm}

E-mails: {\tt c\_hwang@postech.ac.kr, joonho0@snu.ac.kr,
skim@phya.snu.ac.kr, jaemo@postech.ac.kr}

\end{center}

\vspace{1.5cm}

\begin{abstract}

Instanton partition functions of 5d $\mathcal{N}=1$ gauge theories are Witten indices
for the ADHM gauged quantum mechanics with $(0,4)$ SUSY. We derive the integral contour
prescriptions for these indices using the Jeffrey-Kirwan method, for gauge theories with
hypermultiplets in various representations. The results can be used to study various 4d/5d/6d QFTs. In this paper,
we study 5d SCFTs which are at the UV fixed points of 5d SYM theories. In particular,
we focus on the $Sp(N)$ theories with $N_f\leq 7$ fundamental and $1$ antisymmetric
hypermultiplets, living on the D4-D8-O8 systems. Their superconformal indices
calculated from instantons all show $E_{N_f+1}$ symmetry enhancements. We also
discuss some aspects of the 6d SCFTs living on the M5-M9 system. It is crucial to understand
the UV incompleteness of the 5d SYM, coming from small instantons in our problem.
We explain in our examples how to fix them. As an aside, we derive the index for
general gauged quantum mechanics with $(0,2)$ SUSY.

\end{abstract}

\end{titlepage}

\renewcommand{\thefootnote}{\arabic{footnote}}

\setcounter{footnote}{0}

\renewcommand{\baselinestretch}{1}

\tableofcontents

\renewcommand{\baselinestretch}{1.2}

\section{Introduction and summary}

Instantons play important roles in understanding non-perturbative physics of gauge theories.
In this paper, we study the partition function for multi-instantons in gauge theories
preserving $8$ SUSY, on Omega-deformed $\mathbb{R}^4$ or $\mathbb{R}^4\times S^1$
\cite{Nekrasov:2002qd,Nekrasov:2003rj}. These partition functions, first studied to
understand the Seiberg-Witten solutions \cite{Seiberg:1994rs} of the 4d $\mathcal{N}=2$
gauge theories, turn out to have much wider applications. In particular,
\cite{Nekrasov:2002qd,Nekrasov:2003rj} considered circle uplifts of the 4d partition
functions to 5d. They are Witten indices which capture
the BPS spectrum of 5d SYM theories. The usefulness of this partition function in
the 4d and 5d SUSY partition functions on curved spaces was also shown in
\cite{Pestun:2007rz,Hama:2012bg} and
\cite{Kim:2012ava,Kim:2012tr,Lockhart:2012vp,Kim:2012qf,Kim:2012gu,
Kim:2013nva,Qiu:2013aga,Nieri:2013yra}.

Instanton partition functions for the pure $\mathcal{N}=2$ theories with classical gauge
groups were studied in \cite{Nekrasov:2004vw}, with generalizations \cite{Shadchin:2005mx}
to matter hypermultiplets in some representations. These partition functions are given by
contour integrals. To the best of our knowledge, a systematic derivation of the
contour has not been available, although working prescriptions are known with examples.
The contour issue becomes subtle when there are hypermultiplets in
matrix-valued or higher representations of the gauge group. Gauge theories with such
matters are important for many reasons.
For instance, they appear in various quiver gauge theories, engineered by D-branes or
M-theory branes. Adjoint, bi-fundamental, and even tri-fundamental representations
\cite{Gaiotto:2009we} appear. In this paper, they appear in the 5d SYM
descriptions of 5d/6d superconformal field theories. In particular, we are interested in
the 5d SCFTs which admit low energy descriptions by 5d SYMs after relevant deformations
\cite{oai:arXiv.org:hep-th/9608111,Morrison:1996xf,Intriligator:1997pq}.

In this paper, we address a few technical or conceptual issues related to the 5d instanton
partition functions. The subjects that we discuss are: (1) contour choice in
Nekrasov's instanton partition function, and more generally indices of gauged
quantum mechanics with $(0,2)$ SUSY; (2) the physics of 5d SCFTs from
instanton partition functions, addressing various properties of the UV fixed points
such as $E_n$ symmetry enhancements; (3) the role of small instantons
in UV incomplete 5d gauge theories, and the meaning of Nekrasov's partition functions there.
We explain these issues below, also summarizing our results.

\hspace*{-.65cm}{\bf Indices of SUSY gauged quantum mechanics:}
On the technical side, we would like to clarify a step which was left somewhat
incomplete in the literature, concerning the choice of integral
contour when the matter hypermultiplet is in various representation of the gauge group.
In the ADHM quantum mechanics description of instantons,
hypermultiplets with higher rank ($\geq 2$) representations in the gauge theory yield
bosonic degrees in the mechanics. The contour integrand encounters poles coming from
these bosons. The question is how the contour goes around these poles.
For the poles coming from the vector multiplets, \cite{Nekrasov:2002qd} stated the famous
contour prescription. For some gauge theories with matrix-valued hypermultiplets, such as the
4d $\mathcal{N}=2^\ast$ theory with $U(N)$ gauge group and its 5d uplift, \cite{Nekrasov:2002qd}
also found the prescription.\footnote{The $U(N)$ result of
\cite{Nekrasov:2002qd}, given by the sum of particular residues, was later rederived in
\cite{oai:arXiv.org:1110.2175} using what is called `Higgs branch localization' nowadays, which
never refers to a contour integral at all.} Also, \cite{Hollands:2011zc} explains the working
contour prescription with tri-fundamental hypermultiplets of $SU(2)^3$ or bi-fundamental
hypermultiplets of $Sp(1)\times SO(4)$. \cite{Nekrasov:2013xda} recently
studied the quiver gauge theories with many $SU(n)$ gauge groups and matters in the
adjoint/bi-fundamental representations. Our general derivation of the contour will explain
or reconfirm these results. See the last four paragraphs of section 2.3 for the
contour prescription.

In fact, a similar problem was recently solved for the indices for the circle
compactified 2d gauge theories \cite{Benini:2013nda,Benini:2013xpa,Gadde:2013dda}. This is
a SUSY partition function on a torus, called an elliptic genus. Following their derivation
in the context of the gauged quantum mechanics, we derive a similar contour for the instanton
partition function, with new aspects which do not have 2d analogues. The ADHM quantum mechanics
for instantons is formally obtained by a 1d reduction of 2d $\mathcal{N}=(0,4)$ gauge theories.
This is also called $\mathcal{N}=4$B SUSY quantum mechanics. Exceptionally for 5d maximal SYM,
the ADHM quantum mechanics preserves $\mathcal{N}=(4,4)$ SUSY. Regarding our
$(0,4)$ mechanics as a $(0,2)$ system, we derive the general form of the $\mathcal{N}=(0,2)$
index and apply it to our ADHM quantum mechanics.

The zero modes appearing in the 2d path integral of
\cite{Benini:2013nda,Benini:2013xpa,Gadde:2013dda} live on tori, while the path integral
for our quantum mechanics has zero modes living on cylinders. The non-compactness of
the zero mode space is a new aspect in the quantum mechanical index. This could be
subtle because noncompact moduli develop a continuum in the spectrum above the BPS states.
In the context of instanton quantum mechanics, the noncompact direction corresponds to the
Coulomb branch moduli space through which instantons can `escape to infinity.'
For instance, in D-brane realization, this is the direction in which Dp-branes (string theory
uplift of instantons) move away from D(p+4)-branes. These degrees do not represent any
degrees of freedom in the 5d QFT, but
enter while one attempts to engineer the UV incomplete instanton quantum mechanics by a UV
complete ADHM quantum mechanics. When the 5d SYM has a 5d UV fixed point
\cite{oai:arXiv.org:hep-th/9608111,Morrison:1996xf,Intriligator:1997pq},
the instantons typically cannot move away in this noncompact direction. This is because the
1-loop effects provide linear growths in the instanton masses in the Coulomb branch, and
confine them. So in our index, the zero mode integral is convergent in the asymptotic
regions of the cylinder. There are interesting exceptions to this confinement, which we
study in detail. The 5d gauge theories which uplift to the 6d SCFTs on $S^1$ also have more
nontrivial asymptotic behaviors, which we explain in section 3.

\hspace*{-.65cm}{\bf 5d superconformal field theories:}
While we make a general study on a large class of 5d SCFTs classified in
\cite{Intriligator:1997pq}, we shall also focus on a special class of them in section 4,
which we explain in some detail here.
We shall study the 5d $Sp(N)$ $\mathcal{N}=1$ gauge theories coupled to
$N_f$ fundamental hypermultiplets and one anti-symmetric hypermultiplet.
From string theory, this system is engineered by $N$ D4-branes near $N_f$ D8-branes and
an O8-plane. At $0\leq N_f\leq 7$, these systems were used to predict the existence of
a class of 5d SCFTs at the UV fixed point \cite{oai:arXiv.org:hep-th/9608111}. Although
the SYM theory at small coupling only exhibits $SO(2N_f)\times U(1)$ global symmetry,
$E_{N_f+1}$ symmetry enhancements were predicted at the strong coupling fixed point.
At $N_f=8$, the 5d SYM is a suitable circle reduction of the 6d $(1,0)$ SCFT living on
the M5-M9 system. We mainly study 5d SCFTs with $N_f\leq 7$ in this paper, although
we also explain the case with $N_f=8$ in section 3.4.2. More detailed studies on the
6d SCFT related to the case with $N_f=8$ will appear elsewhere \cite{Kim:2014dza}.

Recently, the indices for 5d SCFTs at $N_f\leq 7$ were studied
starting from \cite{Kim:2012gu}. They studied the Nekrasov's partition function,
and also the superconformal index \cite{Kinney:2005ej} which one can calculate using the
former. \cite{Kim:2012gu} mainly discussed the $Sp(1)$ theories. Since the
anti-symmetric representation of $Sp(1)$ is neutral, they naturally considered the systems
with $N_f$ fundamental hypermultiplets only. For $N_f\leq 5$, the superconformal indices
they calculate showed the $E_{N_f+1}$ symmetry enhancement. For $N_f=6,7$, where we expect $E_7, E_8$ symmetries, the calculations of \cite{Kim:2012gu} did not exhibit the spectra
with these symmetries. Our studies began by a rather small motivation to understand the
correct indices at $N_f=6,7$. Here we note that \cite{Hayashi:2013qwa} computed the $E_7$
index from topological strings by Higgsing the 5d $T_4$ theory, and
\cite{Bao:2013pwa,Hayashi:2013qwa} computed the $E_6$ index from
the 5d $T_3$ theory.

We start by observing that, even if the $Sp(1)$ anti-symmetric
hypermultiplet decouples in the perturbative gauge dynamics, the details
of the ADHM instanton calculus in \cite{Shadchin:2005mx,Kim:2012gu} depend on whether
one includes this matter contribution or not. Such phenomenon is somewhat well known.
For instance, consider a $U(1)$ gauge theory with adjoint
hypermultiplet. As $U(1)$ adjoint is neutral, one may think that this partition function
is totally trivial. Instead,
\cite{Nekrasov:2003rj} captures the $U(1)$ small instantons' contributions to the index,
whose singular configurations see the perturbatively decoupled fields.
In string theory, this index counts the marginal bound states of single D4-brane and D0-branes.
Similar reasoning can be given to our problem with an $Sp(1)$ anti-symmetric hypermultiplet.
These non-perturbative couplings are possible due to the small instantons
which violate the perturbative intuitions of 5d SYM. Namely, the instanton quantum
mechanics given by a SUSY sigma model is incomplete, as the instanton moduli space
develops small instanton singularities. To get a complete description,
one uses the ADHM gauged quantum mechanics. The last mechanics is often motivated from
string theory, most naturally on Dp-D(p+4) systems. It contains extra degrees of freedom,
which have to be carefully treated to compute the QFT observable correctly.

From the brane perspective, the perturbatively decoupled $Sp(1)$ antisymmetric hypermultiplet
contains the degrees of freedom for D4-branes moving along the worldvolume of D8-O8.
So the related degrees in the ADHM quantum mechanics represent D0-branes moving away from
the D4's, bound only to the D8-O8. In other words, the $Sp(1)$ partition
function with an antisymmetric hypermultiplet not only counts the $4+1$ dimensional
BPS states living on the D4-branes, but also captures the $8+1$ dimensional bound
states for D0-D8-O8. This phenomenon was observed in \cite{Kim:2012gu} at one instanton order.
In this paper, we call these `4 dimensional' and `8 dimensional' particles, respectively.
The index for the 8d and 4d particles will factorize. Here, \cite{Kim:2012gu} further assumed
that the effect of including the
$Sp(1)$ antisymmetric hypermultiplet in the instanton calculus was exactly providing this extra
factorized 8d superparticle index and nothing else, to all orders in the instanton number. As
their main interest was the 5d QFT and not these 8d particles, \cite{Kim:2012gu} then
omitted the antisymmetric hypermultiplet and proceeded with the instanton calculus
having $N_f$ fundamental hypermultiplets only.
We find that naively discarding the antisymmetric hypermultiplet's
effect in the instanton calculus is not always the same as discarding the
factorized 8d index. The naive expectation of \cite{Kim:2012gu} turns out to be correct for
$0\leq N_f\leq 5$, precisely when their indices exhibit $E_{N_f+1}$ symmetry enhancement.
However, for $N_f=6,7$, we find that
\begin{equation}\label{decoupling}
  \frac{Z({\rm with\ antisymmetric})}{Z({\rm without\ antisymmetric})}
  \neq Z({\rm 8d\ particle})\ .
\end{equation}
So one has to include the antisymmetric hypermultiplet
in the calculation, and then divide by the 8d index which can be
computed separately. This procedure completely
restores the $E_7$ and $E_8$ symmetries in the superconformal index.

We can understand in many ways why \cite{Kim:2012gu} got good results by simply
discarding the antisymmetric hypermultiplet at low $N_f$. Firstly, for $0\leq N_f \leq 4$,
the 5d system has a good 4d limit, which yields either asymptotically free or conformal QFT.
Since 4d QFT is well defined, any instanton calculus there should be free of
ambiguities, especially concerning small instantons.\footnote{In 4d, small instantons are singular
configurations contributing to the path integral, called constrained instantons
\cite{'tHooft:1976fv,Affleck:1980mp}. They have nothing to do with the ill-defined nature of the QFT.}
So including the decoupled $Sp(1)$ anti-symmetric hypermultiplet should not affect the QFT dynamics,
and in particular the Seiberg-Witten solution. This implies
\begin{equation}\label{4d-expectation}
  \frac{Z_{\rm 4d}({\rm with\ antisymmetric})}{Z_{\rm 4d}({\rm without\ antisymmetric})}
  =({\rm independent\ of\ Coulomb\ VEV})\ .
\end{equation}
Suppose that the 5d version (\ref{decoupling}) of the relation holds. Then the right hand
side is independent of the Coulomb VEV, as the decoupled 8d particles do not see the 5d gauge
group. So it will not affect the 4d Seiberg-Witten solution, as required by
(\ref{4d-expectation}). We may understand the success of \cite{Kim:2012gu} at $N_f\leq 4$
from the constraints that the 4d QFT is complete.

There is another way to understand the result of \cite{Kim:2012gu} at $N_f\leq 5$,
and actually to improve it to $N_f\leq 6$ using a different string theory completion
of the same nonlinear sigma model.
We are studying 5d SCFTs with 1 dimensional Coulomb branch and $E_{N_f+1}$ symmetry.
There are two ways of engineering them from string theory \cite{Intriligator:1997pq}.
We start by explaining two (generally inequivalent) classes of 5d rank $N$ SCFTs
engineered by sting theory. Both admit relevant deformations to $Sp(N)$ SYM
with $N_f$ fundamental hypermultiplets. The first class comes with an extra antisymmetric
hypermultiplet, while the second class does not. The first class has a UV
interacting fixed point for $N_f\leq 7$, and the second class has one for $N_f\leq 2N+4$.
The first can be engineered either by D4-D8-O8 \cite{oai:arXiv.org:hep-th/9608111}, or M-theory on certain singular Calabi-Yau 3-folds \cite{Intriligator:1997pq}, by taking
low energy decoupling limit. The second class is engineered by M-theory on a different type
of singular CY$_3$'s \cite{Intriligator:1997pq}. At $N=1$, the two classes are expected
to yield the same 5d SCFTs.
The descriptions using M-theory on CY$_3$ become identical for the two classes, as the two
CY$_3$'s become the same. However, the D4-D8-O8 description provides a different string
theory background for the first class. In both descriptions, the Hilbert spaces at low energy are expected to
factorize into the sectors of 5d QFT and the rest. The QFT sector is expected to be the same
in any descriptions, but the extra sectors are not. These aspects
descend to the ADHM quantum mechanics descriptions of instantons.

The D-brane description of $Sp(1)$ theories exists for $N_f\leq 7$,
while the M-theory description exists for $N_f\leq 2N+4=6$. The ADHM quantum mechanics used in
\cite{Kim:2012gu} (without degrees coming from `$Sp(1)$ antisymmetric hypermultiplet')
can be understood as the latter. This ADHM system is rather simple for $N_f\leq 5$, as
studied in \cite{Kim:2012gu}. At $N_f=6$,
there exists a continuum in the ADHM quantum mechanics which comes from the extra degrees
in the string theory. So we find $Z_{\rm QM}^{N_f=6}=Z_{\rm QFT}^{N_f=6}Z_{\rm extra}$, where
$Z_{\rm QM}$ refers to the index of ADHM quantum mechanics, and $Z_{\rm extra}\neq 1$
to the extra states' contribution. See section 3.4.1 for its form.
$Z_{\rm QFT}^{N_f=6}$ is the 5d SCFT index. In fact we find that $Z_{\rm QFT}^{N_f=6}$
computed from two different ADHM quantum mechanics are the same.
This shows that $Sp(1)$ theory with $N_f\leq 6$ can be studied as \cite{Kim:2012gu},
taking into account the subtle $Z_{\rm extra}$ factor. For $N_f=7$ with $E_8$, only the
D-brane engineering should work. The D-brane approach also generalizes to
$N_f=8$, which has an interesting 6d UV fixed point.\footnote{When we
talk about the extra contribution, this is the contribution we obtain from an extra
decoupled sector irrelevant for the QFT, for instance in the string theory engineering,
in the decoupling limit $M_{\rm pl}\rightarrow \infty$. }

So to summarize, it is crucial to carefully follow the string theory guidance
rather than the perturbative QFT intuition, when one makes nonperturbative studies
on the UV incomplete theories.
We shall also study other 5d SCFTs classified in \cite{Intriligator:1997pq} which
exhibit similar phenomena. Separating out the factors from string theory which are
irrelevant for QFT is the key step, which we shall explain with many examples.

\hspace*{-.65cm}{\bf Partition functions of UV incomplete theories:}
Recently, 5d gauge theories played important roles in
understanding many enigmatic SCFTs in 6
\cite{Douglas:2010iu,Lambert:2010iw,Bern:2012di,Papageorgakis:2014dma} and 5 dimensions.
The 5d Yang-Mills theories that one uses to compute the CFT
observables are non-renormalizable, at least apparently. They are low energy effective
descriptions of UV SCFTs with relevant deformations (5d SCFT) or circle compactifications
(6d SCFT). The instanton partition functions on $\mathbb{R}^4\times S^1$ are often related
to other SCFT observables, given by SYM partition functions on curved
5-manifolds such as $S^5$ or $S^4\times S^1$
\cite{Kim:2012ava,Kim:2012tr,Lockhart:2012vp,Kim:2012qf,Kim:2012gu,Kim:2013nva,
Qiu:2013aga,Nieri:2013yra}.
So it is important to develop a more abstract and intrinsic notion of the instanton partition
functions of 5d SYMs as those of 5d/6d SCFTs on Omega-deformed
$\mathbb{R}^4\times S^1$ or $\mathbb{R}^4\times T^2$, not referring to the descriptions
one uses to compute them. The computation of our 5d SYM observable on $\mathbb{R}^4\times S^1$
boils down to the 1d path integral for the instanton quantum mechanics \cite{Nekrasov:2002qd}.
The UV incompleteness of the original 5d SYM leaves a remnant on the instanton quantum mechanics,
by exhibiting small instanton singularities. We UV-complete it to ADHM gauged quantum mechanics,
with extra UV degrees, which reduces to the instanton quantum mechanics in the limit of strong
gauge coupling (or equivalently low energy) in the `Higgs branch' of the mechanics.
Doing computations this way, one should either find a method to decouple the extra UV
degrees, or should separately compute the extra stringy contribution and divide it.
After all, our $Z_{\rm QFT}$ is such an intrinsic partition function for the
higher dimensional CFTs.

A simple example is the $U(1)$ theory with one adjoint hypermultiplet. This is naively a free QFT
in 5d, but \cite{oai:arXiv.org:1110.2175,Lockhart:2012vp,Kim:2012qf,Kim:2013nva} could get
the spectrum of circle compactified 6d $(2,0)$ theory physics of the free tensor multiplet
from small instantons. So this is clearly an extra UV input beyond the naive 5d SYM, but here
we do not acquire extra states' contribution from string theory. Similarly, the
$U(N)$ $\mathcal{N}=1^\ast$ theory that was used to study the circle compactified non-Abelian
$(2,0)$ theory could also contain such ambiguities resolved by string theory considerations.
Our example of $Sp(1)$ gauge theory with one antisymmetric hypermultiplet exhibits
a more noticeable subtlety of the 5d gauge theory. This is partly because the subtlety of the
5d SYM theory becomes manifest in a regime ($N_f\geq 6$) that 4d QFT limit does not
exist. So it should be clear that this is really the subtlety of 5d
non-renormalizable gauge theory. We hope
that the nontrivial examples studied in this paper could help clarify the usefulness
and subtleties of the 5d gauge theories in studying higher dimensional
quantum field theories.

The rest of this paper is organized as follows. In section 2, we derive the index
of general gauged quantum mechanics with $(0,2)$ SUSY, and apply it to the ADHM
gauged quantum mechanics with $(0,4)$ SUSY. In section 3, we explain examples
which have 5d or 6d UV fixed points. In particular in section 3.4,
we explain how to factor out the extra contribution to the index from
the string theory states, decoupled from the QFT. In section 4, we explain the 5d SCFT indices
on the D4-D8-O8 system, from our SYM theories with $N_f\leq 7$. We show the $E_{N_f+1}$
symmetry enhancements of these superconformal indices.

\hspace*{-.65cm} {\bf Note added:} Our sections 2.2 and 2.3 on the general index
have an overlap with \cite{Hori:2014tda}. Also, \cite{Cordova:2014oxa} discusses the
same index as ours presented in section 2.3.

\section{Instantons and their indices}

We consider 5d $\mathcal{N}=1$ SYM, with $U(N)$,
$SO(N)$, $Sp(N)$ gauge groups. We are interested in self-dual
instantons which satisfy $F_{mn}=\star_4F_{mn}=\frac{1}{2}\epsilon_{mnpq}F_{pq}$ on spatial
$\mathbb{R}^4$ ($m,n,p,q=1,2,3,4$). Let us decompose the 8 SUSY into $Q_{\alpha}^A$, $\bar{Q}_{\dot\alpha}^A$, where $\alpha=1,2$, $\dot\alpha=\dot{1},\dot{2}$ are for the doublets
of $SU(2)_l\times SU(2)_r=SO(4)$ rotating $\mathbb{R}^4$, and $A=1,2$ is for the doublet of
$SU(2)_R$ R-symmetry. $Q^A_\alpha$, $\bar{Q}^A_{\dot\alpha}$ are subject to
symplectic-Majorana conditions. The SUSY algebra is given by
\begin{equation}
  \{Q^A_M,Q^B_N\}=P_\mu(\Gamma^\mu C)_{MN}\epsilon^{AB}+i\frac{4\pi^2 k}{g_{YM}^2}C_{MN}\epsilon^{AB}
  +i{\rm tr}(v\Pi)C_{MN}\epsilon^{AB}\ .
\end{equation}
$M,N=1,2,3,4$ are Dirac spinor indices, $C_{MN}$ is the charge conjugation matrix,
$k$ is the instanton number, $v$ is the scalar VEV in the Coulomb branch, $\Pi$ is the electric charge.
The topological $U(1)_I$ charge of self-dual instantons is given by
\begin{equation}
  k=\frac{1}{8\pi^2}\int_{\mathbb{R}^4}{\rm tr}(F\wedge F)\in\mathbb{Z}_+\ .
\end{equation}
These instantons preserve $4$ real SUSY $\bar{Q}^A_{\dot\alpha}$. They can also form marginal
bound states with perturbative particles with electric charges, namely the W-bosons and their
superpartners. The bound states preserve the same $4$ SUSY, with the BPS mass
\begin{equation}
  M=\frac{4\pi^2 k}{g_{YM}^2}+{\rm tr}(v\Pi)\ ,
\end{equation}
provided that the signs of the electric charges are properly chosen. The signs should be
chosen with ${\rm tr}(v\Pi)\geq 0$. (\sk{till here})

\subsection{ADHM quantum mechanics with $\mathcal{N}=(0,4)$ SUSY}

The zero modes of self-dual instantons can be described by the ADHM data \cite{Atiyah:1978ri}
for classical gauge groups, subject to the ADHM constraint equations. In 5d SYM,
the moduli space approximation of these instantons is given by a supersymmetric sigma
model with the target space given by the instanton moduli space. The partition function
of \cite{Nekrasov:2002qd} can be understood as that of this mechanical system.
The moduli space has small instanton singularities, implying that the quantum mechanics
description is incomplete. It admits a UV completion by a gauged SUSY quantum mechanics,
often called gauged linear sigma model (GLSM). The ADHM data provide the dynamical degrees,
and the ADHM constraints are realized as the vanishing condition of the D-term potential.
The incomplete mechanical system is obtained at low energy, or equivalently when quantum
mechanical gauge coupling is infinite. When the 5d gauge theory and instantons are engineered
by D-branes in string theory, the ADHM mechanics is the D-brane quantum mechanics.
As we shall explain below, the UV completion includes extra degrees
of freedom, on top of the degrees relevant for QFT.

We first explain the ADHM data for various classical gauge groups. The $SU(N)$
gauge group is replaced by $U(N)$, which has no effect in the classical gauge
theory viewpoint (as the overall $U(1)$ plays no role in constructing self-dual solutions).
For each 5d gauge group, $G=U(N),Sp(N),SO(N)$, the quantum mechanical gauge group $\hat{G}$
for $k$ instantons is given by $\hat{G}=U(k),O(k),Sp(k)$, respectively.
The ADHM data consists of the following matrices: $q_{\dot\alpha}$ which is in bi-fundamental
representation of $G\times\hat{G}$, and
$a_{\alpha\dot\beta}=\frac{1}{\sqrt{2}}a_m\sigma^{m}_{\alpha\dot\beta}$
which assumes a matrix-valued representation of $\hat{G}$. Here,
$\sigma^m=(i\vec\tau,{\bf 1})$. $q_{\dot\alpha}$ with given $\dot\alpha$ is
an $N\times k$ matrix for $U(N)$, an $N\times 2k$ matrix for $SO(N)$, and an $2N\times k$
matrix for $Sp(N)$. For $G=U(N),Sp(N),SO(N)$, $a_m$ is in the adjoint, symmetric, and antisymmetric representation of $\hat{G}=U(k),O(k),Sp(k)$, respectively. In the last two
cases with $Sp(N),SO(N)$, $q_{\dot\alpha}$ and $a_m$ are subject to suitable reality conditions.
Together with superpartners, their quantum mechanics action preserves $4$ SUSY $\bar{Q}^A_{\dot\alpha}$
of the self-dual instantons. They form $\mathcal{N}=(0,4)$ SUSY, i.e. 2d $(0,4)$ SUSY
reduced to 1d. It comes with an $SO(4)=SU(2)_r\times SU(2)_R$ `R-symmetry.'

Let us explain their supermultiplet structures.
Together with the fermionic superpartners $\psi^A$, $\lambda^A_{\alpha}$, the above ADHM
degrees form the following multiplets under the $(0,4)$ SUSY
\begin{equation}\label{ADHM-multiplet}
  (q_{\dot\alpha},\psi^A)\ ,\ \ (a_{\alpha\dot\beta}, \lambda_\alpha^A)\ .
\end{equation}
The scalars are in the $({\bf 2},{\bf 1})$ representations of the $SO(4)=SU(2)_r\times SU(2)_R$.
These are $(0,4)$ hypermultiplets. The UV quantum mechanics also
comes with a vector multiplet for $\hat{G}$, which consists of a worldline gauge field $A_t$,
a scalar $\varphi$, fermions $\bar\lambda_{\dot\alpha}^A$ and $3$ auxiliary fields
$D_{\dot\alpha\dot\beta}=D_{\dot\beta\dot\alpha}$. Although $A_t$, $D$ do not possess physical
degrees, $\varphi$, $\bar\lambda^A_{\dot\alpha}$ do. They are extra degrees present only in
the UV theory. Their action is given by
\begin{eqnarray}\label{(0,4)-action}
  L&=&\frac{1}{g_{QM}^2}{\rm tr}\left[\frac{1}{2}(D_t\varphi)^2
  +\frac{1}{2}(D_ta_m)^2+D_tq_{\dot\alpha}D_t\bar{q}^{\dot\alpha}
  +\frac{1}{2}[\varphi,a_m]^2-(\varphi\bar{q}^{\dot\alpha}-\bar{q}^{\dot\alpha}v)
  (q_{\dot\alpha}\varphi-vq_{\dot\alpha})\right.\nonumber\\
  &&+\frac{1}{2}(D^I)^2-D^I\left((\tau^I)^{\dot\alpha}_{\ \dot\beta}
  \bar{q}^{\dot\beta}q_{\dot\alpha}+\frac{1}{2}(\tau^I)^{\dot\alpha}_{\ \dot\beta}
  [a^{\dot\beta\alpha},a_{\alpha\dot\beta}]-\zeta^I\right)
  +\frac{i}{2}(\bar\lambda^{A\dot\alpha})^\dag D_t\bar\lambda^{A\dot\alpha}
  +\frac{i}{2}(\lambda^A_\alpha)^\dag D_t\lambda^A_\alpha\nonumber\\
  &&+i(\psi^A)^\dag D_t\psi^A+\sqrt{2}i\left((\bar\lambda^{A\dot\alpha})^\dag
  \bar{q}^{\dot\alpha}\psi^A\!-\!(\psi^A)^\dag q_{\dot\alpha}\bar\lambda^{A\dot\alpha}\right)
  +(\psi^A)^\dag(\psi^A\varphi-v\psi^A)\nonumber\\
  &&\left.+\frac{1}{2}(\bar\lambda^{A\dot\alpha})^\dag
  [\varphi,\bar\lambda^{A\dot\alpha}]
  -\frac{1}{2}(\lambda^A_\alpha)^\dag [\varphi,\lambda^A_\alpha]
  \!-\!i(\lambda^A_\alpha)^\dag(\sigma^m)_{\alpha\dot\beta}
  [a_m,\bar\lambda^{A\dot\beta}]\right]\ .
\end{eqnarray}
$v$ is the Coulomb VEV of the 5d SYM, and $\zeta^I$ for $I=1,2,3$ are FI parameters.
For instance, this action can be obtained by starting from the $\mathcal{N}=(4,4)$ ADHM
instanton mechanics for 5d maximal SYM, and truncating to the $(0,4)$ system. We used the
notation of \cite{oai:arXiv.org:1110.2175}.\footnote{However, we put relative minus signs
on Yukawa couplings involving $\varphi$, due to a convention change here.} Also, one can
possibly add 1 dimensional Chern-Simons term
\begin{equation}\label{QM-CS}
  L_{CS}=\kappa(\varphi+A_t)\ .
\end{equation}
This is induced by the Chern-Simons term of 5d SYM at level $\kappa$
\cite{Kim:2008kn,Collie:2008vc}. Our $(0,4)$ supersymmetry transformation can also
be obtained from \cite{oai:arXiv.org:1110.2175} by discarding half of the $(4,4)$ SUSY.
In our discussion below, we shall only need
\begin{equation}\label{hyper-SUSY}
  \bar{Q}^{A\dot\alpha}\Phi_{\dot\beta}=\sqrt{2}\delta^{\dot\alpha}_{\dot\beta}\Psi^A
\end{equation}
for any hypermultiplet of the form $(\Phi^{\dot\alpha},\Psi^A)$. $g_{QM}$ is the quantum
mechanical gauge coupling. The strong coupling limit $g_{QM}\rightarrow\infty$, or equivalently
the low energy limit, of this theory is the quantum mechanics of instantons in the `Higgs branch,'
where the mechanical hypermultiplet degrees in (\ref{ADHM-multiplet}) are nonzero \cite{Aharony:1997th}. Namely, due to the interaction
$|q_{\dot\alpha}\varphi|^2$ on the first line (say, at $v=0$), the mass for $\varphi$ in the
Higgs branch is $g_{QM}|q_{\dot\alpha}|$, and the mass for the hypermultiplet fields in the
Coulomb branch is $g_{QM}|\varphi|$. Both become large in the strong coupling limit, implying
a low energy decoupling in this system. We are interested in the Higgs branch system.

It is also helpful to use the notation of $(0,2)$ fields, by
regarding $Q\equiv-\bar{Q}^{1\dot{2}},Q^\dag$ as the $\mathcal{N}=(0,2)$ supercharges.
The above $(0,4)$ multiplets decompose into the following $(0,2)$ multiplets,
\begin{eqnarray}\label{multiplet-decompose-1}
  {\rm vector}\ (A_t,\varphi,\bar\lambda^A_{\dot\alpha})&\rightarrow&
  {\rm vector}\ (A_t,\varphi,\bar\lambda^1_{\dot{1}},\bar\lambda^2_{\dot{2}})\ +\
  {\rm Fermi}\ (\bar\lambda^1_{\dot{2}},\bar\lambda^2_{\dot{1}})\nonumber\\
  {\rm hyper}\ (\phi^{\dot\alpha},\psi^A)&\rightarrow&{\rm chiral}\ (\phi^{\dot{1}},\psi^1)\ +\
  {\rm chiral}\ (\bar\phi_{\dot{2}}=\bar\phi^{\dot{1}},\bar\psi_2)\ .
\end{eqnarray}
We should take $\phi^{\dot{1}}$ and $\bar\phi_{\dot{2}}=\bar\phi^{\dot{1}}$
to be the complex scalars of the chiral multiplet, since from (\ref{hyper-SUSY}) they
transform nontrivially under the chiral supercharge $Q=-\bar{Q}^{1\dot{2}}$. The charge
$J\equiv J_r+J_R$ will play important roles in understanding the structure of the index,
where $J_r$, $J_R$ are the Cartans of $SU(2)_r$, $SU(2)_R$. The Cartans are
defined so that objects with upper $\dot{1},\dot{2}$ component have $\pm\frac{1}{2}$ eigenvalues
for $J_r$, and objects with upper $1,2$ component have $\pm\frac{1}{2}$ eigenvalues
for $J_R$. So the scalars $\phi^{\dot{1}},\bar\phi^{\dot{1}}$ in the $(0,2)$ chiral
multiplets coming from 5d vector multiplet have $J=+\frac{1}{2}$.

If the 5d SYM has hypermultiplets, there are more zero modes in the instanton background,
which can be determined by an index theorem. They are fermionic degrees in the UV incomplete
supersymmetric sigma model, and no more normalizable bosonic zero modes appear. In
the UV complete ADHM gauged quantum mechanics, the degrees of freedom depend on the representation
of the 5d hypermultiplet in $G$, and also possibly on the string theory engineering. There
may be extra UV bosonic degrees in the ADHM quantum mechanics. In appendix A, we explain
a few examples of ADHM mechanical degrees coming from the 5d hypermultiplets. (Related discussions
can be found in \cite{Shadchin:2005mx} in the context of equivariant indices of these mechanical
systems.)
The 5d hypermultiplets induce mechanical hypermultiplets and/or Fermi multiplets.
The hypermultiplet in quantum mechanics takes the following form \cite{Douglas:1996uz},
\begin{equation}\label{hyper-hyper}
  (\Phi^A,\Psi_{\dot\alpha})\ ,\ \
  \bar{Q}^{A\dot\alpha}\Phi_B=\sqrt{2}\delta^A_B\Psi^{\dot\alpha}\ .
\end{equation}
Its representation under $G\times\hat{G}$ depends on the 5d hypermultiplet:
see appendix A. An important difference from the
hypermultiplets (\ref{ADHM-multiplet}) in the ADHM data is that the scalars in
(\ref{hyper-hyper}) are in the $({\bf 1},{\bf 2})$ representation  of
$SU(2)_r\times SU(2)_R$. These are called twisted hypermultiplets.
There could also be various
`Fermi multiplets,' whose on-shell degree is only a complex fermion $\Psi$.
The representation of $\Psi$ in $G\times\hat{G}$ depends on 5d hypermultiplets.
The $(0,2)$ decompositions of these multiplets are
\begin{equation}
  {\rm hyper}\ (\phi^A,\psi_{\dot\alpha})\rightarrow
  {\rm chiral}\ (\phi^2,\psi^{\dot{2}})\ +\
  {\rm chiral}\ (\bar\phi_1=-\bar\phi^{2},\bar\psi_{\dot{1}})\ ,
\end{equation}
and $\Psi$ reduces to a $(0,2)$ Fermi multiplet.
Again $\phi^2$, $\bar\phi_1\sim\bar\phi^2$ are chosen because
$Q$ acts on them nontrivially. The scalars have $J=-\frac{1}{2}$. The $(0,2)$
Fermi multiplets $\Psi$ come with holomorphic potentials $E_\Psi(\phi)$, $J_\Psi(\phi)$,
whose on-shell values are determined by $(0,4)$ SUSY. We shall explain them in appendix A.

Including all these degrees, it suffices for us to know the off-shell
$(0,2)$ action and SUSY to compute the index. Apart from the 1d Chern-Simons
term (\ref{QM-CS}), these can be obtained by a 1d reduction of a suitable 2d $(0,2)$
action and SUSY, e.g. given in the appendices of \cite{Benini:2013nda,Benini:2013xpa}.

\subsection{$(0,2)$ indices for instantons}

The full partition function of 5d SYM takes the form of $Z=Z_{\rm pert}Z_{\rm inst}$, where
$Z_{\rm pert}$ is the perturbative partition function which acquires contribution only
from the W-bosons and superpartners. $Z_{\rm inst}$ acquires contribution from instantons,
and takes the form of $Z_{\rm inst}=\sum_{k=0}^\infty Z_kq^k$ with $Z_0=1$. In the ADHM
quantum mechanics, we will study a Witten index $Z_{\rm QM}^k$ for $k$ instantons,
which counts states
preserving a pair of mutually Hermitian conjugate supercharges $Q,Q^\dag$, chosen
among $\bar{Q}^A_{\dot\alpha}$. The index $Z_{\rm QM}\equiv\sum_{k=0}^\infty Z_{\rm QM}^kq^k$ is
essentially $Z_{\rm inst}$, up to a possible multiplicative factor which we call $Z_{\rm extra}$
that depends on the string theory embedding of the QFT system. We explain $Z_{\rm extra}$ in
section 3.4 with examples. Following
\cite{Nekrasov:2002qd}, we choose $Q\equiv\bar{Q}^1_{\dot{1}}=-\bar{Q}^{1\dot{2}}$ and
$Q^\dag\equiv\bar{Q}^2_{\dot{2}}=\bar{Q}^{2\dot{1}}$. The index for the ADHM mechanics
is defined by
\begin{equation}\label{qm-index}
  Z_{\rm QM}^k(\epsilon_1,\epsilon_2,\alpha_i,z)={\rm Tr}\left[(-1)^Fe^{-\beta\{Q,Q^\dag\}}
  e^{-\epsilon_1(J_1+J_R)}e^{-\epsilon_2(J_2+J_R)}e^{-\alpha_i\Pi_i}e^{-z\cdot F}\right]\ .
\end{equation}
$J_1,J_2$ are the Cartans of $SO(4)$, rotating two orthogonal 2-planes of $\mathbb{R}^4$.
In the $SU(2)_l\times SU(2)_r$ notation, the two Cartans $J_l,J_r$ are given by
$J_l=\frac{J_1-J_2}{2}$, $J_r=\frac{J_1+J_2}{2}$. $J_R$ is the Cartan
of the $SU(2)_R$ R-symmetry. $\alpha_i$ is the chemical potential for the electric charges
in the Coulomb branch, where $i$ runs from $1$ to the rank of $G$. All other flavor
symmetries are collectively called $F$, conjugate to the chemical potential which we
call $z$. $\beta$ is the standard regulator of the Witten index, which does not appear in
$Z_{\rm QM}^k$. $Z_{\rm QM}^k$ admits a supersymmetric path integral representation, in which
the time direction is compactified with circumference $\beta$.

The trace ${\rm Tr}$ is taken over the Hilbert space, acquiring contributions from the states
annihilated by $Q,Q^\dag$ only. Since this sector is often attached to the continuum, with which
the computation of the index is very tricky, we comment on how we compute this index. Firstly,
there could be continua coming from the non-compact
hypermultiplet scalars. These flat directions can be lifted by turning on generic $\alpha_i$,
$z$, $\epsilon_{1,2}$ which effectively provide masses to these fields. We may keep these
chemical potentials and study the spectrum of this system: for instance, the electric charge
chemical potentials $\alpha_i$ can be treated this way, in the Coulomb branch of the 5d QFT.
However, some parameters can also be regarded as IR regulators.
For instance, $\epsilon_{1,2}$ are conjugate to the angular momentum of particles on $\mathbb{R}^4$,
which one may want to turn off while studying the internal degeneracy of single particle states.
To study the last degeneracy, one first computes from (\ref{qm-index}) the multiparticle index
$Z_{\rm QM}=\sum_{k=0}^\infty Z_{\rm QM}^k$ at finite $\epsilon_{1,2}$. Then one computes the
single particle index $f(q,\epsilon_{1,2},\alpha,z)$ using the relation $Z_{\rm QM}=
\exp\left[\sum_{n=1}^\infty\frac{1}{n}f(q^n,n\epsilon_{1,2},n\alpha,nz)\right]$. $f$ takes the
form of $f=\frac{(\rm numerator)}{4\sinh\frac{\epsilon_1}{2}\sinh\frac{\epsilon_2}{2}}$,
where the numerator has a finite limit $\epsilon_{1,2}\rightarrow 0$. The denominator comes from
the center-of-mass zero modes of a single particle, which causes infra-red divergence of the
multiparticle path integral at $\epsilon_{1,2}=0$. So extracting the numerator of $f$, we finally
obtain the IR finite single particle index. (However, it is completely fine to keep $\epsilon_{1,2}$
in $f$, to capture the spin quantum numbers.) Other chemical potentials $z$ can be treated in either
sense, depending on one's problem. We would rather explain how we treat them in a case-by-case
manner later. Here, we should really stress that the way one treats the chemical potentials
depends very much on one's final observable of interest. For instance, in the
partition functions \cite{Kim:2012gu,Lockhart:2012vp,Kim:2012qf,Kim:2013nva,Qiu:2013aga,Nieri:2013yra}
on curved 5-manifolds, the index (\ref{qm-index}) is used as building blocks where the Coulomb VEV
$\alpha$ is integrated over.

Even after turning on all possible chemical potentials, there still exists a continuum
attached to our BPS sector, coming from the vector multiplet scalar $\varphi$. This continuum
is not lifted by chemical potentials because $\varphi$ is neutral in all global symmetries.
If the gauge group $\hat{G}$ is a product of $U(n)$ factors, turning on nonzero Fayet-Iliopoulous
(FI) parameters to give masses to these fields, eliminating the continuum. In this
setting, the Witten index (\ref{qm-index}) can be computed without a continuum.
When $\hat{G}$ contains other non-Abelian groups, the continuum cannot be lifted
this way. In this case, the computation with a continuum is very tricky, and we would like to
sketch how we shall obtain the correct index in section 3, for the instanton quantum mechanics.

Had there been no continuum, the Witten indices for gapped theories generically do not change as
one changes the continuous parameters of the theory, such as the coupling $g_{QM}^2$, and also
as one changes the regulator $\beta$. For instance, it would be easiest to compute such
indices in the weak coupling regime. In our mechanical model, $g_{QM}^2$ has dimension of (time)$^{-3}$,
so the weak coupling regime of our path integral is defined by $g_{QM}^2\beta^3\ll 1$.
If there is a continuum attached to the BPS sector, the index counting BPS states is defined at
$\beta\rightarrow\infty$, and at the actual value of $g_{QM}^2$ that one is interested in. Changing
their values would yield a change in the index due to an extra contribution from the attached continuum.
For instance, the index may even have non-integral coefficients due to the continuum's
contribution.

In our ADHM quantum mechanics, we are interested in the regime with `large coupling'
$g_{QM}$ in which the instanton dynamics decouples from the extra degrees in the UV description
(including $\varphi$ which causes the continuum). If there is a specific energy
scale $E$ in one's observable, the decoupling regime would be $g_{QM}^2\gg E^3$. In our problem,
$E$ is identified as the energy scale of our BPS states which is proportional to other mass scales
(like $g_{\rm YM}^{-2}$ or the Coulomb VEV of 5d SYM), but is independent of $g_{QM}^{\frac{2}{3}}$
which is much larger than the mass parameters appearing in $E$. So keeping
the decoupling condition $g_{QM}^2\gg E^3$ obeyed, we continuously decrease $g_{QM}^2\beta^3$
from $\infty$ to $0$, by changing $\beta$. In the last limit $g_{QM}^2\beta^3\rightarrow 0$, one can
essentially do the weak-coupling computation, which one casually writes as ``$g_{QM}\rightarrow 0$.''
However, note that the change of $g_{QM}^2\beta^3$ can be made strictly within the decoupling regime
$g_{QM}^2\gg E^3$, only changing $\beta$ from $\infty$ to $0$. So the index computed this way will
exhibit a factorization $Z_{\rm QM}=Z_{\rm QFT}Z_{\rm extra}$ due to decoupling.
Now as one moves from the regime $g_{QM}^2\beta^3\gg 1$ where the BPS index is
defined, to the regime $g_{QM}^2\beta^3\ll 1$ in which computation is easy, the Coulomb branch
continuum will provide extra contribution to the index. However, since the index can be
computed strictly within the decoupling limit, the unwanted change of the index will only affect
$Z_{\rm extra}$ and not $Z_{\rm QFT}$. (Note that the decoupled QFT sector is expected not
to have a continuum at all, after turning on all possible chemical potentials.)
So supposing that one can identify $Z_{\rm extra}$ factor correctly, one can obtain
$Z_{\rm QFT}$ which is unaffected by the change of $g_{YM}^2\beta^3$.
Identifying $Z_{\rm extra}$ is not always straightforward, but in many problems this is reasonably
easy. In particular, understanding the symmetries and a little bit of bulk dynamics, one can often
deduce the chemical potential dependence of $Z_{\rm extra}$. Just with this knowledge, one can
often unambiguously factor out this bulk contribution $Z_{\rm extra}$ with fractional coefficients,
without laboriously computing the continuum contribution. See section 3 for how this can be
done in many examples.

Now with these understood, all the dimensionful parameters below will be made dimensionless
by multiplying suitable powers of $\beta$. So for instance, the limit ``$g_{QM}^2\rightarrow 0$''
will actually mean $g_{QM}^2\beta^3\rightarrow 0$, reached within the decoupling
limit of the system.

The index (\ref{qm-index}) can be regarded formally as a circle reduction of the
elliptic genus partition function of a 2d $\mathcal{N}=(0,4)$ gauge theory on $T^2$.
The multiplets that we explained in section 2.1 uplift to those in 2d, where
we uplift $A_t,\varphi$ to the 2d vector potential. In particular, our index
(\ref{qm-index}) has the same structure as the 2d $(0,2)$
elliptic genus, regarding $Q\equiv-\bar{Q}^{1\dot{2}},Q^\dag$ as the $\mathcal{N}=(0,2)$
SUSY. So $Z^k_{\rm QM}$ can be computed closely following \cite{Benini:2013nda,Benini:2013xpa}.
Following \cite{Benini:2013nda}, we shall
first present the `naive' computation, highlight the subtlety and then explain the proper
derivation. The index does not depend on continuous parameters of the theory preserving
$Q,Q^\dag$, in the sense explained above, and also with extra caveats explained below which
has to do with the FI parameters. So we can tune these parameters to the values
which ease the computation. The action is multiplied by the gauge coupling
$\frac{1}{g_{QM}^2}$. We replace $\frac{1}{g_{QM}^2}$ by
$\frac{1}{e^2}$ for the gauge kinetic term, and by $\frac{1}{g^2}$ for the matter kinetic term,
and send $e,g\rightarrow 0$, following \cite{Benini:2013nda,Benini:2013xpa}. (As explained above,
this is in fact $\beta\rightarrow 0$ limit with dimensionful $e,g$ kept sufficiently large
for decoupling.) It naively appears that this will yield a steep Gaussian integral
around zero modes. So one first identifies all the zero modes, and keep them fixed and integrate
over the non-zero modes first. After that one should integrate over the zero modes exactly.

The zero modes are given by the holonomy of the gauge field $A_\tau$ on the temporal circle
$S^1$, and the scalar $\varphi$ in the vector multiplet. Here and below, we multiply a
suitable power of $\beta$ to these variables, as well as all other variables appearing in
the index calculus, to make them dimensionless. In particular, the rescaled eigenvalues of
$A^I_\tau$ have period $2\pi$. So there are $r$ complex eigenvalues
$\phi^I=\varphi^I+iA_\tau^I$ living on cylinders, where $r$ is the rank of $\hat{G}$. Note that
in \cite{Benini:2013nda,Benini:2013xpa}, the analogous zero modes for 2d gauge theories are the
holonomies $A_1+iA_2$, whose eigenvalues live on tori. Also, there are fermionic zero modes
from $\bar\lambda^{\dot{1}}_1$, $\bar\lambda^{\dot{2}}_2$.

We first explain the 1-loop determinant obtained by integrating over massive modes
\cite{Nekrasov:2002qd,Shadchin:2005mx,Benini:2013nda,Benini:2013xpa}. A $(0,2)$ chiral
multiplet $\Phi$ in representation $R_\Phi$ of $\hat{G}$ contributes to a factor of
\begin{equation}\label{chiral-1-loop}
  Z_{\Phi}=\prod_{\rho\in R_\Phi}\frac{1}{2\sinh
  \left(\frac{\rho(\phi)+J\epsilon_++F\cdot z}{2}\right)}\ .
\end{equation}
$\rho$ runs over the weights in $R_\Phi$,
$J=J_r+J_R$ is the sum of Cartans of $SU(2)_r\times SU(2)_R$, $F$ collectively denotes
the rest of global charges, in our case for $G$ and $SU(2)_l$, and possibly extra flavor
symmetries. $z$ also collectively denotes their chemical potentials,
$\alpha$, $\epsilon_-$, $m$. All charges in the arguments of $\sinh$'s are those of
the complex scalar of the chiral
multiplet. A $(0,2)$ Fermi multiplet $\Psi$ contributes to a factor of
\begin{equation}\label{fermi-1-loop}
  Z_{\Psi}=\prod_{\rho\in R_\Psi}2\sinh\left(\frac{\rho(\phi)+J\epsilon_++Fz}{2}\right)\ .
\end{equation}
In particular, the $(0,2)$ Fermi multiplet $\bar\lambda^1_{\dot{2}},\bar\lambda^2_{\dot{1}}$
from the $(0,4)$ vector multiplet yields
\begin{equation}\label{vector-1-loop}
  \prod_{\alpha\in{\rm adj}(\hat{G})}2\sinh\left(\frac{\alpha(\phi)+2\epsilon_+}{2}\right)
\end{equation}
where $\alpha$ is the weight of the adjoint representations of $\hat{G}$: it runs over
all roots as well as Cartans. Physically, these represent the complex ADHM constraints.
The contribution of the vector multiplet $A_t,\varphi,\bar\lambda^{1}_{\dot{1}},
\bar\lambda^2_{\dot{2}}$ takes the same form as that of a Fermi multiplet. Since
$\bar\lambda^{1}_{\dot{1}},\bar\lambda^2_{\dot{2}}$ carry charges $J=0$, $F=0$,
the determinant is given by
\begin{equation}
  Z_V=\prod_{\alpha\in{\rm root}}2\sinh\frac{\alpha(\phi)}{2}
  \prod_{I=1}^r\frac{d\phi_I}{2\pi i}\ .
\end{equation}
We multiplied the integral measure for $\phi_I$ in foresight.
The index formula that we will get is
\begin{equation}
  Z=\frac{1}{|W|}\oint e^{\kappa{\rm tr}(\phi)}Z_{\textrm{1-loop}}=
  \frac{1}{|W|}\oint e^{\kappa{\rm tr}(\phi)}Z_V\prod_\Phi Z_\Phi\prod_\Psi Z_\Psi\ ,
\end{equation}
with the contour to be derived below. $W$ is the Weyl group of $\hat{G}$.
(\ref{chiral-1-loop}), (\ref{fermi-1-loop}), (\ref{vector-1-loop}) can be obtained
by taking the `$q\rightarrow 0$ limit' of (2.13), (2.14), (2.15) in \cite{Benini:2013xpa},
ignoring the overall `vacuum energy' factor of the form $q^\#$.\footnote{Zero modes are
related by $\phi_{\rm ours}=2\pi iu_{\rm theirs}$. Also, we multiplied $-1$ to their (2.13),
and multiplied $(-1)^r$ to their (2.15). As $(0,2)$
chiral multiplets are paired in our $(0,4)$ systems, the first $-1$ is invisible at least in
our examples. The $(-1)^r$ factor may produce an overall sign difference with the final result
of \cite{Benini:2013nda,Benini:2013xpa} for odd $r$, but this will make various formulae for
instantons simpler.  Anyway, there are often extra overall signs descending from the 5d SYM,
so the full sign issue cannot be answered within mechanics. We shall just illustrate with
various examples in section 3 what the overall signs should be.} When the gauge group $\hat{G}$ is disconnected,
one should turn on discrete holonomies which change the above formula. We shall encounter
them in our example with $Sp(N)$ instantons, as $\hat{G}=O(k)$ is disconnected. We shall
explain the effect of discrete holonomies in sections 3.2 with this example, following
\cite{Kim:2012gu}.

There are various dangerous regions in the zero mode space, some analogous to those of
the 2d index \cite{Benini:2013nda,Benini:2013xpa} and some being intrinsic to the quantum
mechanics. Let us explain the latter first, and then the former.
After integrating over non-zero modes, we should perform an integral over the
cylinders. The integral is over a noncompact region.
For the $r$ eigenvalues $\phi_I$ ($I=1,2,\cdots,r$), we take
$-\Lambda_1\leq\varphi_I={\rm Re}(\phi_I)\leq\Lambda_2$, with large IR cut-offs
$\Lambda_1,\Lambda_2>0$. After we compute the integral over all the modes, we can take
these cutoffs back to infinity. As we shall see, the limit $\Lambda_1,\Lambda_2\rightarrow\infty$
will not be singular, but may sometimes leave boundary contributions which have interesting
interpretations.

Now we discuss the second kind of dangerous regions in the zero mode space,
whose structure is similar to those of \cite{Benini:2013nda,Benini:2013xpa} in 2d.
It is legitimate to perform the Gaussian integration over the nonzero modes with fixed
$\phi=\varphi+iA_\tau$ only when the non-zero modes carry large masses after the deformation
$e,g\rightarrow 0$. This assumption fails near the points $\phi=\phi_\ast$ where
the integrand diverges. These dangerous points are provided by the poles from
the chiral multiplet determinant, at $\rho(\phi_\ast)+J\epsilon_++Fz=0$. Thus the
path integral should be done more carefully near $\phi_\ast$. Following
\cite{Benini:2013nda,Benini:2013xpa}, we keep the zero modes of the auxiliary field
$D$ and carefully take the $e,g\rightarrow 0$ limit.
With given nonzero $D$, the only determinant that changes is $Z_\Phi$, which is
\begin{equation}\label{chiral-D}
  Z_\Phi(\phi,\epsilon_+,z,D)=\prod_{\rho\in R_\Phi}\prod_{n=-\infty}^\infty
  \frac{-2\pi in+\rho(\bar\phi)+J\bar\epsilon_++F\bar{z}}
  {|2\pi i n+\rho(\phi)+J\epsilon_++Fz|^2+i\rho(D)}\ .
\end{equation}
The calculus which takes into account these dangerous regions was done (for 2d theories)
in \cite{Benini:2013nda} when the gauge group has rank $1$, and then in \cite{Benini:2013xpa}
for the gauge group with general rank. In this subsection, we repeat the analysis of
\cite{Benini:2013nda} for the rank $1$ case in our quantum mechanics version,
taking care of small differences in 1d. In the next subsection we obtain the result
with general rank. At rank $1$, all weights are replaced by the charge $Q_i$ of the $i$'th mode.

We take $e,g$ small but finite, and keep also the $D$ zero modes. The path integral
then reduces to an integral of the form
\begin{equation}
  Z=\int_{\mathbb{R}} dD\int_M d^2\phi f_{e,g}(\phi,\bar\phi,D)
  \exp\left(-\frac{D^2}{2e^2}-i\zeta D\right)\ .
\end{equation}
$\zeta$ is the Fayet-Iliopoulos parameter that one can turn on if $\hat{G}$ contains
$U(1)$ factors. $M$ is the space of the zero mode $\phi$,
$f_{\rm e,g}$ is obtained by the path integral except $\phi,D$ (and also integrated over
the zero modes of vector multiplet fermion). As in \cite{Benini:2013nda}, the
$g\rightarrow 0$ limit exists for any $\phi$, as long as $e$ is not zero, since
the coupling of chiral multiplet scalars with $D$ yields a potential
of the form $e^2(|\phi|^2-\zeta)^2$. Now before sending the $e\rightarrow 0$ limit, we identify
the dangerous regions of the $\phi$ integral following \cite{Benini:2013nda}. We call $M_{\rm sing}$
the set of all pole locations $\phi_\ast$, and let $\Delta_\varepsilon$ with small $\varepsilon$
to be the $\varepsilon$-neighborhood of $M_{\rm sing}$. We divide the integral over $\phi$ as
\begin{equation}
  \int_{M\setminus\Delta_{\varepsilon}}d^2\phi+\int_{\Delta_\varepsilon}d^2\phi\ .
\end{equation}
Following \cite{Benini:2013nda}, we take the $e\rightarrow 0$ limit in a way that
the second integration in $\Delta_\varepsilon$ does not contribute. This can be done
if $\varepsilon$ is sent to zero much faster than $e^\#$ with a positive number $\#$,
so that the small volume factor
of $\Delta_\varepsilon$ dominates over the divergent behavior as $e\rightarrow 0$. See
section 3.2.1 of \cite{Benini:2013nda} for the precise scaling of the limit
$e,\varepsilon\rightarrow 0$.

Thus we are left with
\begin{equation}
  Z=\lim_{e,\varepsilon\rightarrow 0}\int_{\mathbb{R}}dD\int_{M\setminus\Delta_\varepsilon}
  d^2\phi\ f_{\rm e}(\phi,\bar\phi,D)\exp\left(-\frac{D^2}{2e^2}-i\zeta D\right)\ .
\end{equation}
The factor $f_{\rm e}=\lim_{g\rightarrow 0}f_{\rm e,g}(\phi,\bar\phi,D)$ is given
in the $e\rightarrow 0$ limit by \cite{Benini:2013nda}
\begin{eqnarray}
  f_{\rm e}(\phi,\bar\phi,D)&\stackrel{e\rightarrow 0}{\longrightarrow}&
  \int d\lambda_0 d\bar\lambda_0\left\langle\int d^2x\lambda\sum_iQ_i\bar\Psi_i\phi_i
  \int d^2x\bar\lambda\sum_iQ_i\psi_i\bar\phi_o\right\rangle_{\rm free}\nonumber\\
  &=&h(\phi,\bar\phi,D,\epsilon_+,z)g(\phi,\bar\phi,D,\epsilon_+,z)\ ,
\end{eqnarray}
where $\lambda_0$ is the zero mode of the $(0,2)$ gaugino
($\bar\lambda^1_{\dot{1}},\bar\lambda^2_{\dot{2}}$ in our ADHM mechanics),
\begin{equation}
  g(\phi,\bar\phi,D,\epsilon_+,z)=e^{\kappa{\rm tr(\phi)}}Z_{\rm v}(\phi)\prod_\Phi
  Z_{\Phi}(\phi,\bar\phi,D,\epsilon_+,z)\prod_{\Psi}Z_\Psi(\phi,\epsilon_+,z)
\end{equation}
and
\begin{equation}
  h(\phi,\bar\phi,D,\epsilon_+,z)=c\sum_{i,n}
  \frac{Q_i^2}{\left(|2\pi in+Q_i\phi+J\epsilon_++Fz|^2+iQ_iD\right)
  \left(-2\pi in+Q_i\bar\phi+J\bar\epsilon_++F\bar{z}\right)}\ .
\end{equation}
The constant $c$ is taken to be $c=i/\pi$ following \cite{Benini:2013nda}, with
comments in the footnote 5 on the sign in mind. Here, one can show that
\begin{equation}\label{total-derivative}
  h(\phi,\bar\phi,D,\epsilon_+,z)g(\phi,\bar\phi,D,\epsilon_+,z)=
  \frac{1}{\pi D}\frac{\partial}{\partial\bar\phi}g(\phi,\bar\phi,D,\epsilon_+,z)\ ,
\end{equation}
again following \cite{Benini:2013nda}.

Now consider the integral over $D$ along $\mathbb{R}$.  For convenience, we deform
the contour slightly away from the real line \cite{Benini:2013nda}, to one of
the following two contours $\Gamma_\pm$, $D\in\mathbb{R}+i\delta$ with
$0<\pm\delta<\varepsilon^2$. Our final result will be independent of the sign
of $\delta$. So using (\ref{total-derivative}), one obtains
\begin{equation}\label{partition-integral}
  Z=-\lim_{e,\varepsilon\rightarrow 0}\int_{\Gamma_\pm} \frac{dD}{2\pi iD}
  \exp\left(-\frac{D^2}{2e^2}-i\zeta D\right)\oint_{\partial\Delta_\varepsilon+\partial M_0
  +\partial M_{\infty}}\frac{d\phi}{2\pi i}\ g(\phi,D,\epsilon_+,z)\ ,
\end{equation}
where $\Gamma_\pm$ is the deformed $D$-contour with positive/negative $\delta$,
respectively. We also used $\partial(M\setminus\Delta_\varepsilon)=-\partial
\Delta_\varepsilon-\partial M_0-\partial M_{\infty}$, where $\partial M_{0,\infty}$
are the boundary of the cylindrical region $M$ at $\varphi=-\Lambda_1,\Lambda_2$.
The orientations are all counterclockwise around the poles for $\partial\Delta_\varepsilon$,
around $z=e^{\phi}=0$ for $\partial M_0$, and around $w=e^{-\phi}=0$ for $\partial M_{\infty}$.
The poles of $D$ in the measure are located at $D=0$ and
\begin{equation}
  D=\frac{i}{Q_i}\left|2\pi in+Q_i\phi+J\epsilon_++Fz\right|^2\ .
\end{equation}
Let us consider the above integral expression for each patch of $\partial\Delta_\varepsilon
+\partial M_0+\partial M_{\infty}$. The patch can either surround a pole $\phi_\ast$, or
can be at $\varphi=-\Lambda_1,\Lambda_2$. We first consider the former. With small $\varepsilon$,
the pole for $D$ (apart from $D=0$) which is closest to the real axis is located at
$D=iQ_i\varepsilon^2$. This pole approaches the real axis of the $D$ plane in the
$e,\varepsilon\rightarrow 0$ limit. If the sign of $Q_i$ is opposite to the sign of
the contour shift parameter $\delta$,
then the closest pole approaching the real axis does not hit the $D$ integration contour.
If the two signs are the same, the pole $D=iQ_i\varepsilon^2$ will cross the contour.
Let us denote by $\partial\Delta_\varepsilon^{(+)}$ the union of small contours
surrounding poles with $Q_i>0$, and by $\partial\Delta_\varepsilon^{(-)}$ the union of
small contours surrounding poles with $Q_i<0$. Firstly, consider the case with
$\phi\in\partial\Delta_\varepsilon^{(\mp)}$ and $D\in\Gamma_\pm$ (upper/lower signs correlated).
Then there are no poles, including $D=iQ_i\varepsilon^2$, which cross $\Gamma_\pm$ as
$\varepsilon\rightarrow 0$. So we can relax the condition $|\delta|<\varepsilon^2$ and keep
$\delta={\rm Im}(D)$ fixed as $\varepsilon\rightarrow 0$. So one obtains
\begin{equation}\label{vanishing}
  -\lim_{e,\varepsilon\rightarrow 0}\int_{\Gamma_{\pm}} \frac{dD}{2\pi iD}
  \exp\left(-\frac{D^2}{2e^2}-i\zeta D\right)\oint_{\partial\Delta_\varepsilon^{(\mp)}}
  \frac{d\phi}{2\pi i}\ g(\phi,D,\epsilon_+,z)=0\ ,
\end{equation}
since $g(\phi,D,\epsilon_+,z)$ with given nonzero $D$ is a bounded integrand, while the
$d\phi$ integral region shrinks as $\varepsilon\rightarrow 0$. On the other hand,
consider the case with $\phi\in\partial\Delta_\varepsilon^{(\pm)}$ and $D\in\Gamma_\pm$.
The poles $D=iQ_i\varepsilon^2$ approach to zero in the $\varepsilon\rightarrow 0$ limit
and cross the contour $\Gamma_\pm$. To compute the integral avoiding these crossings,
we deform the contour $\Gamma_\pm$ to $\Gamma_\pm=\Gamma_\mp\mp C_0$, where $C_0$ is
a small contour surrounding $D=0$ counterclockwise \cite{Benini:2013nda}. Following the
similar argument which led to (\ref{vanishing}), one obtains
\begin{eqnarray}
  &&-\lim_{e,\varepsilon\rightarrow 0}\int_{\Gamma_{\pm}} \frac{dD}{2\pi iD}
  \exp\left(-\frac{D^2}{2e^2}-i\zeta D\right)\oint_{\partial\Delta_\varepsilon^{(\pm)}}
  \frac{d\phi}{2\pi i}\ g(\phi,D,\epsilon_+,z)\\
  &&=\pm\lim_{e,\varepsilon\rightarrow 0}\int_{C_0} \frac{dD}{2\pi iD}
  \exp\left(-\frac{D^2}{2e^2}-i\zeta D\right)\oint_{\partial\Delta_\varepsilon^{(\pm)}}
  \frac{d\phi}{2\pi i}\ g(\phi,D,\epsilon_+,z)=
  \pm\oint_{\partial\Delta_\varepsilon^{(\pm)}}
  \frac{d\phi}{2\pi i}\ g(\phi,\epsilon_+,z)\nonumber
\end{eqnarray}
where $g(\phi,\epsilon_+,z)\equiv g(\phi,D=0,\epsilon_+,z)$ is the holomorphic
integrand in $\phi$. Thus the $\partial\Delta_\varepsilon$ part of the integral
(\ref{partition-integral}), which we call $Z^\pm_{\partial\Delta_\varepsilon}$ depending
on the $D$ contour choice $\Gamma_\pm$, is given by
\begin{equation}
  Z_{\partial\Delta_\varepsilon}^+=\sum_{Q_i>0}R_i\ \ \ ,\ \ \ Z_{\partial\Delta_\varepsilon}^-=-\sum_{Q_i<0}R_i\ .
\end{equation}
$R_i$ is the residue of $g(\phi,\epsilon_+,z)$ at the $i$'th pole.

\begin{figure}[t!]
  \begin{center}
    \includegraphics[width=6cm]{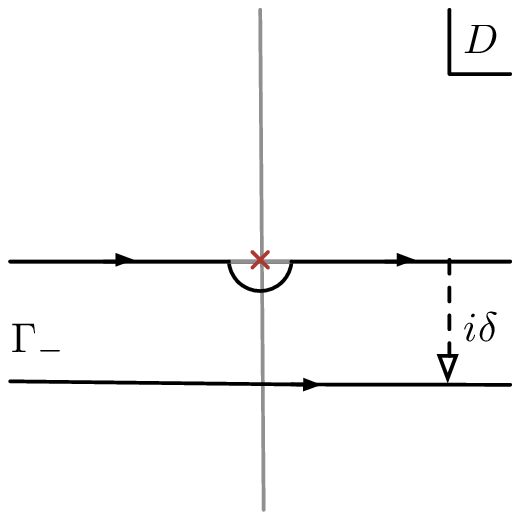}\hspace{1.5cm}
    \includegraphics[width=6cm]{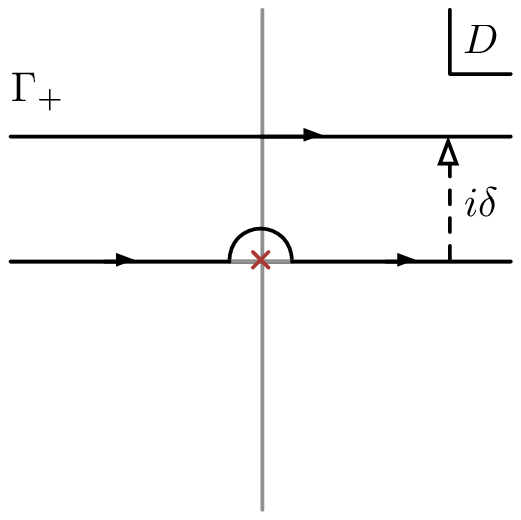}
\caption{Deformations of the contours $\Gamma_\pm$ to compute $f_\pm(0)$}\label{Gamma-deform}
  \end{center}
\end{figure}

We finally consider the contribution to (\ref{partition-integral}) from the patch
$\phi\in\partial M_0+\partial M_\infty$. If $\phi$ is on one of these contours
with large IR cutoffs $\Lambda_1,\Lambda_2$, $\varphi$ is very large so that
$D$ appearing in the denominator of (\ref{chiral-D}) is negligible. So we
replace $g(\phi,D,\epsilon_+,z)$ in (\ref{partition-integral}) by
holomorphic $g(\phi,\epsilon_+,z)=g(\phi,D=0,\epsilon_+,z)$ and obtain
\begin{equation}
  \oint_{\partial M_0+\partial M_\infty}\frac{d\phi}{2\pi i}g(\phi,\epsilon_+,z)
  =R_0+R_\infty\ ,
\end{equation}
where $R_0$ and $R_\infty$ are the residues of $\frac{dz}{z}g(\phi,\epsilon_+,z)$ in
$z\equiv e^\phi=0$ and $w\equiv z^{-1}=0$, respectively. The contribution to
the partition function (\ref{partition-integral}) is given by
\begin{equation}
  Z^\pm_{\partial M_0+\partial M_\infty}=-(R_0+R_\infty)\lim_{e,\varepsilon\rightarrow 0}
  \int_{\Gamma_\pm}\frac{dD}{2\pi iD}\exp\left(-\frac{D^2}{2e^2}-i\zeta D\right)
  \equiv -(R_0+R_\infty)\lim_{e,\varepsilon\rightarrow 0}f_\pm(e\zeta)\ .
\end{equation}
The function $f_\pm(e\zeta)\equiv\int_{\Gamma_\pm}\frac{dD}{2\pi iD}\exp
\left(-\frac{D^2}{2e^2}-i\zeta D\right)$ can be computed as follows. First consider
\begin{equation}
  \frac{df_\pm}{d\zeta}=-\frac{1}{2\pi}\int_{\Gamma_\pm}dD\
  \exp\left(-\frac{D^2}{2e^2}-i\zeta D\right)=-\sqrt{\frac{e^2}{2\pi}}
  \exp\left(-\frac{\zeta^2 e^2}{2}\right)\ ,
\end{equation}
which is the same for both contour choices $\Gamma_\pm$. So both $f_\pm(e\zeta)$
are given in terms of the error function ${\rm erf}(x)$ defined by
\begin{equation}
  {\rm erf}(x)=\frac{2}{\sqrt{\pi}}\int_0^x dt\ e^{-t^2}\ .
\end{equation}
Namely, one obtains
\begin{equation}
  f_\pm(e\zeta)=f_\pm(0)-\frac{1}{2}{\rm erf}\left(\frac{e\zeta}{\sqrt{2}}\right)\ .
\end{equation}
$f_\pm(0)$ can be obtained from the definition of $f_\pm$,
\begin{equation}
  f_\pm(0)=\int_{\Gamma_\pm} \frac{dD}{2\pi iD}\ \exp\left(-\frac{D^2}{2e^2}\right)=
  \mp\frac{1}{2}+\int_{-\infty}^\infty dD\ {\rm Pr}\left[\frac{1}{D}e^{-\frac{D^2}{2e^2}}\right]
  =\mp\frac{1}{2}\ ,
\end{equation}
where ${\rm Pr}$ denotes the principal value. These computations are done after deforming the
contours $\Gamma_\pm$ to those which are almost at the real axis and go around the pole at
$D=0$ in two different ways, as shown in Fig.~\ref{Gamma-deform}.
So we find that
\begin{equation}
  f_\pm(e\zeta)=\mp\frac{1}{2}-\frac{1}{2}{\rm erf}\left(\frac{e\zeta}{\sqrt{2}}\right)\ .
\end{equation}
Note that these functions satisfy $f_-(e\zeta)-f_+(e\zeta)=1$.
The contribution to the partition function from the two residues at the infinities
of the cylinder is thus
\begin{equation}\label{infinity-contour}
  Z^\pm_{\partial M_0+\partial M_\infty}=
  \frac{1}{2}(R_0+R_\infty)\left[\pm 1+\lim_{e,\varepsilon\rightarrow 0}
  {\rm erf}\left(\frac{e\zeta}{\sqrt{2}}\right)\right]\ .
\end{equation}
Note that $Z^\pm_{\partial M_0+\partial M_\infty}$ is the only part of our computation
which depends on (the sign of) the FI parameter $\zeta$. In examples with no $U(1)$
factors in $\hat{G}$, such as $\hat{G}=Sp(N),SO(N)$, one might think that this part
is ambiguous. However, note that the impossibility of turning on nonzero $\zeta$ is
closely related to the Weyl symmetry of the last groups. The same Weyl symmetry would
map the poles $\phi=-\infty$ and $\infty$, which implies $R_0+R_\infty=0$. So whenever
one cannot turn on FI parameters, one would find $Z^\pm_{\partial M_0+\partial M_\infty}=0$.

Adding all the contributions $Z^\pm_{\partial\Delta_\varepsilon}$ and
$Z^\pm_{\partial M_0+\partial M_\infty}$, the index $Z$ is given by
\begin{eqnarray}
  Z^+&=&\sum_{Q_i>0}R_i+\frac{1}{2}(R_0+R_\infty)\left[1+\lim_{e,\varepsilon\rightarrow 0}
  {\rm erf}\left(\frac{e\zeta}{\sqrt{2}}\right)\right]\nonumber\\
  Z^-&=&-\sum_{Q_i<0}R_i-\frac{1}{2}(R_0+R_\infty)\left[1-\lim_{e,\varepsilon\rightarrow 0}
  {\rm erf}\left(\frac{e\zeta}{\sqrt{2}}\right)\right]\ .
\end{eqnarray}
The sign of $\delta$ does not affect the result because
\begin{equation}
  Z^+-Z^-=\sum_{Q_i>0}R_i+\sum_{Q_i<0}R_i+R_0+R_\infty=0\ ,
\end{equation}
where the last equation holds since the sum of all residues on the cylinder
is zero if one includes those at infinities. So from now on we call $Z=Z^+=Z^-$.
The FI term $\zeta$ \textit{does} affect the result. For instance, at $e\zeta=-\infty$,
$0$ and $+\infty$, one obtains
\begin{equation}
  Z(e\zeta=0)=\sum_{Q_i>0}R_i+\frac{1}{2}(R_0+R_\infty)\ ,\ \
  Z(e\zeta=\infty)=-\sum_{Q_i<0}R_i\ ,\ \
  Z(e\zeta=-\infty)=\sum_{Q_i>0}R_i\ .
\end{equation}
Since $\zeta$ is a parameter of the theory, the index can
depend on it only when there is a continuum contribution from the Coulomb
branch. In particular, $Z(e\zeta)$ depends continuously
on $e\zeta$, so that $Z(e\zeta)$ expanded in the fugacities cannot generally have
integral coefficients. This is also expected in general, with continuum contributions.
The point $\zeta=0$ is where the Coulomb branch with nonzero $\phi$ meets the Higgs branch.
Nonzero $\zeta$ generates a mass gap for the Coulomb branch degrees $\varphi$ with the mass
proportional to $e^2\zeta$, so that the continuum cannot affect the Witten index. Since we
computed the index in the $e\rightarrow 0$ limit, the above result as a function of finite
$e\zeta$ generates vanishing mass gap $e^2\zeta\rightarrow 0$. The indices with finite gaps
are thus obtained in the $e\zeta\rightarrow\pm\infty$ limit. Indeed in section 3.4, we
shall illustrate that $Z(e\zeta=\pm\infty)$ have integral coefficients, as
expected from general considerations.

To ease the analysis with higher rank in the next subsection, we rephrase the rank $1$
result as follows. The functions $f_\pm(e\zeta)$ satisfy
$f_\pm(e\zeta=\mp\infty)=0$.\footnote{It is easy to understand the vanishing of these
quantities. $\zeta$ dependence comes from $e^{-i\zeta D}=e^{\zeta\delta}e^{-i\zeta{\rm Re}(D)}$
in the integrand. Firstly, $e^{-i\zeta {\rm Re}(D)}$
always provides a large destructive interference in both $\zeta\rightarrow\pm\infty$ limits.
When $\zeta\delta <0$, the limit provides another small factor $e^{-|\zeta\delta|}$,
explaining $f_\pm(e\zeta=\mp\infty)=0$. However, when $\zeta\delta>0$,
$e^{\zeta\delta}$ provides a large factor which competes with the destructive interference,
balancing at $f_\pm(e\zeta=\pm\infty)=1$.} So when $\delta$
and $\zeta$ have opposite signs, the contribution (\ref{infinity-contour}) from the IR
cutoff contours vanishes. One thus obtains
\begin{equation}\label{rank-1-simple}
  Z(\zeta)={\rm sign}(-\zeta)\sum_{{\rm sign}(-\zeta)Q_i>0}R_i\ ,
\end{equation}
which does not refer to the contributions from residues at infinities.

The possible difference between
the Witten indices in two limits $e\zeta=\pm\infty$,
\begin{equation}
  Z(\zeta>0)-Z(\zeta<0)=R_0+R_\infty\ ,
\end{equation}
implies a wall crossing of the index across $\zeta=0$, at which
the system of our interest meets a continuum from the Coulomb branch.

All discussions till here go through for general quantum mechanical index with $(0,2)$
SUSY. Now let us consider the instanton partition function with our $(0,4)$ mechanics.
We find that $\zeta$ dependence of the 5d SYM index is unacceptable. This is because
the $\zeta$ dependence incurs a strange dependence of $Z$ on $\epsilon_+$,
conjugate to the Cartan of the diagonal combination of $SU(2)_r\times SU(2)_R$. Nekrasov's
partition function counts half-BPS states of the 5d $\mathcal{N}=1$ gauge theory,
preserving $4$ Hermitian supercharges. Their short multiplets break
neither $SU(2)_r$ nor $SU(2)_R$ symmetry. So although our index only refers to $2$ SUSY,
it should be an even function of $\epsilon_+$ if it only captures the spectrum of these
half-BPS states with a further refinement with $\epsilon_+$. Since both $\zeta,\epsilon_+$
break $SU(2)_R$, the non-invariance of the partition function in the sign flip of $\zeta$
effectively induces the non-invariance under the sign flip of $\epsilon_+$. So if
$Z(\zeta\lessgtr 0)$ are same, it is an even function of
$\epsilon_+$. However, if $R_0+R_\infty\neq 0$, this measures the failure of $Z(\zeta>0)$
and $Z(\zeta<0)$ being even in $\epsilon_+$. Thus, $\zeta$ dependence and
$\epsilon_+\rightarrow-\epsilon_+$ asymmetry is unreasonable if this index is counting
half-BPS bound states of instantons and W-bosons in the 5d super-Yang-Mills theory.
In 5d SCFTs, this is also a consequence of the superconformal symmetry which contains
$SU(2)_R$.

To find the resolution to this puzzle, we should understand the true nature of $Z$.
The possible wall crossing happens because the Coulomb branch continuum appears
at $\zeta=0$. From our ADHM quantum mechanics, the Coulomb branch degrees appear only
when we go to the UV complete gauge theory description of the instanton quantum mechanics.
So there may appear contribution to the index from the extra UV degrees of the ADHM
quantum mechanics which do not belong to the QFT Hilbert space.
This can either be a fractional contribution from the continuum, or
integral contribution coming from marginal bound states involving the
extra stringy states. In any case, these have to factorize with the
true QFT index,
\begin{equation}
  Z=Z_{\rm QFT}Z_{\rm extra}\ ,
\end{equation}
since the field theories that we shall study from string theory are obtained
by taking suitable decoupling limits. The half-BPS bounds of the 5d SYM is captured
by the $Z_{\rm QFT}$ factor.

Coming back to the $\zeta$ dependence and $\epsilon_+\rightarrow-\epsilon_+$ asymmetry,
the wall crossing of the index should happen only in $Z_{\rm extra}$, but not in
$Z_{\rm QFT}$. Note that the extra BPS bounds involving the bulk degrees are well defined
only with nonzero $\zeta$, so this sector does not need to respect the
$\epsilon_+\rightarrow-\epsilon_+$ symmetry. So as a resolution of the puzzle phrased above,
we claim that all the $SU(2)$ asymmetric should go to $Z_{\rm extra}$, and $Z_{\rm QFT}$
is invariant under it. This shall be supported with many examples in section 3.4.

When $R_0=R_\infty=0$ separately, the continuum from $\varphi$ is lifted by
quantum effects. See section 3 for classifications and examples. In this case, the index is
independent of the continuous parameters of the theory, and its coefficients are integers.
When $R_0+R_\infty=0$ but $R_0=-R_\infty\neq 0$, there is no wall crossing but
still is a continuum from $\varphi$. In this case, since a continuum is attached
to the Higgs branch, the index may have fractional coefficients. However,
these non-integral contributions will all go to $Z_{\rm extra}$, and not to $Z_{\rm QFT}$.

\subsection{$(0,2)$ indices with higher rank gauge groups}

Now we study the index with gauge symmetry $\hat{G}$ of general rank $r$.
In the $k$ instanton quantum mechanics, $r=k$ for $\hat{G}=U(k),Sp(k)$, while
$r=\left[\frac{k}{2}\right]$ for
$O(k)_+$ and $r=\left[\frac{k-1}{2}\right]$ for $O(k)_-$.

The path integral again can be computed in the same $e\rightarrow 0$ and $g\rightarrow 0$ limit.
Again the limit should be taken with care. We again start by
identifying the zero modes in the limit, do Gaussian integration over non-zero modes which
yield product of various $Z_{\rm v}$, $Z_{\Phi}$, $Z_{\Psi}$ factors, and then integrate
over the zero modes. The last integral becomes a multi-variable contour integral, determined
by a careful computation similar to the previous subsection. Following \cite{Benini:2013xpa},
we explain the general structure of the index for the connected group $\hat{G}$ in this section,
and explain examples with disconnected $O(k)$ group in section 3 following
\cite{Kim:2012gu}. The zero modes consist of $r$ holonomy eigenvalues of $A_\tau$ along
the temporal circle with circumference $\beta$, and also the $r$ eigenvalues of the scalar
$\varphi$ in the vector multiplet. The $r$ complex eigenvalues $\phi_I=\varphi_I+iA_\tau$
with $I=1,2,\cdots,r$  locally form $r$ copies of cylinders. More precisely, let
$h_{\mathbb{C}}$ be the complexified Cartan subalgebra and $Q^\vee$ the coroot lattice. Then defining
$M=h_{\mathbb{C}}/Q^\vee$, the space of zero modes is given by $M/W$, where $W$ is the Weyl
group. The `naive' integrand
\begin{equation}
  g(\phi_I,\epsilon_+,z)=Z_{\rm v}\prod_{\Phi}Z_\Phi\prod_\Psi Z_\Psi
\end{equation}
again diverges at various points $\phi=\phi_\ast$,
from the chiral multiplet factors $Z_\Phi$. Each chiral multiplet
$\Phi_i$ defines a hyperplane $H_i$ defined by $\rho_i(\phi)+J_i\epsilon_++F_iz=0$,
where $\rho_i$ is a weight vector in $R_{\Phi_i}$ representation of $\hat{G}$. Using the variables
$z_I\equiv e^{\phi_I}$, $r$ copies of cylinders map
to $\mathbb{C}^r$, where the two infinities of a cylinder map to the origin and
infinity of $\mathbb{C}$. Since the space of zero mode is again noncompact, we have to
introduce an IR cutoff for large $\varphi_I$ and then remove the regulator after the path integral.

We shall explain the expression for the higher rank index in terms of the residue sum
shortly. As in the rank $1$ index, one should perform a careful path integral near the above
hyperplanes (including infinities of the cylinders). Just as in the rank $1$ case,
one slightly shifts the integration contours for $r$ eigenvalues of $D$ field,
from $\mathbb{R}^r$ to $\mathbb{R}^r+i\delta$, where $\delta$ is an $r$ dimensional real
vector in $h$. In the case of 2d index, the expressions appearing after the different choices
of $\delta$ are apparently different, but they should turn out to be the same after taking into
account that suitable `sums of residues' are zero for a meromorphic function.
$\delta$ is constrained by fixing a vector $\eta\in h^\ast$ and demanding $\eta(\delta)>0$
\cite{Benini:2013xpa}, which was enough to specify the index as a suitable residue sum.
In quantum mechanics,  $\zeta$ dependence may appear due to the poles at infinities,
but we already know that there are simple choices of $\delta$
(and thus $\eta$) which provides an expression for the index without referring to the residues
at infinities. In (\ref{rank-1-simple}) and footnote 6, we have seen that the
contributions from infinities vanish when the direction of the contour shift and
the sign of $\zeta$ are correlated, $\zeta\delta<0$. In the higher rank case, the same
arguments can be given for $\zeta(\delta)<0$. So to realize the last condition,
we choose the vector $\eta$ to be $\eta=-\zeta$ in all expressions
below whenever necessary, and work without worrying about the residues at infinities
of the cylinder. When the gauge group $\hat{G}$ does not contain any overall
$U(1)$ factor, such as $Sp(k)$ or $O(k)$, the sum of two residues from the two infinities
$\rho(\varphi)\rightarrow\pm\infty$ of a cylinder always vanishes, from the Weyl symmetry
$\varphi_I\rightarrow-\varphi_I$ of $\hat{G}$. In this case the index will not depend on
the choice of $\eta$, just like the 2d index of \cite{Benini:2013xpa}.

The contour choice for the $r$ variables $\phi_I$ for our quantum mechanical index follows
from the same consideration as \cite{Benini:2013xpa}, assuming $\eta=-\zeta$ below whenever
necessary. \cite{Benini:2013xpa} phrased the `contour' by specifying the set of residues that
have to be summed over, with sign factors. The $r$ complex variables encounter
a `pole' when $n\geq r$ hyperplanes among $H_i$ meet at a point $\phi=\phi_\ast$. When
$n=r$, the intersection is called non-degenerate.
If $n>r$, it is called degenerate. Hyperplane arrangements are called projective when the $n$
charge vectors ${\bf Q}(\phi_\ast)\equiv\{Q_i|\phi\in H_i\}$ responsible for the pole
$\phi_\ast$ are contained in a half-space of $h^\ast$. The results of \cite{Benini:2013xpa}
apply straightforwardly when all hyperplane arrangements are projective. In all kinds
of instanton calculus from the $(0,4)$ systems, we find that the projective condition is
met. So without repeating the arguments of \cite{Benini:2013xpa}, we shall simply state the result.

Near the pole $\phi_\ast$, one can Laurent expand the integrand $Z_{\textrm{1-loop}}$, in
negative powers of $Q_i(\phi-\phi_\ast)$. The nonzero residues are obtained only from
the `simple pole' parts, which are linear combinations of the functions of the form
\begin{equation}
  \frac{1}{Q_{j_1}(\phi-\phi_\ast)\cdots Q_{j_r}(\phi-\phi_\ast)}\ .
\end{equation}
$Q_{j_1},\cdots,Q_{j_r}$ are chosen in ${\bf Q}(\phi_\ast)$. The so-called Jeffrey-Kirwan residue
JK-Res, that is relevant for writing down our index, also refers to the auxiliary vector
$\eta$. JK-Res is defined by \cite{Benini:2013xpa}
\begin{equation}\label{JK-Res}
 \textrm{ JK-Res}({\bf Q}_\ast,\eta)\frac{d\phi_1\wedge\cdots\wedge d\phi_r}
 {Q_{j_1}(\phi)\cdots Q_{j_r}(\phi)}=\left\{\begin{array}{ll}
 \left|\det(Q_{j_1},\cdots,Q_{j_r})\right|^{-1}&{\rm if}\
 \eta\in{\rm Cone}(Q_{j_1},\cdots,Q_{j_r})\\
 0&{\rm otherwise}\end{array}\right.\ .
\end{equation}
`Cone' denotes the cone spanned by the $r$ independent vectors. Namely,
$\eta\in{\rm Cone}(Q_1,\cdots,Q_r)$ if $\eta=\sum_{i=1}^r a_iQ_i$ with
positive coefficients $a_i$. Although this definition apparently looks over-determining
JK-Res as a linear functional, it is known to be consistent: see \cite{Benini:2013xpa}
and references therein.  As explained above, $\eta$ specifies
how the $D$ contour is deformed to the imaginary direction $+i\delta$ by
demanding $\eta(\delta)>0$. The index is given by \cite{Benini:2013xpa}
\begin{equation}
  Z=\frac{1}{|W|}\sum_{\phi_\ast}\textrm{JK-Res}({\bf Q}(\phi_\ast),\eta)
  \ Z_{\textrm{1-loop}}(\phi,\epsilon_+,z)\ .
\end{equation}
Note that the result depends on the choice of $\eta$ when the residue sum at the
two infinities of a cylinder (spanned by any $\rho(\phi)$ appearing in $Z_{\textrm{1-loop}}$)
does not vanish. In this case, the above expression is understood with $\eta=-\zeta$. So $Z$
in this case is a piecewise constant function of $\zeta$.

We note that, for non-Abelian gauge group $\hat{G}$, the co-vector $\zeta$ is restricted
to be along the overall $U(1)$ factors only. $\eta$ is chosen in \cite{Benini:2013xpa} not
to coincide with the weights $Q_i$ associated with the poles. (More generally $\eta$ is
taken not to lie at the boundary of the `chambers.' See \cite{Benini:2013xpa} for explanations.)
So one might think that it would be troublesome to impose $\eta=-\zeta$ if $\zeta$ is aligned
along the forbidden direction for $-\eta$, e.g. being proportional to a weight in the problem.
For Abelian theories,
such as $U(1)^r$ theories, $\zeta$ can be a generic vector in $h^\ast$ so that $\zeta$ on
the boundary of a chamber is potentially a wall-crossing point. $\zeta$ can be
displaced from such a value in the Abelian theories, and one obtains different results across
the wall by setting $\eta=-\zeta$ for this displaced $\zeta$. However, for $U(r)$, its
FI term is along a fixed direction on $h^\ast=\mathbb{R}^r$, proportional to $(1,1,\cdots,1)$.
This might be at the boundary of chambers. For instance, the weight
$(1,1,\cdots,1)$ appears in the rank $r$
totally symmetric representation of $U(r)$, or a totally antisymmetric representation of it.
If $\zeta$ is at the boundary, then one can remove the ambiguity by slightly shifting $\eta$
away from $-\zeta(1,1,\cdots,1)$. Different shifts may leave $\eta$ in different chambers.
However, since $-\zeta(1,\cdots,1)$ is a Weyl invariant point of $U(r)$,
these different chambers  map to one another by Weyl reflection. Due to this symmetry,
different shifts of $\eta$ should yield the same result.

In all examples that we study, we find that the Jeffrey-Kirwan residue rule is
equivalent to the following prescription, which is well known in the instanton calculus
for some theories. We shall perform the contour integral
over $r$ variables $\phi_I$ one by one. The contour integral takes the form of
\begin{equation}
  \frac{1}{|W|}\oint\prod_{I=1}^r\frac{d\phi_I}{2\pi i}\
  Z_{\rm v}(\phi,\epsilon_+,z)\prod_\Phi Z_\Phi(\phi,\alpha,\epsilon_+,z)
  \prod_\Psi Z_\Psi(\phi,\alpha,\epsilon_+,z)\ .
\end{equation}
The poles from a chiral multiplet factor takes the form
\begin{equation}\label{pole-chiral}
  \frac{1}{2\sinh\left(\frac{Q_i(\phi)+J_i\epsilon_++F_iz}{2}\right)}\sim
  \frac{1}{e^{Q_i(\phi)}-e^{-J_i\epsilon_+}e^{-F_iz}}\ ,
\end{equation}
where $Q_i$ is the charge vector of the chiral multiplet $\Phi_i$.
So the pole one picks up for the $z_I=e^{\phi_I}$ variables are determined
by $r$ different equations of the kind $e^{Q_i(\phi)}=(e^{-\epsilon_+})^{J_i}e^{-F_iz}$.
In the instanton calculus, the value of charge $J$ conjugate to $\epsilon_+$ is always
positive when the quantum mechanical chiral multiplet comes from the 5d SYM theory's
vector multiplet (namely, the ADHM degrees). On the other hand, one always finds that
$J<0$ for the mechanical chiral multiplet coming from 5d hypermultiplets. in $(0,4)$ language,
the sign of $J$ charge is different for hypermultiplet and twisted hypermultiplet.
At this point, we temporarily treat the $e^{-\epsilon_+}$ factors appearing in the
$Z_\Phi$'s from the 5d vector multiplet (1d hyper) and those from the 5d hypermultiplet
(1d twisted hyper) independently.
Namely, we substitute $e^{-\epsilon_+}\rightarrow t\ll 1$ for the measure coming from
5d vector multiplet, and $e^{-\epsilon_+}\rightarrow T\gg 1$ for the measure from 5d
hypermultiplet. In this setting, the pole for $e^{Q_i(\phi)}$ appearing in (\ref{pole-chiral})
is inside the unit circle.

With these understood, the alternative residue prescription which turns out to give the same
result is obtained by regarding each integral variable $z_I$ as living on the unit circle
on the complex plane. Then we integrate over these variables one by one, for which
we have to pick up poles inside the contour and sum over their residues (assuming $t\ll 1$,
$T\gg 1$).
After the residue sums of all $r$ integrals, we set $t,T$ back to the same value
$t=T=e^{-\epsilon_+}$. This yields the same result as the index obtained by the sum
of JK-Res. Of course, to see the agreement most clearly, one should choose $\eta$
carefully, related to the order of integral for $\phi_1,\phi_2,\cdots,\phi_r$ in our
alternative
prescription. Whenever one encounters a pole at $z_I=0$, one does not include their
residues, as part of our prescription. Also, one occasionally encounters poles which
are not clearly inside or outside the unit circle with $t\ll 1$, $T\gg 1$ only.
Here, one may choose other fugacities arbitrarily to shift such poles away from
the unit circle. The arbitrary shift will not affect the result, as we will illustrate.
So far, this is the prescription which works when the index has zero residue sums at
infinities over a cylinder. For some $U(k)$ instanton calculus for which the sign of
FI term matters, we chose
$\eta=-\zeta$ and summed over JK-Res. In our alternative prescription, $\zeta$ dependence
appears as a failure of the index to be an even function of $\epsilon_+$. So the two
different indices are obtained by either running through the above prescription as explained
above, or alternatively taking $t\gg 1$ and $T\ll 1$ temporarily and
going through the same unit circle contour prescription. This yields results which are related to
each other by flipping $\epsilon_+\rightarrow-\epsilon_+$, or equivalently $\zeta\rightarrow-\zeta$.

In the rank $1$ case, it is immediate that the alternative prescription yields the same result
as the result of section 2.2. This is because (\ref{pole-chiral}) for rank $1$ is given by
\begin{equation}
  \frac{1}{z^{Q_i}-t^{J_i}e^{-F_iz}}\ \ ,\ \ \ \frac{1}{z^{Q_i}-T^{J_i}e^{-F_iz}}
\end{equation}
for the chiral multiplet originating from 5d vector and hypermultiplet,
respectively. The rule in section 2.2 was to sum over the residues with $Q_i>0$.
The poles to be kept are
\begin{equation}
  z=t^{J_i/Q_i}e^{-F_i/Q_iz}\ ,\ \ z=T^{J_i/Q_i}e^{-F_i/Q_iz}\ \ \ \ \
  (Q_i>0)\ .  \label{poleprescript}
\end{equation}
With $t\ll 1,T\gg 1$, these are all inside the unit circle $|z|<1$,
since $J_i\gtrless 0$ respective for 1d
hypermultiplets and twisted hypermultiplets. So this agrees with our unit
circle contour prescription. In fact the temporary relaxation $e^{-\epsilon_+}\rightarrow t<1$
and $e^{-\epsilon_+}\rightarrow T>1$ can be understood as pushing all poles with nonzero JK-Res
inside the unit circles. So even for the higher rank case, this prescription is quite heuristic
but we are not aware of a general proof that the two are equivalent. We shall just provide
comparisons of the two rules with higher rank examples in sections 3. Although the final result
is the same, the latter prescription picks more poles and residues in the intermediate stage
compared to the JK-Res rule: the extra residues however all cancel out in pairs,
as explained in sections 3.1 and 3.3.

We seek for such an alternative rule because this is known and widely used
in the instanton calculus, starting from \cite{Nekrasov:2002qd}.
Comparisons of the two rules above in the next sections will thus rigorously justify the existing
prescriptions from the JK-Res rules. When there are subtleties due to the poles at the infinities
of the cylinders, the JK residue rule also justifies various vague steps of the existing
prescriptions. We also note that temporarily substituting $t<1$ and $T>1$ for the
vector/hypermultiplet poles is essentially the `$i\epsilon$' and `$-i\epsilon$'
prescriptions given to the vector and hypermultiplet poles, observed in \cite{Hollands:2011zc}.
We can rephrase our alternative prescription as picking all the poles/residues inside the unit
circles from the 5d vector multiplets, and picking all of them outside the unit circles from
the 5d hypermultiplets, assuming $e^{-\epsilon_+} \ll 1$.

\section{Examples}

Since 5d SYM is non-renormalizable, it is important to pick theories which are related to
consistent quantum systems. In this section, we explain various classes of 5d SYM theories
related to interesting QFTs. Some examples are explained in more detail in the subsections.

Among others, 5d SYMs and their partition functions could be useful as follows.
\begin{enumerate}
\vspace{-.35cm}
\item Compactifying the 5d theory on a small $S^1$, we can study the effective action
of 4d $\mathcal{N}=2$ SYM in their Coulomb phase from the instanton partition functions.
The theories in this class should be asymptotically free or conformal in 4d.

\item Some 5d SYM theories are relevant deformations of 5d SCFTs.
A class of such 5d Yang-Mills theories was studied in \cite{Intriligator:1997pq}.
In this, the 1-loop correction to the coupling matrix is nonzero, with non-negative
eigenvalues everywhere, in the Coulomb branch. Then one can take the bare coupling to
infinite, yielding a 5d SCFT at the origin of the Coulomb branch. They also admit string
theory engineerings. One setting is the M-theory on singular Calabi-Yau 3-folds. This is
often dual to type IIB $(p,q)$ 5-brane webs \cite{Aharony:1997bh}. D4-D8-O8 systems also
generate a class of 5d SCFTs \cite{oai:arXiv.org:hep-th/9608111}.

\item Some 5d SYM theories are obtained at low energy by circle compactification of
6d $(2,0)$ or $(1,0)$ superconformal field theories. A necessary condition for the 5d SYM
in this class is to have vanishing 1-loop correction to the coupling matrix in the Coulomb
branch.
\vspace{-.35cm}
\end{enumerate}
The first class has various examples. The second class is partly classified in
\cite{Intriligator:1997pq}. There are known examples in the third class, some of which
we explain in detail below. We now explain the above three cases in some detail.

In 4d, the non-positivity of the 1-loop beta function for the gauge group $G$ demands
\begin{equation}
  c_2(G)-\sum_iC(R_i)\geq 0\ ,
\end{equation}
where $i$ runs over hypermultiplets, $C(R)$ is given by ${\rm tr}_{R}(T^aT^b)=C(R)\delta^{ab}$.
For instance, $C({\rm adj}(G))=c_2(G)$ with  $c_2(G)$ being the dual Coxeter number, which are given by
$c_2(SU(N))=N$, $c_2(Sp(N))=N+1$, $c_2(SO(N))=N-2$. Also, $C({\rm fund})=\frac{1}{2}$
for fundamental representations of classical groups. When the gauge theory has
the simple gauge group $G=ABCD$, the allowed matter contents are
(we list the number of hypermultiplets in various representations):
\begin{enumerate}
\vspace{-.35cm}
\item $SU(N)$: $N_f\leq 2N$ fundamental; $1$ antisymmetric and $N_f\leq N+2$
fundamental; $1$ symmetric and $N_f\leq N-2$ fundamental; $1$ antisymmetric and $1$ symmetric; $2$ antisymmetric and $N_f\leq 4$ fundamental; $1$ adjoint.

\item $Sp(N)$: $N_f\leq N+2$ fundamental; $1$ antisymmetric and $N_f\leq 4$ fundamental;
$1$ adjoint.

\item $SO(N)$: $N_f\leq N-2$ fundamental; $1$ adjoint.
\vspace{-.35cm}
\end{enumerate}
When the number of fundamental hypermultiplets saturates the inequality,
or in case of $SU(N)$ group with $1$ symmetric and $1$ antisymmetric hypermultiplets,
1-loop beta function vanishes.

In \cite{Intriligator:1997pq}, the following gauge groups and the matter contents are
found for the 5d gauge theories which could admit nontrivial UV fixed points:
\begin{enumerate}
\vspace{-.35cm}
\item $Sp(N)$ theories can come with either $n_A=0,1$ antisymmetric hypermultiplet.
When $n_A=1$, there can be $N_f$ fundamental hypermultiplets with $0\leq N_f\leq 7$.
When $n_A=0$, there can be $N_f\leq 2N+4$ fundamental hypermultiplets. Exceptionally at
$Sp(1)$, the theories with $n_A=0$ are identical to theories with $n_A=1$,
so $N_f\leq 7$ is allowed.

\item $SU(N)$ theories can come with bare Chern-Simons term at level $\kappa$.
If the theory has $N_f$ fundamental hypermultiplets, $\kappa$ is integral if $N_f$ is
even, while $\kappa$ is half an odd integer if $N_f$ is odd. 5d UV fixed point exists
if $N_f+2|\kappa|\leq 2N$. When $N\leq 8$, one can have $1$ antisymmetric and $N_f$
fundamental hypermultiplets if $N_f+2|\kappa|\leq 8-N$. At $N=4$, there can be $2$
antisymmetric hypermultiplets with $N_f=\kappa=0$. The case with $N=2$ is exceptional
as the $SU(2)$ Chern-Simons term is zero. This should be treated as an $Sp(1)$ theory,
admitting $N_f\leq 7$ fundamental hypers.

\item $SO(N)$ theories can come with $n_V\leq N-4$ hypermultiplet in the vector (fundamental)
representation. For $N\leq 12$, there can be $n_S\leq 2^{6-N/2}$ spinor and $n_V\leq N-4$
vector hypermultiplets at even $N$, and $n_S\leq 2^{5-(N-1)/2}$ and $n_V\leq N-4$ at odd $N$.
\vspace{-.35cm}
\end{enumerate}
\cite{Intriligator:1997pq} engineers many 5d SCFTs from M-theory on singular Calabi-Yau 3-folds.
One can also engineer some QFTs using type IIB
5-brane webs \cite{Aharony:1997bh}. $Sp(N)$ theories at $n_A=1$ and $N_f\leq 7$ fundamental
hypermultiplets can be engineered by D4-D8-O8 systems \cite{oai:arXiv.org:hep-th/9608111}.
We should stress that the classification of \cite{Intriligator:1997pq} is not most general.
Namely, \cite{Intriligator:1997pq} demanded the absence of `Landau pole'
singularities on the Coulomb branch, and obtained the above classification. However,
some Landau poles are given precise interpretations, and are further required by string
dualities. For instance, product gauge groups are
also allowed \cite{Bergman:2012kr}, which goes beyond the above classification.

Weakly coupled 5d SYMs for circle compactified 6d SCFTs should have zero 1-loop correction
to the coupling constant in the Coulomb branch. The coupling constant thus remains
constant throughout the moduli space. This `microscopic' coupling is identified with the
radius of the compactified circle of the 6d SCFT. 5d maximal SYM is the simplest example,
related to the circle compactified 6d $(2,0)$ SCFT. Another example is the $Sp(N)$
theory with $1$ antisymmetric and $8$ fundamental hypermultiplets, related
to the circle compactified 6d SCFT for the M5-M9 system. These two examples will be briefly
discussed in this paper, to explain how to compute their partition functions.

A tricky point of the index calculus in 1d gauge theories is the continuum coming
from the vector multiplet scalar $\varphi$. In $(0,2)$ and $(0,4)$ mechanics (but not in $(2,2)$
mechanics), classical continuum can be lifted by 1-loop effects.
This can be measured by the  asymptotic behaviors
of the integrand $g(\phi,\alpha,\epsilon_+,z)$, since $V(\varphi)=-\log g$
provides the effective potential for $\varphi$.
When $g$ does not vanish at large $\varphi_I$, becoming a constant,
then $V(\varphi)$ also approaches to a constant. This implies a true quantum
a continuum, without attractive force. On the other hand, if $g$ vanishes at large $\varphi_I$,
then $V(\varphi)$ becomes large and provides attractive force to $\varphi_I$. In this case,
the continuum is lifted. When $g$ diverges at large $\varphi_I$, this would imply that
$\varphi_I$ acquires repulsive force away from $\varphi=0$. The last case
will not appear in the examples that we discuss.

Let us explain the asymptotic behaviors in the context of the ADHM mechanics for 5d/6d SCFTs.
Since different eigenvalues $\varphi_I$ stand for identical multi-instanton particles, we expect
on physical grounds that it
suffices to study the behaviors of one $\varphi_I$ going to $\pm\infty$, while other
eigenvalues are fixed. Equivalently, it should be sufficient to investigate
the asymptotic behaviors for the rank $1$ case.
We use the formulae for $g$ for the $U(N)$ or $Sp(N)$ theories explained in
sections 3.1, 3.2, 3.3, or use the formulae of \cite{Shadchin:2005mx} that we
did not record in this paper for other theories. In particular, the ADHM quantum mechanics
sees the continuum for the following 5d SYM:
\begin{enumerate}
\vspace{-.35cm}
\item $U(N)$ theories with Chern-Simons level $\kappa$ and $N_f=2N-2|\kappa|$ fundamental
hypers

\item $Sp(N)$ theories with $N_f=2N+4$ fundamental hypers, with $n_A=0$

\item $Sp(N)$ theories with $N_f=8$ fundamental and $n_A=1$ antisymmetric hypers

\item $SO(N)$ theories with $n_V=N-4$ fundamental (vector) hypers

\item All 5d $\mathcal{N}=1^\ast$ theories.
\vspace{-.35cm}
\end{enumerate}
In all but the first case, the residue sum $R_0+R_\infty$ is zero. In the first case,
$R_0=0$ and $R_\infty\neq 0$ when $\kappa>0$, $R_0\neq 0$ and $R_\infty=0$ when
$\kappa<0$, and both $R_0\neq R_\infty$ are nonzero when $\kappa=0$.

All but the third and fifth cases are in the 5d SCFTs classified by \cite{Intriligator:1997pq}.
The interpretations of these continua are given in the following subsections, in terms of
bulk decoupled states. In the third and fifth cases, the 5d SYMs describe circle
compactified 6d SCFTs. The string theory uplifts of instantons can move away from the QFT,
developing continua. The fifth case (with $U(N)$, $SO(2N)$ gauge groups) is the
D0-D4 system, for the circle compactified 6d $(2,0)$ SCFT. The third example is the
D0-branes in the background of $N$ D4-branes, $8$ D8-branes, and an O8-plane, for
the circle compactified 6d $(1,0)$ SCFT of the M5-M9 system. Since the D8-brane charges
cancel, the type I' dilaton remains constant transverse to the 8-branes. This implies
that D0-brane mass remains constant as it moves farther away from the D8-O8. So they
can escape the D4-D8-O8 system, developing a continuum.

For the 4d asymptotically free or conformal theories with simple gauge groups $G$,
their 5d uplifts can sometimes develop poles at infinities. For instance,
$SO(N)$ theory is allowed with $N_f\leq N-2$ fundamental hypermultiplets in 4d.
Their 5d uplifts do not have poles at infinities when $N_f\leq N-5$, exhibit
simple/double/triple poles at $N_f=N-4$, $N-3$, $N-2$, respectively.
We have not made further studies on these 4d examples.

In the remaining part of this section, we shall discuss some examples in more detail.

\subsection{5d $\mathcal{N}=1^\ast$ theories}

We first discuss the SYM theory with one adjoint hypermultiplet, with gauge group
$G=U(N),Sp(N),SO(N)$. The fields of their ADHM quantum mechanics
is explained in appendix A. The contour integral of the index takes the form of
\begin{equation}
  \frac{1}{k!}\oint\left[\prod_{I=1}^k\frac{d\phi_I}{2\pi i}\right]Z_{\rm vec}
  (\phi,\alpha,\epsilon_{1,2})Z_{\rm adj}(\phi,\alpha,\epsilon_{1,2},m)\ .
\end{equation}
$Z_{\rm vec}$ is the 1-loop determinant for the quantum mechanical modes
which come from the 5d gauge theory's vector multiplet.
For brevity, we shall introduce the notation $2\sinh{(a\pm b)} \equiv 2\sinh{(a+b)}\cdot 2\sinh{(a-b)}$ and so on.
For $G=U(N)$ and $\hat{G}=U(k)$,
it is given by
\begin{eqnarray}\label{U(N)-vector}
  Z_{\rm vec}\!&\!=\!&\!\frac{\prod_{\alpha\in{\rm root(\hat{G})}}2\sinh\left(\frac{\alpha(\phi)}{2}\right)
  \cdot\prod_{\alpha\in{\rm adj(\hat{G})}} 2\sinh\left(\frac{\alpha(\phi)+2\epsilon_+}{2}\right)}
  {\prod_{\hat\rho\in{\rm fund(\hat{G})}}\prod_{\rho\in{\rm fund(G)}}
  2\sinh\left(\frac{\pm(\hat\rho(\phi)-\rho(\alpha))+\epsilon_+}{2}\right)
  \prod_{\alpha\in{\rm adj}(\hat{G})}
  2\sinh\left(\frac{\alpha(\phi)\pm \epsilon_- +\epsilon_+}{2}\right)}\\
  &=\!&\!\frac{\prod_{I\neq J}2\sinh\left(\frac{\phi_{IJ}}{2}\right)
  \cdot\prod_{I,J=1}^k 2\sinh\left(\frac{\phi_{IJ}+2\epsilon_+}{2}\right)}
  {\prod_{I=1}^k\prod_{i=1}^N
  2\sinh\left(\frac{\phi_I-\alpha_i+\epsilon_+}{2}\right)\cdot
  2\sinh\left(\frac{\alpha_i-\phi_I+\epsilon_+}{2}\right)\prod_{I,J=1}^k
  2\sinh\left(\frac{\phi_{IJ}+\epsilon_1}{2}\right)
  \!\cdot\!2\sinh\left(\frac{\phi_{IJ}+\epsilon_2}{2}\right)} \nonumber
\end{eqnarray}
where $\phi_{IJ}\equiv\phi_I-\phi_J$, and `adj' in the product means that all modes in
the adjoint representation including Cartans are included.
For $G=Sp(N),\hat{G}=O(k)_+$ and $G=SO(N),\hat{G}=Sp(k)$, with reality conditions
on mechanical degrees, it is given by
\begin{equation}
  Z_{\rm vec}=\frac{\prod_{\alpha\in{\rm root(\hat{G})}}2\sinh\left(\frac{\alpha(\phi)}{2}\right)
  \cdot\prod_{\alpha\in{\rm adj(\hat{G})}}
  2\sinh\left(\frac{\alpha(\phi)+2\epsilon_+}{2}\right)}
  {\prod_{\hat\rho\in{\rm fund(\hat{G})}}\prod_{\rho\in{\rm fund(G)}}
  2\sinh\left(\frac{\hat\rho(\phi)-\rho(\alpha)+\epsilon_+}{2}\right)
  \prod_{\rho\in{\rm R}(\hat{G})}2\sinh\left(\frac{\rho(\phi)+\epsilon_1}{2}\right)
  \cdot 2\sinh\left(\frac{\rho(\phi)+\epsilon_2}{2}\right)}\ ,
\end{equation}
where $R$ is symmetric/antisymmetric representation of $O(k)_+$/$Sp(k)$, respectively.
The result for $O(k)_-$ is more complicated \cite{Kim:2012gu}, which we review in section 3.3.
$Z_{\rm adj}$ is the 1-loop determinant for the quantum mechanical modes coming from
the 5d theory's adjoint hypermultiplet. For $G=U(N), \hat{G}=U(k)$, it is given by
\begin{equation}\label{U(N)-hyper}
  Z_{\rm adj}=\frac{\prod_{I=1}^k\prod_{i=1}^N
  2\sinh\left(\frac{\phi_I-\alpha_i+m}{2}\right)\cdot
  2\sinh\left(\frac{\alpha_i-\phi_I+m}{2}\right)\prod_{I,J=1}^k
  2\sinh\left(\frac{\phi_{IJ}\pm m-\epsilon_-}{2}\right)}
  {\prod_{I,J=1}^k2\sinh\left(\frac{\phi_{IJ}\pm m-\epsilon_+}{2}\right)},
\end{equation}
and for other groups,
\begin{equation}
  Z_{\rm adj}=\frac{\prod_{\hat\rho\in{\rm fund(\hat{G})}}\prod_{\rho\in{\rm fund(G)}}
  2\sinh\left(\frac{\hat\rho(\phi)-\rho(\alpha)+m}{2}\right)
  \prod_{\rho\in{\rm R}(\hat{G})}2\sinh\left(\frac{\rho(\phi)\pm m-\epsilon_-}{2}\right)}
  {\prod_{\alpha\in{\rm adj(\hat{G})}} 2\sinh\left(\frac{\alpha(\phi)\pm m-\epsilon_+}{2}\right)}
\end{equation}
where $R$ is chosen in the same way for each group as in $Z_{\rm vec}$.

Now we consider the contour integral. One can show that the sums of two residues at
the infinities of cylinders are always zero, so $\eta$ can be arbitrarily chosen without
referring to $\zeta$. Here we study the well-known case with $G=U(N)$, $\hat{G}=U(k)$
\cite{Nekrasov:2002qd}. For $k$ instantons, the covector space $h^\ast$ for charges is
$\mathbb{R}^k$. We choose $\eta=(1,1,\cdots,1)=e_1+e_2+\cdots+e_k$. Let us first explain
all possible choices of $k$ charges $\{Q_{i_1},Q_{i_2},\cdots,Q_{i_k}\}$ satisfying
$\eta\in{\rm Cone}(Q_{i_1},\cdots,Q_{i_k})$.
They determine ${\bf Q}(\phi_\ast)$ for poles with nonzero JK-Res, both in non-degenerate cases ($n=k$) and in
degenerate cases ($n>k$) where the $k$ charges form a subset of ${\bf Q}(\phi_\ast)$.

Possible $Q_i$'s are $\{\pm e_I\}$ from the fundamental/anti-fundamental weights,
and $\{e_I-e_J\}$ from adjoint. With $\eta$ having all positive components, we first
note that $-e_I$ can never be chosen in the $k$ charges which contain $\eta$ in
their cone. Using the Weyl invariance of
$U(k)$ which permutes $k$ $e_I$'s, it suffices to show that $Q_1=-e_1=(-1,0,\cdots,0)$
cannot be chosen. Suppose that we can. Then we should choose the remaining $k-1$
charge vectors which satisfy
\begin{equation}\label{JK-rule-out}
  \eta=(1,1,\cdots,1)=(-a_1,0,\cdots,0)+\sum_{I=2}^ka_IQ_I
\end{equation}
with $a_1,a_2,\cdots,a_k>0$. For this to be true, at least one of the
$k-1$ $Q_I$'s should have positive first component, which we take to be $Q_2$.
$Q_2=e_1$ is impossible, because then $Q_1,Q_2$ are linearly dependent.
Other choices are $Q_2=e_1-e_I$ for $I\neq 1$, which we take with $I=2$
using Weyl symmetry. Then nonzero second component of $a_1Q_1+a_2Q_2=(a_2-a_1,-a_2)$
requires that $Q_3=e_2-e_3$ up to Weyl reflection, and so on. This step repeats,
until one finds (up to Weyl reflections) all the $k$ vectors given by
\begin{equation}
  (Q_1,Q_2,\cdots,Q_k)=(-e_1,e_1-e_2,e_2-e_3,\cdots,e_{k-1}-e_k),
\end{equation}
for the first $k-1$ components of (\ref{JK-rule-out}) to be positive. Then
one finds that the last component of (\ref{JK-rule-out}) is $-a_k<0$, arriving at
a contradiction.

So we choose $k$ charges among $\{e_I\}$ and $\{e_I-e_J\}$ only.
Using the arguments similar to the previous paragraph based on positivity and linear
independence, the allowed charges are given as follows. Firstly, there should be one
or more charges chosen among $\{e_I\}$, since the latter set $\{e_I-e_J\}$ only
generates $k-1$ dimensional subspace of $\mathbb{R}^k$. Let us choose $p(\leq k)$ of them,
which we can take to be $e_1,e_2,\cdots,e_p$ again up to Weyl reflections. For each
chosen $e_{I}$ with $1\leq I\leq p$, the other charges can be divided into $p$
groups, each group containing exactly one $e_I$.
As an example, let us pick the group containing $e_1$ and explain its structure, as other groups will be similar.
First, charges of the type $e_1-e_{J_a}$ should not be selected.
Once we choose both $e_1$ and $e_1-e_{J_a}$, say $J_a = k$, we require
\begin{equation}
	(a_1+a_2,0, \cdots,0,-a_2)+\sum_{I=3}^{k} a_I Q_I = (1,1,\cdots,1).
\end{equation}
Since the first component can always be a unity by adjusting $a_1$,
we can simply drop it so that the problem gets reduced to picking $k-2$ additional charges
in which $-e_k$ has been chosen. The previous paragraph showed that this is impossible.
Second, there are charges of the form $e_{J_a}-e_1$ with $J_a\neq 2,\cdots,p$.
$J_a$'s have to be different from $2,\cdots,p$, since otherwise there will be
a linearly dependent combination of charge vectors. We can say that these
make a tree graph, with a branch $e_{J_a}-e_1$ attached to $e_1$. Then,
with $e_1$ and (possibly more
than one) $e_{J_a}-e_1$ chosen, one can further find $Q_i$ vectors which branch out
from one of $e_{J_a}$'s, taking the form of $e_{K_b}-e_{J_a}$. $K_b$ are again different
from all the subscripts which appeared so far ($I=1,2,\cdots,p$, $J_a$'s), to avoid linear
relations among selected $Q$
vectors. This procedure can be repeated, attaching adjoint charge vectors to $e_{K_b}$,
and so on. This forms a tree graph originating from $e_1$. The same tree graph can be
formed starting from $e_2$. It starts from $e_{L_c}-e_2$, with $L_c$ being different
from all indices that appeared so far. In this way, we can make $p$ possible trees with
$k$ charges. This tree structure will be further constrained below,
by considering whether there actually exist poles which refer to these charges in
${\bf Q}(\phi_\ast)$.

\begin{figure}[t!]
  \begin{center}
    \includegraphics[width=5cm]{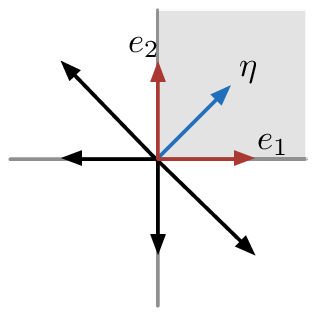}
    \includegraphics[width=5cm]{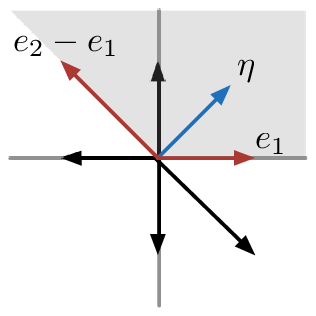}
    \includegraphics[width=5cm]{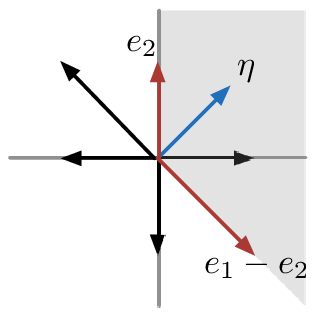}
\caption{Choice of charge vectors for $U(N)$ index at $k=2$ with $\eta=(1,1)$}\label{U(N)k=2-charge}
  \end{center}
\end{figure}
For instance, for $k=2$ with $\eta=(1,1)$,
the selected charge vectors are
\begin{equation}
  \{e_1;e_2\}\ ,\ \ \{e_1,e_2-e_1\}\ ,
\end{equation}
and other charges obtained from above by permuting $e_I$'s: here $\{e_2,e_1-e_2\}$.
These can also be immediately found from Fig.~\ref{U(N)k=2-charge}. For $k=3$ with $\eta=(1,1,1)$,
one finds
\begin{equation}
  \{e_1;e_2;e_3\}\ ,\ \ \{e_1,e_2-e_1;e_3\}\ ,\ \ \{e_1,e_2-e_1,e_3-e_1\}\ ,\ \
  \{e_1,e_2-e_1,e_3-e_2\}\ ,
\end{equation}
and others obtained by permuting $e_I$'s. For $k=4$ with $\eta=(1,1,1,1)$,
one finds
\begin{align}
  &\{e_1;e_2;e_3;e_4\}\ ,\ \{e_1,e_2-e_1;e_3;e_4\}\ ,\  \{e_1,e_2-e_1,e_3-e_1;e_4\},\\
  &\{e_1,e_2-e_1,e_3-e_2;e_4\}\ ,\ \{e_1,e_2-e_1;e_3,e_4-e_3\}\ ,\
  \{e_1,e_2-e_1,e_3-e_1,e_4-e_1\},\nonumber\\
  &\{e_1,e_2-e_1,e_3-e_1,e_4-e_2\}\ ,\  \{e_1,e_2-e_1,e_3-e_2,e_4-e_3\}\nonumber
\end{align}
and their Weyl reflections.

Now we consider the pole $\phi_\ast$ whose ${\bf Q}_\ast$ forms a tree that we just
explained (non-degenerate), or contains it (degenerate). The poles $\phi_\ast$
that actually arise from the integrand are labeled as follows, which we shall prove below by
induction. The poles will be labeled by the colored Young diagrams \cite{Nekrasov:2002qd}.
For each element $Q_i$ in the chosen $\{Q_1,Q_2,\cdots,Q_n\}$, we assign a hyperplane
equation which constrains $\phi_\ast$. When $Q_i$ is one of the fundamental weights,
$\{e_I\}$, then one should impose an equation of the form
\begin{equation}\label{U(N)-hyperplane1}
  \phi_I-\alpha_i+\epsilon_+=0\ ,
\end{equation}
with $i=1,\cdots,N$. When $Q_i$ belongs to the type of $e_I-e_J$, one should impose
one of the following equations,
\begin{equation}\label{U(N)-hyperplane2}
  \phi_I-\phi_J+\epsilon_1=0\ ,\ \ \phi_I-\phi_J+\epsilon_2=0\ ,\ \
  \phi_I-\phi_J-\epsilon_++m=0\ ,\ \ \phi_I-\phi_J-\epsilon_+-m=0\ ,
\end{equation}
where the first two come from $Z_{\rm vec}$ and the latter two come from $Z_{\rm adj}$.
When $n>k$, $n-k$ of them should be redundant for deciding $\phi_\ast$. So we first pick
$k$ independent hyperplane equations which we shall use to define $\phi_\ast$. Since
we are interested in the poles with nonzero JK-Res, there should be at least one
choice $\{Q_1,Q_2,\cdots,Q_k\}$ in ${\bf Q}_\ast$ which contains $\eta$ in their cone.
We work with $k$ hyperplane equations picked in this way, whenever necessary.

The `Young diagram rule' first states that there are no poles with nonzero JK-Res
which refer to last two types of hyperplane equations in
(\ref{U(N)-hyperplane2}) (containing $m$).
Namely, \cite{Nekrasov:2002qd} asserts that the poles coming from the 5d hypermultiplet
measure $Z_{\rm adj}$ can be completely ignored when classifying relevant JK-Res. Then
\cite{Nekrasov:2002qd} focuses on the hyperplanes (\ref{U(N)-hyperplane1}) and the first
two types of hyperplanes in (\ref{U(N)-hyperplane2}), all coming from $Z_{\rm vec}$. The
set of hyperplanes from the poles of $Z_{\rm vec}$ with nonzero residues are
classified by the $N$-colored Young diagrams with $k$ boxes. A colored Young
diagram consists of $N$ Young diagrams $Y=(Y_1,\cdots,Y_N)$
which satisfy $|Y_1|+\cdots+|Y_N|=k$, where $k_i=|Y_i|$ is the number of boxes of the
Young diagram. Each box in the diagram $(Y_1,\cdots,Y_N)$ corresponds to a hyperplane
among (\ref{U(N)-hyperplane1}) and the first two of (\ref{U(N)-hyperplane2}).
We explain how $n_i\geq k_i$ hyperplanes are chosen for a given Young diagram
$Y_i$. Firstly, assign to each of the $k_i$ boxes one of the $k_i$ variables $\phi_{I_{1}},\cdots,\phi_{I_{k_i}}$. Let us say that $\phi_{I_1}$ maps to the box
at the upper-left corner. The corresponding hyperplane is given by
\begin{equation}
  \phi_{I_1}-\alpha_i+\epsilon_+=0\ .
\end{equation}
Then, consider all possible pairs of boxes one can form in $Y_i$, by grouping
horizontally attached boxes or vertically attached boxes.
For a horizontal pair, with $\phi_{I_1}$ and $\phi_{I_2}$ mapping to the left
and right box respectively, we assign the hyperplane
\begin{equation}
  \phi_{I_2}-\phi_{I_1}+\epsilon_1=0\ .
\end{equation}
For a vertical pair, with $\phi_{I_1}$ and $\phi_{I_2}$ mapping to the upper
and lower box respectively, we assign the hyperplane
\begin{equation}
  \phi_{I_2}-\phi_{I_1}+\epsilon_2=0\ .
\end{equation}
One obtains at least $k_i$ independent hyperplanes this way. For instance,
the diagram $Y_i=\Yvcentermath1\scriptsize{\young(12,3)}$ defines $k_i=3$ hyperplanes
\begin{equation}
  \phi_1-\alpha_i+\epsilon_+=0\ ,\ \ \phi_{21}+\epsilon_1=0\ ,\ \
  \phi_{31}+\epsilon_2=0\ ,
\end{equation}
while the diagram $Y_i=\Yvcentermath1\scriptsize{\young(12,34)}$ with $k_i=4$
defines $n_i=5>k_i$ hyperplanes
\begin{equation}\label{young-example}
  \phi_1-\alpha_i+\epsilon_+=0\ ,\ \ \phi_{21}+\epsilon_1=0\ ,\ \
  \phi_{31}+\epsilon_2=0\ ,\ \ \phi_{43}+\epsilon_1=0\ ,\ \ \phi_{42}+\epsilon_2=0\ .
\end{equation}
In all hyperplane assignments, one can easily see that $n_i\geq k_i$ equations
determine unique $\phi_\ast$ and never over-determine it. Repeating the process for
all $N$ Young diagrams, one picks $n=\sum_{i=1}^Nn_i\geq k$ independent hyperplanes.
By taking a look, one can convince oneself that the chosen ${\bf Q}(\phi_\ast)$ is
always projective. For instance, the $5$ charges responsible for (\ref{young-example})
are $e_1,e_{21},e_{31},e_{43},e_{42}$ on $\mathbb{R}^4$. They are projective, since
they are contained in the half-space $x_4+\epsilon(x_2+x_3)+\epsilon^2 x_1>0$ with small
enough $\epsilon$. The mapping of $\phi_I$ variables to the $k$ boxes of $Y$ can be done
in a unique way, by eating up the Weyl symmetry factor $\frac{1}{k!}$.

To derive the above `colored Young diagram rules,' we shall make the
inductive argument. Firstly, we show that this is true at $k=1$. At $k=1$, there is
no pole from $Z_{\rm adj}$ so that ignoring all possible poles from $Z_{\rm adj}$
is trivially true. Then, we only have to choose the pole value of single $\phi$
variable. By the JK residue rule, or equivalently the rank $1$ residue choice
rule of section 2.2, this is given by choosing one of the $N$ equations
$\phi_I-\alpha_i+\epsilon_+=0$ for the pole. These choices correspond to $N$
different colored Young diagrams with $1$ box, confirming the rule at $k=1$.

Now assume that the `Young diagram rule' is true at rank $k-1$. To use induction,
we pick the $k$ independent hyperplane equations with $\eta\in{\rm Cone}(Q_1,\cdots,Q_k)$
in ${\bf Q}(\phi_\ast)$. Here, recall the `tree structure' of these $k$ charge vectors.
Apart from the case with $k$ independent trees without any branches from $e_I$'s,
corresponding to the colored Young diagram in which each $Y_i$ contains only a
single box, there are always one or more charge vectors of the form $Q=e_I-e_J$ which
are at the end of a tree (not having further branches attached to them).
The hyperplane equation $Q(\phi)+\cdots=0$ with such a $Q$ is the only one which refers
to $\phi_I$ coordinate among the $k$ hyperplane equations.
Using the Weyl symmetry, we take $\phi_k$ to be such a coordinate which appears only
once in the $k$ hyperplane equations. So we take the set of $k$ hyperplanes to be
\begin{equation}
  (k-1\ {\rm hyperplanes\ referring\ to}\ \phi_1,\cdots,\phi_{k-1}\ {\rm only})
  \cap H_k\ .
\end{equation}
$H_k$ is the only hyperplane whose equation contains the $\phi_k$ coordinate,
so that the other $k-1$ hyperplanes refer to $\phi_1,\cdots,\phi_{k-1}$ only.
The charges appearing in these $k-1$ hyperplanes are those for the $U(k-1)$.
Also, the integrand which contains $\phi_1,\cdots,\phi_{k-1}$ but not $\phi_k$ is
the integrand for the $k-1$ instantons in the $U(N)$ theory. Finally, if
$\{Q_1,Q_2,\cdots,Q_k\}$ is the set of charges
which contains $k$ dimensional $\eta^{(k)}=(1,\cdots,1)$ in their cone, then
$\{Q_1,\cdots,Q_{k-1}\}$ contains $\eta^{k-1}=(1,\cdots,1,0)$ in
$\mathbb{R}^{k-1}\subset\mathbb{R}^k$. This is obvious from the fact that the charge
$Q=\phi_k-\phi_J$ is at the end of the tree, so dropping it yields a tree with $k-1$ charges.
So the poles on $\mathbb{R}^k$ with nonzero JK-Res are obtained by first studying the poles
on $\mathbb{R}^{k-1}$ for $\phi_1,\cdots,\phi_{k-1}$ with nonzero JK-Res, and then
determining the values of $\phi_k$ by considering possible $H_k$'s.

\begin{figure}[t!]
  \begin{center}
    \includegraphics[width=5cm]{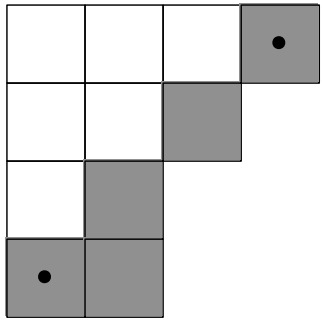}
\caption{Shaded boxes form the border of a Young diagram. Dotted boxes are
at the corners.}\label{border}
  \end{center}
\end{figure}
By the assumption of the induction, $k-1$ dimensional poles with nonzero JK-Res
are classified by colored Young diagrams with $k-1$ boxes, which we call $Y^{(k-1)}$.
We now show that all possible extra hyperplane conditions $H_k$ with nonzero $k$
dimensional JK-Res map to the possibilities of adding one more boxes to $Y^{(k-1)}$ which
makes all possible $Y^{(k)}$'s. Now we collect all possible hyperplane
equations for $H_k$. The equation could be
\begin{equation}
  \phi_k-\alpha_i+\epsilon_+=0
\end{equation}
only if the $i$'th Young diagram $Y_i$ is empty in $Y^{(k-1)}$.
(If $Y_i$ is already occupied with $\phi_J-\alpha_i+\epsilon_+=0$,
then $\sinh\frac{\phi_{IJ}}{2}$ in the numerator of (\ref{U(N)-vector}) vanish.)
This configuration by definition forms a colored Young diagram with $k$ boxes, where
a new nonempty diagram $Y_i$ with one box is created. Other possible equations could be
\begin{equation}\label{last-hyperplane}
  \phi_k-\phi_I+\epsilon_1=0\ ,\ \ \phi_k-\phi_I+\epsilon_2=0\ ,\ \
  \phi_k-\phi_I-\epsilon_+\pm m=0\ .
\end{equation}
Firstly, we explain that the
hyperplanes $\phi_k-\phi_I-\epsilon_+\pm m=0$ yield zero residues. If
$\phi_I$ maps to the box at the upper-left corner of a Young diagram, then
$\phi_I=\alpha_j-\epsilon_+$ for some $j$. Then, the factor
$2\sinh\left(\frac{\phi_k-\alpha_j\pm m}{2}\right)$ in the numerator of
$Z_{\rm adj}$ vanishes so that the pole does not exist. If $\phi_I$ does not map
to the box at the upper-left corner, then there should be a box with $\phi_J$ which is
left-adjacent or upper-adjacent to the box with $\phi_I$, namely $\phi_J=\phi_I-\epsilon_{1,2}$
for either $\epsilon_1$ or $\epsilon_2$. Here we note that there are factors
\begin{equation}
  \prod_{\pm}2\sinh\left(\frac{\phi_{kJ}\pm\epsilon_--m}{2}\right)\cdot
  2\sinh\left(\frac{\phi_{kJ}\pm\epsilon_-+m}{2}\right)
\end{equation}
in the numerator of (\ref{U(N)-hyper}). Inserting either of
$\phi_J=\phi_I-\epsilon_{1,2}$, one finds that the factor
\begin{equation}
  \prod_\pm 2\sinh\left(\frac{\phi_{kI}-\epsilon_+\pm m}{2}\right)
\end{equation}
is always contained in the numerator, which vanishes due to the hyperplane condition
$\phi_{kI}-\epsilon_+\pm m=0$ for one of the two signs. This shows that the corresponding
poles do not exist. Next, we consider the first two types of hyperplanes in
(\ref{last-hyperplane}). The hyperplane of the first two sorts will correspond to adding
a box to $Y^{(k-1)}$ when the box corresponding to $\phi_I$ is at the `border' of
$Y^{(k-1)}$. See Fig.~\ref{border} for what we mean by the boxes at the border of a Young
diagram.
One can show that the first two equations with $I$ not at the border has zero residue,
as follows. The box $\phi_I$ not at the border always has a right-adjacent
and lower-adjacent boxes, which we call $\phi_{J_1}$, $\phi_{J_2}$, respectively.
These variables are determined by the hyperplane equations $\phi_{J_1I}+\epsilon_1=0$
and $\phi_{J_2I}+\epsilon_2=0$. So if $\phi_I$ is not at the border of $Y^{(k-1)}$,
the factor $\sinh\frac{\phi_{kJ_1}}{2}$ or $\sinh\frac{\phi_{kJ_2}}{2}$ in the numerator
of (\ref{U(N)-vector}) vanishes, yielding zero residue.
The remaining hyperplane conditions in (\ref{last-hyperplane}) that are
not ruled out are $\phi_{kI}+\epsilon_{1,2}=0$ with $\phi_I$ at the border. Now using
the `box' language, the box $\phi_k$ may either attach
to two boxes $\phi_I,\phi_J$ of $Y^{(k-1)}$ like $\Yvcentermath1\scriptsize{\young(PI,Jk)}$,
attach to one box at the `corners' of the Young diagram like $\Yvcentermath1\scriptsize{\young(Ik)}$ or $\Yvcentermath1\scriptsize{\young(I,k)}$
(see Fig.~\ref{border}), or
attach to one box $\phi_I$ in the middle of the border of $Y^{(k-1)}$ like
$\Yvcentermath1\scriptsize{\young(P,Ik)}$ and $\Yvcentermath1\scriptsize{\young(PI,:k)}$.
The first three are stacking the $k$'th box to form a colored Young diagram $Y^{(k)}$, while
the last two are not. In the last two cases, the factor
$2\sinh\left(\frac{\phi_{kP}+2\epsilon_+}{2}\right)$ in the numerator of $Z_{\rm vec}$
vanishes so that the corresponding poles do not exist. In the first case, the factor
\begin{equation}
  \frac{\sinh\frac{\phi_{kP}+2\epsilon_+}{2}}{\sinh\frac{\phi_{kI}+\epsilon_1}{2}
  \sinh\frac{\phi_{kJ}+\epsilon_2}{2}}
\end{equation}
partly cancels to keep a simple pole. The second and third cases also develop poles.
So only the first three types of hyperplanes survive, exhausting all possible ways
of putting the $k$'th box to $Y^{(k-1)}$ to make $Y^{(k)}$. This finishes the inductive
proof of the map between poles with nonzero JK-Res and colored Young diagrams.

Having identified the poles, one can compute the JK-Res at these poles. For this,
one expands the integrand in the Laurent
series of $Q_i(\phi-\phi_\ast)$, and the computation boils down to knowing various
${\textrm{JK-Res}}({\bf{Q}}(\phi_\ast),\eta)
\frac{d\phi_1\wedge\cdots\wedge d\phi_k}{Q_{j_1}(\phi-\phi_\ast)\cdots Q_{j_k}
(\phi-\phi_\ast)}$ given by (\ref{JK-Res}). In particular, all JK-Res for the poles labeled by
the colored Young diagrams can be regarded as iterated contour integrals. Firstly,
JK-Res factorizes into $N$ groups, each group mapping to a Young diagram $Y_i$.
Within a given Young diagram $Y_i$, the iterated integral goes in the reverse order
of stacking the boxes. For instance, for the Young diagram $\Yvcentermath1\scriptsize{\young(123,456)}$, the integral over
the relevant pole terms is given by
\begin{align}
  &\textrm{JK-Res}\frac{\bigwedge_{i=1}^6 d\phi_i}
  {\phi_1\phi_{21}\phi_{32}\phi_{41}}\cdot\frac{\phi_{51}}{\phi_{52}\phi_{54}}
  \cdot\frac{\phi_{62}}{\phi_{63}\phi_{65}}
  =\textrm{JK-Res}
  \frac{\bigwedge_{i=1}^6 d\phi_i}{\phi_1\phi_{21}\phi_{32}\phi_{41}}
  \left(\frac{1}{2\phi_{52}}+\frac{1}{2\phi_{54}}\right)
  \left(\frac{1}{2\phi_{63}}+\frac{1}{2\phi_{65}}\right)\nonumber\\
  &=1=
  \frac{1}{(2\pi i)^6}\oint\frac{d\phi_1}{\phi_1} \oint\frac{d\phi_2}{\phi_{21}}
  \oint\frac{d\phi_3}{\phi_{32}} \oint\frac{d\phi_4}{\phi_{41}}
  \oint d\phi_5\frac{\phi_{51}}{\phi_{52}\phi_{54}} \oint d\phi_6
  \frac{\phi_{62}}{\phi_{63}\phi_{65}}\ ,
\end{align}
where $\oint$ for each $\phi_I$ is done around a small
counterclockwise circle surrounding the pole. Such iterated integral
formula holds for all poles labeled by Young diagrams. This yields the following
expression for the $U(N)$ instanton partition function \cite{Flume:2002az,Bruzzo:2002xf}:
\begin{equation}\label{U(N)-final}
  Z_k=\sum_{\sum_i|Y_i|=k}\prod_{i,j=1}^N\prod_{s\in Y_i}
  \frac{\sinh\frac{E_{ij}+m-\epsilon_+}{2}\sinh\frac{E_{ij}-m-\epsilon_+}{2}}
  {\sinh\frac{E_{ij}}{2}\sinh\frac{E_{ij}-2\epsilon_+}{2}}
\end{equation}
where
\begin{equation}\label{Eij}
  E_{ij}=\alpha_i-\alpha_j-\epsilon_1h_i(s)+\epsilon_2(v_j(s)+1)\ .
\end{equation}
Here, $s$ runs over the boxes in the $i$'th Young diagram $Y_i$. $h_i(s)$ is the distance
from the box $s$ to the edge of the right side of $Y_i$ that one reaches by moving to the
right. $v_j(s)$ is the distance from $s$ to the edge on the bottom side of $Y_j$ that one
reaches by moving down \cite{Flume:2002az,Bruzzo:2002xf,oai:arXiv.org:1110.2175}.

Now, we discuss the alternative prescriptions for the $U(N)$ contour integral as
stated at the end of section 2.3. Namely, with the relaxation understood,
$e^{-\epsilon_+}\rightarrow t\ll 1$ and $e^{-\epsilon_+}\rightarrow T\gg 1$, we
take all the $e^{\phi_I}$ variables to live on the unit circles on the complex plane. Multiple
unit circle integrals can be done in any order. In fact, this should be the original
method used by \cite{Nekrasov:2002qd,Bruzzo:2002xf} to get the result (\ref{U(N)-final}).
For the purpose of illustrating how the alternative contour prescription works, we
repeat it for $U(N)$ index at $k=2$.

We first integrate over $z_1=e^{\phi_1}$. We sum over all residues for poles in $|z_1|<1$
inside the unit circle, keeping $z_2$ fixed with $|z_2|=1$. Then we integrate over
$z_2=e^{\phi_2}$, again picking all residues for poles in $|z_2|<1$. The rule excludes
all the poles at the origin, $z_1=0$ or $z_2=0$. The possible poles in this procedure
are shown on the first two columns of Table~\ref{U(2)k=2pole}.
\begin{table}[t!]
\centering
\begin{tabular}{c|c|c|c}
  \hline
  Integral over $z_1$ & Integral over $z_2$ & $(|z_1|,|z_2|)$& Jeffrey-Kirwan\\
  \hline $\phi_1+\epsilon_+-\alpha_i=0$&$\phi_2+\epsilon_+-\alpha_j=0$&$
  (tw_i,tw_j)$&Yes\\
  &$\phi_1-\phi_2+\epsilon_+\pm\epsilon_-=0$&$(tw_i,u^{\mp 1}w_i)^a$&No\\
  &$\phi_1-\phi_2-\epsilon_+\pm m=0$&$(tw_i,tTv^{\mp 1}w_i)^b$&No\\
  &$\phi_2-\phi_1+\epsilon_+\pm\epsilon_-=0$&$(tw_i,t^2u^{\pm 1}w_i)$&Yes\\
  &$\phi_2-\phi_1-\epsilon_+\pm m=0$&$(tw_i,tT^{-1}v^{\pm 1}w_i)$&Yes\\
  \hline
  $\phi_1-\phi_2+\epsilon_+\pm\epsilon_-=0$&$\phi_1+\epsilon_+-\alpha_i=0$
  &$(tu^{\pm 1},u^{\mp 1} w_i)^c$&No\\
  &$\phi_2+\epsilon_+-\alpha_i=0$
  &$(tu^{\pm 1},t w_i)$ &Yes\\ \hline
  $\phi_1-\phi_2-\epsilon_+\pm m=0$&$\phi_1+\epsilon_+-\alpha_i=0$
  & $(T^{-1}v^{\pm 1},tT v^{\mp 1} w_i )^d$&No \\
  &$\phi_2+\epsilon_+-\alpha_i=0$
  & $(T^{-1}v^{\pm 1},t w_i)$&Yes\\
  \hline
\end{tabular}
\caption{Poles for $U(N)$ instantons at $k=2$: $w_i\equiv e^{\alpha_i}$,
$u\equiv e^{-\epsilon_-}$, $v\equiv e^{-m}$.}
\label{U(2)k=2pole}
\end{table}
At a given row, one first chooses an equation from the left column,
which gives the poles for $z_1$ inside the unit contour.
Then one moves on to the second column on the same row,
which gives possible poles for $z_2$ inside its unit contour.
The third column shows the values of $|z_1|,|z_2|$ at the moment we are
going to decide whether the pole is within the unit circle or not.
For $z_1$, it does not necessarily agree with its actual value after the pole
for $z_2$ is selected, since we keep $|z_2| = 1$ while integrating over $z_1$.
Table~\ref{U(2)k=2pole} contains only those selected by the $z_1$ unit contour rule.
Some of them evidently stay inside the $z_2$ unit contour,
while the four cases which are labeled by the superscripts
$a,\cdots,d$ are rather ambiguous with the unit contour rule for $z_2$. One finds that all the
poles which are unambiguously inside the unit contour $T^2=S^1\times S^1$ map
to the poles which are chosen by the JK-Res rule. (Of course, we saw that
some of the residues within this class can be zero, by using extra structures of
the $U(N)$ index.)

As for the four ambiguous cases, whether they are inside or outside the
unit contour for $z_2$ depends on the scales of other fugacities which we did not
specify yet. But one can notice that there always exists a pair of poles at an ambiguous
location of $z_2$. $a,c$ and $b,d$ are such pairs. So the paired poles are either
simultaneously inside or outside the unit contour of $z_2$. When they are outside the
$z_2$ unit contour, they provide no contribution so that the result is consistent
with the JK-Res rule. When they are both inside the $z_2$ unit contour, the two
residues cancel. The sum of two paired residues is actually a result of doing
the contour integral of the form:
\begin{equation}
  \oint \frac{d\phi_2}{2\pi i}\oint \frac{d\phi_1}{2\pi i}
  \frac{f(\phi_1,\phi_2)}{(\phi_1-a)(\phi_1-\phi_2-b)}\ ,
\end{equation}
with $f(\phi_1,\phi_2)$ being regular at $\phi_1=a,\phi_1-\phi_2=b$.
The integral is given by
\begin{equation}\label{pairwise}
  \oint \frac{d\phi_2}{2\pi i}
  \left(\frac{f(a,\phi_2)}{a-b-\phi_2}+\frac{f(\phi_2+b,\phi_2)}{\phi_2+b-a}\right)
  =-f(a,a-b)+f(a,a-b)=0\ .
\end{equation}
The two terms $-f(a,a-b)$ and $f(a,a-b)$ are precisely the pair of residues, such
as $a,c$ and $b,d$ above, guaranteeing cancelation when they are in the unit contour.
This illustrates that the unit contour rule with $e^{-\epsilon_+}\rightarrow t\ll 1$ for hypers
and $e^{-\epsilon_+}\rightarrow T\gg 1$ for twisted hypers yields the same result as the
JK-Res rule: although the unit contour rule may appear to keep more residues, after pairwise
cancelations the two rules become equivalent. We confirmed that similar
pairwise cancelations happen for the $Sp(N)$ indices at $k=4$, as summarized in
section 3.3. In some other cases, such as the $Sp(1)$ index at $k=5$ in section 4.1,
we just used the iterated integral rule along unit contour without checking its
equivalence with the Jeffrey-Kirwan rule. Possibly, one could be able to prove
the equivalence in full generality.

We emphasize here that the above type of pole classification goes through for
$U(N)$ instanton partition functions with other matters. For fundamental hypermultiplets,
there are no extra poles incurred by the hypermultiplets. Then the above arguments can
be reused, simply ignoring all the discussions involving the hyperplanes from $Z_{\rm adj}$.
We also checked that the bi-fundamental hypermultiplets in the $U(n)$ quiver theories
do not provide any poles with nonzero JK-Res at two instanton order, as derived in
\cite{Nekrasov:2013xda}. The absence of poles coming from the hypermultiplet factor
$Z_{\rm adj}$ is an accidental property of the $U(N)$ theory. This simplification does
not happened for the $\mathcal{N}=1^\ast$ theory with other gauge groups.
One should just use the Jeffrey-Kirwan residue rule, or alternatively use
the unit contour integration rule after suitably replacing $e^{-\epsilon_+}$
by $t\ll 1$ and $T\gg 1$. We leave the studies on these indices to the future.

\subsection{$U(N)$ theories with matters and Chern-Simons term}

In this section, we consider the instanton partition function of 5d $U(N)$ SYM,
with $N_f$ fundamental matters and nonzero Chern-Simons term at level $\kappa$. We shall
only consider the theories which are related to the 5d SCFTs at the UV fixed points. The
contour integral has the same structure as what we explained in the previous section,
picking up poles and residues labeled by the colored Young diagrams.
However, there occasionally arise subtleties in this class of theories. The residue sums
at the two ends of cylinders will not be zero when $N_f+2|\kappa|=2N$. The nonzero residues
at the infinity regions of $\varphi_I$ imply a continuum in the ADHM quantum mechanics.
The nonzero sum of two residues at the infinities of a cylinder implies a wall crossing as
the FI parameter changes. These can be naturally understood with the string theory realizations
of these 5d SYMs and the UV SCFTs. Before proceeding, we emphasize that most of the studies in
this subsection and section 3.4.4 are already done in
\cite{Bao:2013pwa,Hayashi:2013qwa,Bergman:2013ala,Bergman:2013aca,Taki:2013vka,Taki:2014pba}.
Mostly, we just reproduce their results, sometimes filling the missing values of $N,N_f,\kappa$ not checked by them, to illustrate the (absence of unphysical) wall crossing issue.

\begin{figure}[t!]
  \begin{center}
    \includegraphics[width=17cm]{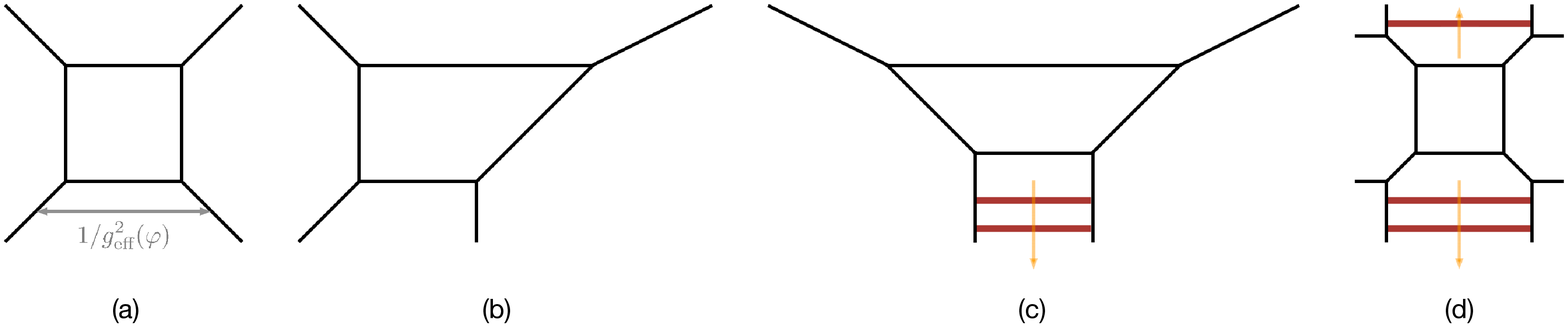}
\caption{(a) 5-brane web for the pure $SU(2)$ theory; (b) $SU(2)$ at $\kappa=1$;
(c) $SU(2)$ at $\kappa=2$; (d) $SU(2)$ with $N_f=4$ at $\kappa=0$.
Horizontal lines are D5-branes on which 5d QFTs live. Red horizontal lines denote D1-branes
which can escape to infinity by developing a continuum.}\label{U(2)-web}
  \end{center}
\end{figure}
\begin{figure}[t!]
  \begin{center}
    \includegraphics[width=9cm]{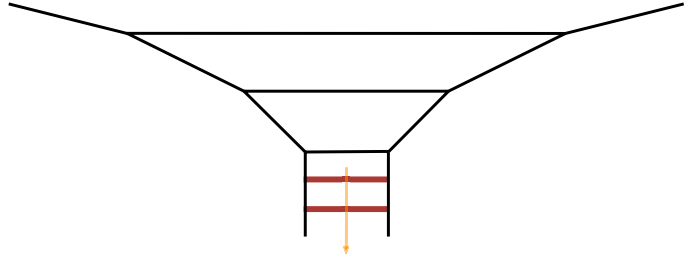}
\caption{5-brane web for the pure $SU(3)$ theory at $\kappa=3$}\label{U(3)-web}
  \end{center}
\end{figure}
In Fig.~\ref{U(2)-web}, various $(p,q)$ 5-brane webs are shown which engineers the
$U(2)$ gauge theory with fundamental hypermultiplets and/or bare Chern-Simons term
\begin{equation}
  S_{\rm CS}=\frac{\kappa}{24\pi^2}\int {\rm tr}\left(A\wedge F\wedge F
  +\frac{i}{2}A^3\wedge F-\frac{1}{10}A^5\right)\ .
\end{equation}
$\kappa$ is integral when the number $N_f$ of fundamental hypermultiplets is even,
and is half an odd integer when $N_f$ is odd. The overall $U(1)$ of the $U(2)$ is
non-dynamical in QFT.
Note that there cannot be a bare Chern-Simons term for the $SU(2)$ theory.
Thus the bare $U(2)$ Chern-Simons term means the mixed Chern-Simons term for the
$U(1)$-$SU(2)$-$SU(2)$, inducing the background $U(1)$ electric charge
to the $SU(2)$ instantons. The two horizontal lines are D5-branes on which the $U(2)$
theory lives. The overall $U(1)$ has infinite inertia, as the
overall displacement of the two D5-branes induce translations of the asymptotic branes.
When $N_f+2|\kappa|=2N$, one finds horizontal D1-branes stretched between the two parallel
vertical lines (NS5-branes). These D1-branes, shown by the red lines in
Fig.~\ref{U(2)-web}(c),(d), can escape up/down from the D5-branes on which the 5d QFT
is defined. This implies that the ADHM quantum mechanics for the D1-D5 system (UV
completing the instanton mechanics) has a continuum
in the Coulomb branch. In the contour integrand  for the instanton index, this continuum
causes a nonzero pole at one or two ends of the cylinder. The case with $N=3,\kappa=3,N_f=0$
is shown in Fig.~\ref{U(3)-web}. So in these examples, the interpretation of the poles at
infinities is the continuum developed by the D1-brane states which can escape. These are
not in the QFT spectrum in the decoupling limit. This issue is studied in
\cite{Bao:2013pwa,Hayashi:2013qwa,Bergman:2013ala,Bergman:2013aca,Taki:2013vka,Taki:2014pba}.

The $k$ instanton partition function, which could possibly include the extra
decoupled $Z_{\rm extra}$ factor when $N_f+2|\kappa|=2N$, is given by the following
contour integral (see, e.g. \cite{Kim:2012gu})
\begin{equation}
  Z_k=\frac{(-1)^{\kappa+\frac{N_f}{2}}}{k!}\oint\left[\frac{d\phi_I}{2\pi i}\right]
  e^{\kappa\sum_{I=1}^k\phi_I}Z_{\rm vec}(\phi,\alpha,\epsilon_{1,2})Z_{\rm fund}(\phi,m_a)
\end{equation}
where $Z_{\rm vec}$ takes the same form as (\ref{U(N)-vector}), and
\begin{equation}\label{U(N)-fund-hyper}
  Z_{\rm fund}=\prod_{I=1}^k\prod_{a=1}^{N_f}2\sinh\left(\frac{\phi_I+m_a}{2}\right)\ .
\end{equation}
The overall sign $(-1)^{\kappa+N_f/2}$ was found in \cite{Bergman:2013ala,Bergman:2013aca}
to be the physically sensible one, from various indirect evidences.  \cite{Bergman:2013ala,Bergman:2013aca} conjectured that it will have to do
with the effect of 5d Chern-Simons term, but its microscopic derivation seems to be
unavailable yet. The pole selection derived from the JK-Res rule is exactly the same as
what we derived for the $\mathcal{N}=1^\ast$ theory in the previous subsection,
labeled by the colored Young diagram. In the previous subsection, we chose
$\eta=(1,\cdots,1)$. Here, we note in foresight that the index may depend on $\zeta$.
The choice of $\eta$ in section 3.1 is for $\zeta<0$. For the theories in this subsection,
the result is given by
\begin{equation}
  Z_k=(-1)^{\kappa+N_f/2}\sum_{\sum_i|Y_i|=k}\prod_{i=1}^N\prod_{s\in Y_i}\frac{e^{\kappa\phi(s)}
  \prod_{l=1}^{N_f}2\sinh\frac{\phi(s)+m_l}{2}}
  {\prod_{j=1}^N2\sinh\frac{E_{ij}}{2}\cdot 2\sinh\frac{E_{ij}-2\epsilon_+}{2}}\ .
\end{equation}
$E_{ij}$ is defined by (\ref{Eij}), and $\phi(s)$ is given by
\begin{equation}
  \phi(s)=\alpha_i-\epsilon_+-(m-1)\epsilon_1-(n-1)\epsilon_2\ ,
\end{equation}
where $s=(m,n)\in Y_i$ with $m,n$ being the vertical and horizontal positions of the box
$s$ from the upper-left corner of $Y_i$ \cite{Kim:2012gu}. When $\zeta>0$, one would have
to choose $\eta=-(1,\cdots,1)$ and use the JK-Res rule. It is easy to get the result
for $\zeta>0$. Since $\zeta\rightarrow-\zeta$ can be undone by the $SU(2)_r$ Weyl
reflection, or the Weyl reflection of the diagonal of $SU(2)_r\times SU(2)_R$,
the sign flip of $\zeta$ is equivalent to that of $\epsilon_+$. So by flipping all
signs of $\epsilon_+$ in the above result, we obtain the index for $\zeta>0$.
The two results will be the same unless $N_f+2|\kappa|=2N$.

At $N_f+2|\kappa|=2N$, $Z_{\rm QM}^k$ factorizes into $Z_{\rm QFT}Z_{\rm extra}$
with nontrivial $Z_{\rm extra}$, and furthermore $Z_{\rm extra}$ exhibits a wall
crossing as $\zeta$ flips sign. Nontrivial $Z_{\rm extra}$ was analyzed and factored
out from $Z_{\rm QM}$ in
\cite{Bao:2013pwa,Hayashi:2013qwa,Bergman:2013ala,Bergman:2013aca,Taki:2013vka,Taki:2014pba}.
We will explain/review these indices and their $\zeta$ dependence in section 3.4.

\subsection{$Sp(N)$ theories}

In this subsection, we study the instanton partition function for the $Sp(N)$
gauge theories with $N_f$ fundamental and $n_A=0,1$ antisymmetric
hypermultiplets.

Let us first write down the contour integral expression. The integral variables are the
zero modes of the ADHM quantum mechanics for the $Sp(N)$ instantons. Part of the zero
modes is the holonomy of $\hat{G}$ on the temporal circle. For $k$ instantons, they
come with $\hat{G}=O(k)$ gauge group. Since $O(k)$ has two components $O(k)_+$ and $O(k)_-$,
one should also turn on discrete holonomies for $e^{iA_\tau}$. These can all be labeled by
the complexified group element $U=e^{\phi}=e^{\varphi+iA_\tau}$, which can be taken as
\cite{Kim:2012gu}
\begin{equation}
  U_+=e^{\phi_+}=\left\{\begin{array}{ll}{\rm diag}
  (e^{\sigma_2\phi_1},\cdots,e^{\sigma_2\phi_n})
  &{\rm for\ even}\ k=2n\\{\rm diag}(e^{\sigma_2\phi_1},\cdots,e^{\sigma_2\phi_n}, 1)
  &{\rm for\ odd}\ k=2n\!+\!1\end{array}\right.
\end{equation}
for $O(k)_+$, and
\begin{equation}
  U_-=e^{\phi_-}=\left\{\begin{array}{ll}{\rm diag}(e^{\sigma_2\phi_1},\cdots,
  e^{\sigma_2\phi_{n-1}},\sigma_3)
  &{\rm for\ even}\ k=2n\\{\rm diag}(e^{\sigma_2\phi_1},\cdots,e^{\sigma_2\phi_n},-1)
  &{\rm for\ odd}\ k=2n\!+\!1\end{array}\right.
\end{equation}
for $O(k)_-$. The above expressions with imaginary $\phi_I$ are the $O(k)_{\pm}$ group elements,
while their complexifications come with $\varphi_I={\rm Re}(\phi_I)$. Below we shall write
$k=2n+\chi$, with $\chi=0,1$. We will get two
intermediate indices $Z_\pm^k$ from the path integral. Each of them is obtained by taking
the complexified holonomy in either $U_\pm$, performing Gaussian integration over non-zero
modes, and then exactly summing or integrating over $U_\pm$ (with
contours explained in section 2.3). The final index is given by \cite{Kim:2012gu}
\begin{equation}
  Z^k=\frac{Z^k_++Z^k_-}{2}\ .
\end{equation}
There is a variation of this result due to
nontrivial $\pi_4(Sp(N))=\mathbb{Z}_2$, which sometimes defines new 5d SCFTs.
With nontrivial $\mathbb{Z}_2$ element, one would have to take \cite{Bergman:2013ala}
\begin{equation}
  Z^k=(-1)^k\frac{Z^k_+-Z^k_-}{2}\ .
\end{equation}
This will be discussed more in section 3.4.4. $Z_\pm^k$ are given by
\begin{equation}
  Z^k_\pm=\frac{1}{|W|}\oint[d\phi]Z_{\rm vec}^\pm Z_{\rm fund}^\pm Z_{\rm anti}^\pm\ .
\end{equation}
The Weyl factors $|W|$ for $O(k)_{\pm}$ are
\begin{align}
    \hspace{-1cm}|W|_{+}^{\chi=0} = \frac{1}{2^{n-1} n!} ,\ |W|_{+}^{\chi=1} = \frac{1}{2^n n!} ,\ |W|^{\chi=0}_{-} =  \frac{1}{2^{n-1}(n-1)!},\ |W|^{\chi=1}_{-} = \frac{1}{2^n n!}.
\end{align}
With the ADHM matter contents explained in appendix A, the integrands are given as follows:
\begin{align}
  \hspace*{-0.5cm}Z_{\rm vec}^+=& \left[\prod_{I<J}^{n} 2\sinh{ \tfrac{ \pm \phi_I \pm \phi_J}{2}} \left(\prod_I^{n} 2\sinh{\tfrac{\pm \phi_I}{2}}\right)^{\chi }\right]
	    \left(\frac{1}{2\sinh{ \frac{\pm \epsilon_- + \epsilon_+}{2}} \, \prod_{i=1}^{N} 2\sinh{ \frac{\pm \alpha_i + \epsilon_+}{2}}} \cdot \prod_{I=1}^{n} \frac{ 2\sinh{ \frac{\pm \phi_I + 2\epsilon_+}{2}}} {2\sinh{ \frac{\pm \phi_I \pm \epsilon_- + \epsilon_+}{2}}}\right)^{\chi}
\nonumber\\
	    &\times \prod_{I=1}^{n} \frac{2 \sinh{\epsilon_+} }{ 2 \sinh{\frac{\pm \epsilon_- + \epsilon_+ }{2}}  2\sinh{ \frac{\pm 2\phi_{I} \pm \epsilon_- + \epsilon_+}{2}} \, \prod_{i=1}^{N} 2\sinh{ \frac{\pm \phi_{I} \pm \alpha_i + \epsilon_+}{2}} }
		\prod_{I < J}^{n} \frac{ 2\sinh{ \frac{\pm \phi_{I} \pm \phi_{J} + 2\epsilon_+}{2} }}{2\sinh{ \frac{\pm \phi_{I} \pm \phi_{J} \pm \epsilon_- + \epsilon_+}{2}}}
\end{align}
from the ADHM data of 5d vector multiplet with $O(k)_+$,
\begin{align}
\hspace*{-0.5cm}Z_{\rm vec}^-=&\left[\prod_{I<J}^{n} 2\sinh{ \tfrac{ \pm \phi_I \pm \phi_J}{2}} \prod_I^{n} 2\cosh{\tfrac{\pm \phi_I}{2}} \right] \times \frac{1}{2\sinh{ \frac{\pm \epsilon_- + \epsilon_+}{2}} \, \prod_{i=1}^{N} 2\cosh{ \frac{\pm \alpha_i + \epsilon_+  }{2}}} \cdot \prod_{I=1}^{n} \frac{ 2\cosh{ \frac{\pm \phi_I + 2\epsilon_+}{2}}} {2\cosh{ \frac{\pm \phi_I \pm \epsilon_- + \epsilon_+  }{2}}}\nonumber\\
	    &\times \prod_{I=1}^{n} \frac{2 \sinh{\epsilon_+} }{ 2 \sinh{\frac{\pm \epsilon_- + \epsilon_+ }{2}}  2\sinh{ \frac{\pm 2\phi_{I} \pm \epsilon_- + \epsilon_+}{2}} \, \prod_{i=1}^{N} 2\sinh{ \frac{\pm \phi_{I} \pm \alpha_i + \epsilon_+}{2}} }
		\prod_{I < J}^{n} \frac{ 2\sinh{ \frac{\pm \phi_{I} \pm \phi_{J} + 2\epsilon_+}{2} }}{2\sinh{ \frac{\pm \phi_{I} \pm \phi_{J} \pm \epsilon_- + \epsilon_+}{2}}}
\end{align}
with $O(k)_-$ when  $k = 2n + 1$;
\begin{align}
\hspace*{-0.5cm}Z_{\rm vec}^-=&\left[\prod_{I<J}^{n-1} 2\sinh{ \tfrac{ \pm \phi_I \pm \phi_J}{2}}  \prod_I^{n-1} 2\sinh{(\pm \phi_I)} \right]   \\
	    &\times \frac{2\cosh{\epsilon_+}}{2\sinh{ \frac{\pm \epsilon_- + \epsilon_+}{2}} \,2\sinh{ (\pm \epsilon_- + \epsilon_+)} \, \prod_{i=1}^{N} 2\sinh{ (\pm \alpha_i + \epsilon_+)}} \cdot \prod_{I=1}^{n-1} \frac{ 2\sinh{ (\pm \phi_I + 2\epsilon_+) } } {2\sinh{ (\pm \phi_I \pm \epsilon_- + \epsilon_+)} } \nonumber \\
	    &\times \prod_{I=1}^{n-1} \frac{2 \sinh{\epsilon_+} }{ 2 \sinh{\frac{\pm \epsilon_- + \epsilon_+ }{2}}  2\sinh{ \frac{\pm 2\phi_{I} \pm \epsilon_- + \epsilon_+}{2}} \, \prod_{i=1}^{N} 2\sinh{ \frac{\pm \phi_{I} \pm \alpha_i + \epsilon_+}{2}} }
		\prod_{I < J}^{n-1} \frac{ 2\sinh{ \frac{\pm \phi_{I} \pm \phi_{J} + 2\epsilon_+}{2} }}{2\sinh{ \frac{\pm \phi_{I} \pm \phi_{J} \pm \epsilon_- + \epsilon_+}{2}}} \nonumber
	\end{align}
with $O(k)_-$ when  $k = 2n$;
\begin{equation}
\hspace*{-1.3cm}Z_{\rm anti}^+=\left( \frac{\prod_{i=1}^{N} 2\sinh{\frac{m \pm \alpha_i}{2}}}{2\sinh{\frac{m \pm \epsilon_+}{2}}} \prod_{I=1}^{n} \frac{2\sinh{\frac{\pm\phi_I \pm m - \epsilon_-}{2}}}{2\sinh{\frac{\pm\phi_I \pm m - \epsilon_+}{2}}} \right)^{\chi} \prod_{I=1}^{n} \frac{2 \sinh{\frac{\pm m - \epsilon_-}{2}} \prod_{i=1}^{N} 2\sinh{\frac{\pm\phi_I \pm \alpha_i - m}{2}} }{2\sinh{\frac{\pm m - \epsilon_+}{2}} \sinh{\frac{\pm 2\phi_I \pm m - \epsilon_+}{2}} } \prod_{I<J}^{n} \frac{2\sinh{\frac{\pm\phi_I \pm \phi_J \pm m - \epsilon_-}{2}}}{2\sinh{\frac{\pm\phi_I \pm \phi_J \pm m - \epsilon_+}{2}}}
\end{equation}
from 5d antisymmetric hypermultiplet for $O(k)_+$;
	\begin{equation}
\hspace*{-0.7cm}Z_{\rm anti}^-=\frac{\prod_{i=1}^{N} 2\cosh{\frac{m \pm \alpha_i}{2}}}{2\sinh{\frac{m \pm \epsilon_+}{2}}} \cdot \prod_{I=1}^{n} \frac{2\cosh{\frac{\pm\phi_I \pm m - \epsilon_-}{2}}}{2\cosh{\frac{\pm\phi_I \pm m - \epsilon_+}{2}}}  \frac{2 \sinh{\frac{\pm m - \epsilon_-}{2}} \prod_{i=1}^{N} 2\sinh{\frac{\pm\phi_I \pm \alpha_i - m}{2}} }{2\sinh{\frac{\pm m - \epsilon_+}{2}} \sinh{\frac{\pm 2\phi_I \pm m - \epsilon_+}{2}} } \cdot \prod_{I<J}^{n} \frac{2\sinh{\frac{\pm\phi_I \pm \phi_J \pm m - \epsilon_-}{2}}}{2\sinh{\frac{\pm\phi_I \pm \phi_J \pm m - \epsilon_+}{2}}}
	\end{equation}
for $O(k)_-$ when  $k = 2n + 1$;
\begin{eqnarray}
Z_{\rm anti}^-&=&\frac{2 \cosh{\frac{\pm m - \epsilon_-}{2}} \prod_{i=1}^{N} 2\sinh{(m \pm \alpha_i)}}{2\sinh{\frac{m \pm \epsilon_+}{2}}\, 2\sinh{(m \pm \epsilon_+)}} \\
&&\times \prod_{I=1}^{n-1} \frac{2\sinh{(\pm\phi_I \pm m - \epsilon_-)}}{2\sinh{(\pm\phi_I \pm m - \epsilon_+)}} \frac{2 \sinh{\frac{\pm m - \epsilon_-}{2}} \prod_{i=1}^{N} 2\sinh{\frac{\pm\phi_I \pm \alpha_i - m}{2}} }{2\sinh{\frac{\pm m - \epsilon_+}{2}} \sinh{\frac{\pm 2\phi_I \pm m - \epsilon_+}{2}} } \cdot
\prod_{I<J}^{n-1} \frac{2\sinh{\frac{\pm\phi_I \pm \phi_J \pm m - \epsilon_-}{2}}}{2\sinh{\frac{\pm\phi_I \pm \phi_J \pm m - \epsilon_+}{2}}}\nonumber
\end{eqnarray}
for $O(k)_-$ when  $k = 2n$;
	\begin{align}\label{Sp(N)-fund-1}
	    Z_{\rm fund}^+=&  \prod_{l=1}^{N_f}\left(\left( 2\sinh{\tfrac{m_l}{2}} \right)^{\chi} \; \prod_{I=1}^{n} 2\sinh{\tfrac{\pm \phi_I + m_l}{2}}\right)
	\end{align}
from $N_f$ fundamental hypermultiplets for $O(k)_+$;
	\begin{align}\label{Sp(N)-fund-2}
	    Z_{\rm fund}^-=& \prod_{l=1}^{N_f}  \left(2\cosh{\tfrac{m_l}{2}}  \prod_{I=1}^{n} 2\sinh{\tfrac{\pm \phi_I + m_l}{2}}\right)
	\end{align}
for $O(k)_-$ when  $k = 2n + 1$;
	\begin{align}\label{Sp(N)-fund-3}
	    Z_{\rm fund}^-=&  \prod_{l=1}^{N_f}  \left(2\sinh{m_l}   \prod_{I=1}^{n-1} 2\sinh{\tfrac{\pm \phi_I + m_l}{2}}\right)
	\end{align}
for $O(k)_-$ when  $k = 2n$. When one considers the index with $n_A=0$,
of course $Z_{\rm anti}^\pm$ factors are dropped from the integrand.

\begin{figure}[t!]
  \begin{center}
    \includegraphics[width=6cm]{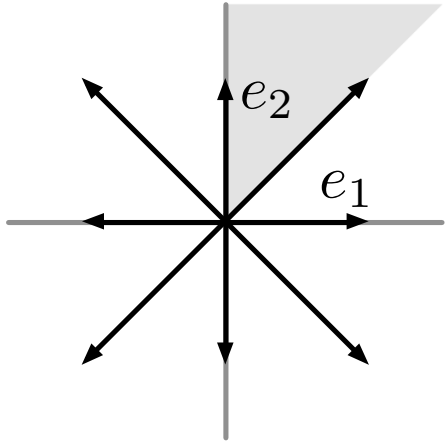}
\caption{The charges for the $Sp(1)$ index at $k=4$. The charges $\pm 2e_1$,
$\pm 2e_2$ are not shown. We chose $\eta$ in the shaded chamber.}\label{sp1-pole}
  \end{center}
\end{figure}
In all of the above integrands, the arguments are written in the form of
$\sinh\left(\frac{Q(\phi)+\cdots}{2}\right)$, where $Q$ is the weight
of the chiral or Fermi multiplet responsible for this factor.
The contour integral is understood as the sum of Jeffrey-Kirwan residues
with a chosen $\eta$. Here, any choice of $\eta$ will provide
the same result. We checked the behavior of poles carefully for the $Sp(1)$
theory with one antisymmetric hypermultiplet, up to $k=4$ instanton order.
The case  with $k=1$ has no integral. The case with $k=2$ either has rank $1$
for $O(2)_+$, where the formulae of section 2.2 applies, or has no integral for $O(2)_-$.
The case with $k=3$ again has at most $1$ integral. The case with $k=4$ has
rank $2$ for $O(4)_+$. In $2$ dimensional $h^\ast$, we take $\eta$ in the shaded
chamber in Fig.~\ref{sp1-pole}. All hyperplane arrangements
are projective, fulfilling the condition posed in \cite{Benini:2013xpa}.
In fact at $k=4$, all poles are non-degenerate, which are trivially
projective.

In the analysis of sections 3 and 4, we used the iterated integrals over
$z_I=e^{\phi_I}$ with $e^{-\epsilon_+}\rightarrow t\ll 1$ (for 1d hypers), $T\gg 1$
(for 1d twisted hypers) replacements.
We have checked the equivalence of the two rules for $Sp(1)$ theory
till $k=4$, similar to what we explained for $U(N)$ $k=2$ in section 3.2.
With $\eta$ chosen in the shaded chamber shown in Fig.~\ref{sp1-pole}, we integrate
over $z_1=e^{\phi_1}$ first and then over $z_2=e^{\phi_2}$. From the integrals
over unit circles, we encounter $372=292+80$ possible poles. $292$ poles are
unambiguously inside the unit circle, and are those kept from the Jeffrey-Kirwan rule.
The $80$ extra poles are ambiguous but show pairwise cancelations,
as explained around (\ref{pairwise}), proving the equivalence.
In $Z_{\textrm{1-loop}}$, some poles are actually absent
because $\sinh$ factors in the numerators vanish at the poles. (Similar phenomena were
repeatedly observed for the $U(N)$ case, while deriving the Young diagram rules.)
Taking these into account, we have $324$ nonzero residues from our unit circle
integrations, and $260$ nonzero Jeffrey-Kirwan residues: $64$ extra
residues from the former cancel pairwise. Finally, identifying $t$ and $T$ at
the final stage, $188$ nonzero poles remain. Similar structures
are found for $Sp(N)$ at $O(4)_+$, although there are more poles.

\subsection{Extra decoupled states and continua}

In this subsection, we explain in various examples how one can factor out
$Z_{\rm extra}$ from the index of ADHM quantum mechanics, and obtain
$Z_{\rm QFT}=\frac{Z_{\rm QM}}{Z_{\rm extra}}$ of our interest. The examples
that we shall mainly discuss are $Sp(N)$ gauge theories with $0\leq N_f\leq 8$ fundamental
and $n_A=1$ antisymmetric hypermultiplets, and $U(N)$ gauge theories with $N_f$ fundamental
hypermultiplets and 5d Chern-Simons level $\kappa$ satisfying $N_f+2|\kappa|\leq 2N$.

\subsubsection{$Sp(N)$ theories for 5d SCFTs}

We first discuss the $Sp(N)$ theories with $N_f\leq 7$ fundamental and $1$
antisymmetric hypermultiplets. The case with $N_f=8$ fundamental hypermultiplets is
discussed in the next subsection separately.
The ADHM quantum mechanics describes the $k$ D0-branes along $0$ direction, $N$ D4-branes along
$01234$ directions, $N_f$ D8-branes and one O8-plane along $0\cdots 8$ directions. The
scalars $\varphi_I$ from the ADHM vector multiplet represent D0-branes' positions along
the $9$ direction, transverse to all D-branes. The general analysis at the beginning of
this section says that there is no pole at infinities of $\varphi_I$. One can expect this,
since $N_f\leq 7$ D8-branes do not completely cancel the charge of the O8-plane,
so that the dilaton runs along the $9$ direction. D0-brane's mass increases linearly
in $\varphi_I$, explaining the absence of the continuum for $\varphi_I$. However, there is
an extra contribution $Z_{\rm extra}$ from D0-branes which are unbound to D4-branes, but
are bound to D8-O8 only. Since the motion of D0's along the worldvolume of D8-O8 is fully
gapped by the chemical potentials $\epsilon_1,\epsilon_2,m$, one could compute the multi-particle
index for the D0-particles in $8+1$ dimensions. These D0-D8-O8 bound states'
index will never refer to the electric charge fugacities $\alpha_i$ on D4. So to detect the
possible $Z_{\rm extra}$ factor, it suffices to examine the expansion of $Z_{\rm QM}$ in
the Coulomb VEV $e^{-\alpha_i}$ with $\alpha_1>\alpha_2>\cdots > 0$, and study the sector which carries zero electric charges. The index can be written as
\begin{equation}
  Z_{\rm QM}(\alpha,\epsilon_{1,2},v,q)=Z^{(0)}(\epsilon_{1,2},v,q)
  Z^{(1)}(\alpha,\epsilon_{1,2},v,q)\ .
\end{equation}
$v=e^{-m}$ is the flavor fugacity rotating
the antisymmetric hypermultiplet. $Z^{(1)}$ is given by
\begin{equation}
  Z^{(1)}=1+\sum_{n_i} Z_{n_i} e^{-n_i\alpha_i}\ .
\end{equation}
One can write
\begin{equation}
  Z_{N_f}^{(0)}=PE\left[f_{N_f}(t,u,v,y_i,q)\right]\equiv\exp\left[\sum_{n=1}^\infty\frac{1}{n}
  f_{N_f}(t^n,u^n,v^n,y_i^n,q^n)\right]\ ,
\end{equation}
where $t=e^{-\epsilon_+}$, $u=e^{-\epsilon_-}$, $v=e^{-m}$, $y_i=e^{m_i/2}$ with
$i=1,\cdots,N_f$. $f_{N_f}$ is the single particle index. One finds
\begin{eqnarray}\label{9d-SYM-inst}
  f_0&=&-\frac{t^2}{(1-tu)(1-t/u)(1-tv)(1-t/v)}\ q\hspace{3cm}{\rm for}\ N_f=0\\
  f_{N_f}&=&-\frac{t^2}{(1-tu)(1-t/u)(1-tv)(1-t/v)}\
  q\chi(y_i)^{SO(2N_f)}_{{\bf 2^{N_f-1}}}\ \ \ \ \ \ {\rm for}\ 1\leq N_f\leq 5\nonumber\\
  f_6&=&-\frac{t^2}{(1-tu)(1-t/u)(1-tv)(1-t/v)}
  \left[q\chi(y_i)_{\bf 32}^{SO(12)}+q^2\right]\nonumber\\
  f_7&=&-\frac{t^2}{(1-tu)(1-t/u)(1-tv)(1-t/v)}\left[q\chi(y_i)_{\bf 64}^{SO(14)}
  +q^2\chi(y_i)_{\bf 14}^{SO(14)}\right]\ .\nonumber
\end{eqnarray}
${\bf 2}^{N_f-1}$ is the chiral spinor representation of $SO(2N_f)$ (in our convention),
whose highest weight state contributes $y_1y_2\cdots y_{N_f}$ to the character.
We have checked these forms of $f_{N_f}$ up to $q^4$ order from the $Sp(1)$ index with all
fugacities kept, and the same result up to $q^3$ from the $Sp(2)$ index. In section \ref{D0D8subsection}, we shall derive these indices from the D0-D8-O8 system,
which proves that $Z^{(0)}_{N_f}=PE[f_{N_f}]$ is indeed $Z_{\rm extra}$.

In the remaining part of this subsection, we show that this $Z^{(0)}$ is precisely what
one expects from the type I' string theory with $N_f$ D8-branes, which should exhibit
$E_{N_f+1}$ gauge symmetry on the 8-branes' worldvolume (from its duality to heterotic
strings \cite{oai:arXiv.org:hep-th/9608111}). To see this, one has to combine $Z^{(0)}$
with the contribution to the index from perturbative type I' string theory.
$N_f$ D8-branes and an O8-plane host massless degrees given by  the 9d SYM theory
with $SO(2N_f)$ gauge group. Nonperturbative enhancement $SO(2N_f)\rightarrow E_{N_f+1}$
is expected from string duality, where $E_{N_f+1}$ includes the D0-brane charge in its
Cartan \cite{oai:arXiv.org:hep-th/9608111}. So the nonperturbative index of the type I'
theory should be that of the 9d $E_{N_f+1}$ SYM theory. In this $8+1$ dimensional setting,
$m$ plays the role of 8d Omega background parameter together with $\epsilon_{1,2}$, providing
an IR regulator of the 8d multi-particle calculus.

Let us explain the perturbative index first. The index of the 9d $SO(2N_f)$ SYM
is defined referring to the same $2$ supercharges that we used to define our instanton
index. The $16$ supercharges preserved by the D8-O8 system can be decomposed according to their
representations of $SO(4)\times SO(4)=SU(2)_l\times SU(2)_r\times SU(2)_R\times SU(2)_F$ symmetry. The first $SO(4)=SU(2)_l\times SU(2)_r$ is the spatial rotation on the common worldvolume of D4-D8-O8. Second $SO(4)=SU(2)_R\times SU(2)_F$ is the rotation on D8-O8 worldvolume transverse to D4. $SU(2)_R$ was the R-symmetry of 5d $\mathcal{N}=1$ theory. $SU(2)_F$ with the chemical
potential $m$ rotates the $Sp(N)$ antisymmetric hypermultiplet. Denoting by $a=1,2$ the doublet
index for $SU(2)_F$, the $16$ supercharges can be written as
\begin{equation}\label{16-SUSY-decompose}
  Q_\alpha^a\ ,\ \ Q_\alpha^A\ ,\ \ \bar{Q}_{\dot\alpha}^a\ ,\ \ \bar{Q}_{\dot\alpha}^A\ ,
\end{equation}
where $\alpha,\dot\alpha,A$ indices are for $SU(2)_l,SU(2)_r,SU(2)_R$ doublets as before.
These supercharges satisfy reality conditions.
9d SYM has half-BPS W-bosons and their superpartners in their BPS spectrum, in the Coulomb
branch where one real scalar is given nonzero VEV. The $SO(2N_f)$ electric charges have
fugacities $y_i\equiv e^{m_i/2}$, which were introduced in 5d SYM as flavor fugacities.
Let us write the $32\times 32$ gamma matrix in 10d as
$(\Gamma^0,\Gamma^9)={\bf 1}_{8}\otimes(\sigma_2,\sigma_1)$,
$\Gamma^i=\gamma^i\otimes\sigma_3$, with $\gamma^i$ given by the $SO(8)$ gamma matrices
($i=1,\cdots,8$). The BPS condition for the half-BPS W-boson is one of
$\Gamma^{09}\epsilon=\pm i\epsilon$ in the 10d chiral Majorana spinor notation.
The SUSY parameter $\epsilon$ satisfies the 10d chirality condition
$\gamma^{1\cdots 8}\otimes\sigma_3\epsilon=\epsilon$. W-bosons' BPS condition says
that $\epsilon$ is either chiral or anti-chiral $SO(8)$ spinors. In our notation,
the preserved supercharges are either $Q^a_\alpha,\bar{Q}^A_{\dot\alpha}$ or $Q^A_\alpha,\bar{Q}^a_{\dot\alpha}$.
Since our indices are always defined using $\bar{Q}^{A=1}_{\dot\alpha=\dot{1}}$ and
$\bar{Q}^{A=2}_{\dot\alpha=\dot{2}}$, the sector which is captured by our index contains
W-bosons preserving the former. The broken supercharges $Q^A_\alpha,\bar{Q}^a_{\dot\alpha}$
provide Goldstone fermion zero modes, which contribute to the single particle index of 9d
W-bosons. The $4$ pairs of fermionic oscillator from these Goldstinos provide a factor
\begin{equation}
  2\sinh\frac{\epsilon_1 }{2}\cdot 2\sinh\frac{\epsilon_2}{2}\cdot
  2\sinh\frac{m+\epsilon_+}{2}\cdot 2\sinh\frac{m-\epsilon_+}{2}=
  \chi^{SO(8)}({\bf 8}_v)-\chi^{SO(8)}({\bf 8}_c)
\end{equation}
to the index, where
\begin{equation}
  \chi^{SO(8)}({\bf 8}_v)=\chi^{SO(8)}({\bf 8}_s)\equiv(t+t^{-1})(u+u^{-1}+v+v^{-1})\ ,\ \ \chi^{SO(8)}({\bf 8}_c)\equiv t^2+2+t^{-2}+(u+u^{-1})(v+v^{-1})
\end{equation}
are the $SO(8)$ characters of the vector, spinor, conjugate spinor representations.
${\bf 8}_v$ and ${\bf 8}_c$ are for the W-bosons $A_\mu$ and superpartner fermions
$\Psi$ in 9d SYM. The index also acquires contribution from $8$ bosonic
zero modes for the translation on $\mathbb{R}^8$. They provide the factor
\begin{equation}
  \frac{1}{\left(2\sinh\frac{\epsilon_1}{2}\cdot 2\sinh\frac{\epsilon_2}{2}\cdot
  2\sinh\frac{m+\epsilon_+}{2}\cdot 2\sinh\frac{m-\epsilon_+}{2}\right)^2}
\end{equation}
in the index. One should also consider $\chi_{\bf adj}^{SO(2N_f)}(y_i)^+$ factor
for the W-bosons, where the $+$ superscript denotes that
only the positive roots contribute to this character. This is because we are counting
only W-bosons and their superpartners in the Coulomb branch of the 9d theory, without
anti-W-bosons or the massless Cartans. So one obtains
\begin{equation}\label{9d-SYM-pert}
  f_{\rm 9d\ SYM}=\frac{\chi_{\bf adj}^{SO(2N_f)}(y_i)^+}
  {2\sinh\frac{\epsilon_1}{2}\cdot 2\sinh\frac{\epsilon_2}{2}\cdot
  2\sinh\frac{m+\epsilon_+}{2}\cdot 2\sinh\frac{m-\epsilon_+}{2}}=
  -\frac{t^2\chi_{\bf adj}^{SO(2N_f)}(y_i)^+}{(1-tu)(1-t/u)(1-tv)(1-t/v)}\ .
\end{equation}
Note that the four factors in the denominator can be understood as the four
complex zero modes on $\mathbb{C}^4=\mathbb{R}^8$, indicating that this is coming
from 8 dimensional particles.

Combining (\ref{9d-SYM-pert}) and (\ref{9d-SYM-inst}) together, we now show
that one obtains the single particle index for the W-bosons of 9d $E_{N_f+1}$ SYM.
One first finds that at $N_f=0$, $E_1=SU(2)$ adjoint decomposes into $3$ states
which have $U(1)_I$ instanton charges $0,+1,-1$, respectively. The latter two are
the non-perturbative enhanced symmetry generators. Adjoint representation of
$E_2=SU(2)\times U(1)$, which is ${\bf 3}+{\bf 1}$ in $SU(2)$, decomposes
in $SO(2)\times U(1)_I$ to two neutral generators, and two non-perturbative
generators carrying $q^{\pm 1}y_1^{\pm 1}$. $E_3=SU(3)\times SU(2)$ contains
the perturbative $SO(4)\times U(1)_I=SU(2)\times SU(2)\times U(1)_I$ in the following
way. The second $SU(2)$ of $SO(4)$ is the same as the $SU(2)$ factor of $E_3$,
while $SU(3)$ adjoint branches to the remaining $SU(2)\times U(1)_I$ irreps as
\begin{equation}
  {\bf 8}\rightarrow{\bf 1}_0+{\bf 3}_{0}+{\bf 2}_1+{\bf 2}_{-1}\ .
\end{equation}
The branching rules of the $E_{N_f+1}$ adjoints, with $N_f\geq 4$, to
$SO(2N_f)\times U(1)_I$ irreps are
\begin{eqnarray}\label{En-adjoint-decompose}
  E_4=SU(5)&:&{\bf 24}\rightarrow{\bf 1}_0+{\bf 15}_0+{\bf 4}_1+\overline{\bf4}_{-1}\nonumber\\
  E_5=SO(10)&:&{\bf 45}\rightarrow{\bf 1}_0+{\bf 28}_0+({\bf 8_s})_1+({\bf 8_s})_{-1}\nonumber\\
  E_6&:&{\bf 78}\rightarrow{\bf 1}_0+{\bf 45}_0+{\bf 16}_1+\overline{\bf 16}_{-1}\nonumber\\
  E_7&:&{\bf 133}\rightarrow{\bf 1}_0+{\bf 66}_0+{\bf 32}_1+{\bf 32}_{-1}
  +{\bf 1}_2+{\bf 1}_{-2}\nonumber\\
  E_8&:&{\bf 248}\rightarrow{\bf 1}_0+{\bf 91}_0+{\bf 64}_1+\overline{\bf 64}_{-1}
  +{\bf 14}_2+{\bf 14}_{-2}\ .
\end{eqnarray}
The subscripts all denote the $U(1)_I$ instanton number. The first ${\bf 1}_0$'s all
denote the generator of the $U(1)_I$, while the next $U(1)_I$ singlets are all
adjoints of $SO(2N_f)$. As explained around (\ref{9d-SYM-pert}), only the positive
roots from the $SO(2N_f)$ adjoints contribute to the index. Among the remaining non-singlets
on the right hand side, only the states which have positive $U(1)_I$ charge will contribute to
the index, as our index counts instantons but not anti-instantons. The instanton
contribution to the 9d $E_{N_f+1}$ SYM index required from (\ref{En-adjoint-decompose})
and the preceding branching rules indeed appear in (\ref{9d-SYM-inst}) for all $N_f$.
So the addition of (\ref{9d-SYM-pert}) and (\ref{9d-SYM-inst}) precisely captures the
contribution from the W-bosons of 9d $E_{N_f+1}$ SYM.

So we conclude that $Z^{(0)}_{N_f}=PE[f_{N_f}]$ with $f_{N_f}$ given by (\ref{9d-SYM-inst})
is precisely the $Z_{\rm extra}$ factor expected from string theory.
This will be reconfirmed in section 3.4.3 by a direct computation of the D0-D8-O8 index,
without assuming type I'-heterotic duality. The index for the 5d SCFT is thus given by
$Z_{\rm QFT}=\frac{Z_{\rm QM}}{Z_{\rm extra}}$, which shall be used in section 4.

Before closing this subsection, we discuss the $Sp(N)$ partition function with
$N_f$ fundamental hypermultiplets at $n_A=0$. This engineers another class of 5d SCFTs,
which can be realized by M-theory on suitable CY$_3$ \cite{Intriligator:1997pq}.
For $Sp(1)$, this should yield the same 5d SCFT indices as those obtained from the
quantum mechanics with $n_A=1$. The only issue is that the two descriptions may have
different $Z_{\rm extra}$ factors. At all $N$, including $N=1$, the condition for
the contour integrand $Z_{\textrm{1-loop}}$ to vanish at $|\varphi|\rightarrow\infty$
is $N_f<2N+4$. $Z_{\textrm{1-loop}}$ approaches a constant asymptotically for $N_f=2N+4$.
So we study the $Sp(1)$ ADHM instanton calculus at $N_f\leq 6$. Apart from the Calabi-Yau
engineering of \cite{Intriligator:1997pq}, one can also realize this system from
branes. Namely, we start from the 5-brane web system on the right side of Fig. \ref{U(2)-web},
corresponding to the $U(2)$ theory with $N_f=4$ quarks. Then put the extra O7-plane and
$4$ D7-plane at the center of the box in this diagram. The O7 changes the gauge groups
from $U(2)$ to $Sp(1)$, and the $4$ quarks provided by D5's reduce to $2$ by orientifolding.
With $4$ more quarks provided by $4$ D7-branes, we have $N_f=6$ quarks in total, realizing
our $Sp(1)$ theory at $N_f=6$.

When $N_f\leq 5$, there is no noncompact moduli in the ADHM mechanics so we expect
$Z_{\rm extra}=1$. This is supported by the analysis
of \cite{Kim:2012gu}. When $N_f=6$, we find $Z_{\rm QM}=Z_{\rm QFT}Z_{\rm extra}$,
where $Z_{\rm QFT}$ is the same QFT partition function that we derived with $n_A=1$, and
\begin{equation}\label{continuum-nA=0}
  Z_{\rm extra}={\rm PE}\left[-\frac{(1+t^2)q^2}{2(1-tu)(1-t/u)}\right]\ .
\end{equation}
This fact was confirmed up to $q^4$ order. Since (\ref{continuum-nA=0})
comes with a fractional coefficient, it clearly has to do with the continuum.
From the brane setting of the previous paragraph, the continuum has to do with the
D1-brane escaping the QFT in Fig. \ref{U(2)-web}. The $q^2$ behavior of the exponent
of (\ref{continuum-nA=0}) is easy to understand, since single instanton is a D1-brane
which is suspended between O7-NS5, which cannot escape to infinity. With this decoupled
factor understood, we confirmed for all $N_f\leq 6$ that $Z_{\rm QFT}$ computed from
the ADHM mechanics with $n_A=0$ and $n_A=1$ are the same, up to $q^4$ order.

\subsubsection{$Sp(1)$ theory for 6d SCFT on M5-M9}

Now we turn to the case with $N_f=8$, for D0-branes probing $N$ D4, $8$ D8's and an O8.
Again the $x^9$ direction is a half-line $\mathbb{R}^+$. The difference from the cases
with $N_f\leq 7$ is that the D8-brane charges completely cancel between $8$ D8's and
one O8. The dilaton asymptotically becomes a constant as one moves away from the brane
system along $x^9$.  So this system uplifts to M-theory on
$\mathbb{R}^{8,1}\times\mathbb{R}^+\times S^1$ at strong coupling. Our 5d SYM is thus
a low energy description of circle compactified 6d $(1,0)$ theory for the M5-M9 system.
In this case, there are poles at infinities of cylinders in $Z_{\textrm{1-loop}}$,
since D0's can move away from the 8-branes with a continuum.
Following the same strategy as the cases with $N_f\leq 7$, we first extract
the $Sp(1)$ neutral $Z^{(0)}$, as this should contain all possible $Z_{\rm extra}$ factors. Again writing
$Z^{(0)}=PE[f]$, $f$ is given by
\begin{eqnarray}\label{Nf=8-neutral}
  f&=&\left[\frac{t(v+v^{-1}-u-u^{-1})}{(1-tu)(1-t/u)}
  -\frac{(t+t^3)(u+u^{-1}+v+v^{-1})}{2(1-tu)(1-t/u)(1-tv)(1-t/v)}\right]\frac{q^2}{1-q^2}\\
  &&-\frac{t^2}{(1-tu)(1-t/u)(1-tv)(1-t/v)}\left[\chi(y_i)^{SO(16)}_{\bf 120}
  \frac{q^2}{1-q^2}+\chi(y_i)^{SO(16)}_{\bf 128}\frac{q}{1-q^2}\right]\ ,\nonumber
\end{eqnarray}
where we checked the $q$ dependence up to 4-instanton order from the $Sp(1)$ theory.
Namely, the above expression is obtained with $\frac{q^2}{1-q^2}\rightarrow q^2+q^4$ and
$\frac{q}{1-q^2}\rightarrow q+q^3$.
So all properties that we show below are proven up to this order. ${\bf 120}$ and ${\bf 128}$
are the adjoint and chiral spinor representations of $SO(16)$.
We now explain the terms in (\ref{Nf=8-neutral}) which should go to $Z_{\rm extra}$.

We first study the second line of (\ref{Nf=8-neutral}). This provides a single particle
index for certain $8+1$ dimensional particles, thus should go to the factorized
$Z_{\rm extra}$ from bulk degrees. Let us first explain what we expect from the string
dualities and heterotic M-theory. Heterotic M-theory was proposed in \cite{Horava:1995qa}
as a strong coupling limit of $E_8\times E_8$
heterotic string theory. It has a low energy limit described by 11d supergravity on
$\mathbb{R}^{9,1}\times I$, where $I=S^1/\mathbb{Z}_2$ is an interval. There are two
fixed planes of the $\mathbb{Z}_2$ action at both ends of $I$, which we call the
M9-planes. Each M9-plane hosts $E_8$ gauge symmetry, having a massless sector of 10d
$E_8$ super-Yang-Mills theory. One can compactify the heterotic M-theory on a small
circle with radius $R$ to $\mathbb{R}^{8,1}\times I$. The circle compactification can
be made with nonzero $E_8\times E_8$ Wilson lines on two 10d SYM theories on M9-planes.
In particular, consider the following Wilson line
\begin{equation}\label{E8-holonomy}
  RA^{E_8}=\left(0,0,0,0,0,0,0,1\right)
\end{equation}
for each $E_8$ SYM. Our convention is to pick $8$ Cartans of $SO(16)\subset E_8$
which rotate $8$ orthogonal $2$-planes of $SO(16)$. The adjoint representation
${\bf 248}$ of $E_8$ decomposes in $SO(16)$ to
\begin{equation}
  {\bf 248}\rightarrow{\bf 120}+{\bf 128}\ .
\end{equation}
The holonomy (\ref{E8-holonomy}) is such that $e^{2\pi iRA}$ leaves ${\bf 120}$ invariant,
while giving $-1$ sign to the spinors. So the compactification with this holonomy yields
a 10d theory with $SO(16)\times SO(16)$ symmetry. This is the type I' string theory on
$\mathbb{R}^{8,1}\times I$, which has two orientifold 8-planes (O8-planes) at the two ends
of $I$. Each O8-plane has
$8$ D8-branes on top of it. A crucial part of this identification is that the
nonperturbative D0-brane physics of type I' theory should enable us to see the
11th circle's KK modes. We will show that the second line of (\ref{Nf=8-neutral}) achieves
it.

Note that the fugacities $q,y_i$ in (\ref{Nf=8-neutral}), especially on the second line,
probe the momentum and $SO(16)$ charges in the background of Wilson
line (\ref{E8-holonomy}). Here, note that the charges of the type I' theory and
the heterotic M-theory are related by \cite{Aharony:1997pm}
\begin{equation}
  k=2P-RA^{E_8}\cdot F^{E_8}=2P-F_8\ ,
\end{equation}
where $k$ is the type I' instanton charge, $P$ is the circle momentum
of heterotic M-theory, $F^{E^8}$ are the $E_8$ charges, $A^{E_8}$ is the holonomy
(\ref{E8-holonomy}). The expression in \cite{Aharony:1997pm} has
more shifts to $k$ on the right hand side, depending on the string winding number, which
is zero for all states captured by $Z^{(0)}$. The fugacities conjugate to $k,F_8$ are
more naturally viewed in the heterotic M-theory as
\begin{equation}
  q^ky_8^{F_8}=q^{2P}(y_8q^{-1})^{F_8}\ .
\end{equation}
If one replaces all $y_8$'s in $Z^{(0)}$ by $y_8q$, this effectively turns
off the background holonomy (\ref{E8-holonomy}). Conversely, shifting an $E_8$ fugacity
$y_8$ by $y_8q^{-1}$, one would obtain the type I' $SO(16)$ fugacity.

We would like to show that the second line of (\ref{Nf=8-neutral}) is what one expects
from the 10d $E_8$ SYM living on the M9-plane, compactified on a circle with Wilson line.
The index of 10d SYM compactified on a circle would be PE of
\begin{equation}\label{E8-expected}
 -\frac{t^2}{(1-tu)(1-t/u)(1-tv)(1-t/v)}\left[\chi^{E_8}_{\bf 248}(y_i)\sum_{P=-\infty}^\infty
 q^{2P}\right]^+
\end{equation}
where $y_i$ are $E_8$ fugacities, and $+$ superscript denotes that one only keeps the
modes whose fugacity factor $q^{2P}\prod_{i=1}^8y_i^{F_8}$ is smaller than $1$: namely,
we only keep BPS states rather than anti-BPS modes. To understand the type I' result,
we first replace $y_8$ by $y_8q^{-1}$, and then keep the BPS modes at $q\ll 1$.
Note that this replacement $y_8\rightarrow y_8 q^{-1}$ temporarily decomposes the
$E_8$ characters into $SO(14)$ characters. With ${\bf 248}\rightarrow{\bf 120}+{\bf 128}$
with $SO(16)$ subgroup understood, one finds
\begin{equation}
  \chi_{\bf 120}^{SO(16)}\rightarrow 1+\chi_{\bf 91}^{SO(14)}
  +(y_8^2+y_8^{-2})\chi_{\bf 14}^{SO(14)}\ ,\ \ \chi_{\bf 128}^{SO(16)}
  \rightarrow y_8\chi_{\bf 64}^{SO(14)}+y_8^{-1}\chi_{\bf \overline{64}}^{SO(14)}\ .
\end{equation}
So here, replacing all $y_8$ by $y_8q^{-1}$, (\ref{E8-expected}) becomes
\begin{equation}
  -\frac{t^2}{(1-tu)(1-t/u)(1-tv)(1-t/v)}\left[\chi^{SO(16)}_{\bf 120}(y_i)
  \frac{q^2}{1-q^2}+\chi^{SO(16)}_{\bf 128}(y_i)\frac{q}{1-q^2}+\chi^{SO(16)+}_{\bf 120}(y_i)\right]
\end{equation}
at $q\ll 1$, where $+$ superscript again denotes contribution from positive roots only.
The third term is what one expect from the 9d perturbative SYM with $SO(16)$ gauge group,
living on the O8-D8 system. The first two terms are the second line of
(\ref{Nf=8-neutral}). So the second line of (\ref{Nf=8-neutral}) is precisely what one
expects from the heterotic M-theory. This proves that the second line of (\ref{Nf=8-neutral})
should go to $Z_{\rm extra}$.

Then in (\ref{Nf=8-neutral}), we consider the term
\begin{equation}\label{11d-SUGRA}
  -\frac{(t+t^3)(u+u^{-1}+v+v^{-1})}{2(1-tu)(1-t/u)(1-tv)(1-t/v)}\frac{q^2}{1-q^2}
\end{equation}
on the first line. The overall coefficient $\frac{1}{2}$ shows that this is
clearly the continuum contribution. In fact, there is no way to turn on the FI term
with $O(k)$ gauge group, so that we cannot decouple the continuum from the Witten
index calculus. Although we do not have an account for the factor $\frac{1}{2}$,
in a way similar to \cite{Yi:1997eg,Sethi:1997pa}, we can derive all the dependence
on the fugacities from the continuum states in our problem, and further argue that
this term should go to $Z_{\rm extra}$.

To show this, we investigate the 11d supergravity spectrum on
$\mathbb{R}^{8,1}\times S^1\times\mathbb{R}^+$. The continuum is formed by the states
which propagate along $\mathbb{R}^+$. The $\mathbb{R}^8\times S^1$ part of the space has
a fully gapped spectrum,
either by having compact space or by having nonzero chemical potentials
for the rotations. So the gapped part of the spectrum can be computed by investigating
the supergravity multiplet, setting aside an overall fractional coefficient
which can only be determined by knowing the dynamics along $\mathbb{R}^+$ (and our
deformations in the index computation).
The factor $\frac{q^2}{1-q^2}$ simply reflects the fact that the KK modes of the 11d gravity
on circle with different $P$ have same spin contents in 10d. So the $t,u,v$ dependence of this
term can be computed from the 10d type I' supergravity. Also, since we are only
paying attention to the $\mathbb{R}^8$ part of the spectrum, we can replace $\mathbb{R}^+$
by $I=S^2/\mathbb{Z}_2$ and apply T-duality along this direction, after which the well
known type I supergravity spectrum will be relevant. The type I supergravity contains
a dilaton $\phi$, RR 2-form $C_2$, graviton $g_{\mu\nu}$,
dilatino $\lambda$, and the gravitino $\psi_\mu$. All of them are in the following
representation of $SO(8)$, rotating $\mathbb{R}^8$:
\begin{equation}\label{graviton}
  ({\bf 1}\oplus{\bf 28}\oplus{\bf 35}_v)_{\rm boson}\oplus({\bf 8}_s\oplus
  {\bf 56}_s)_{\rm fermion}=({\bf 8}_v\otimes{\bf 8}_v)_{\rm sym}\oplus
  ({\bf 8}_v\otimes{\bf 8}_c)\oplus({\bf 8}_c\otimes{\bf 8}_c)_{\rm anti}\ .
\end{equation}
The $SU(2)^4$ characters of ${\bf 8}_v,{\bf 8}_s,{\bf 8}_c$
on the right hand sides (with $(-1)^F$ signs for ${\bf 8}_s$, ${\bf 8}_c$) are
\begin{eqnarray}
  \chi({\bf 8}_v)&=&(t+t^{-1})(u+u^{-1}+v+v^{-1})\\
  \chi({\bf 8}_c)&=&-t^2-2-t^{-2}-(u+u^{-1})(v+v^{-1})\nonumber\\
  \chi({\bf 8}_s)&=&-(t+t^{-1})(u+u^{-1}+v+v^{-1})\ .\nonumber
\end{eqnarray}
From this, one can compute the index for the right hand side of (\ref{graviton}).
Note that the symmetrized and anti-symmetrized characters are given by
$\frac{f(t,u,v)^2\pm f(t^2,u^2,v^2)}{2}$, where $f$ is the character of
${\bf 8}_{v,c}$ appearing in the (anti)symmetrization. Multiplying this with
the factor $\frac{t^4}{(1-tu)^2(1-t/u)^2(1-tv)^2(1-t/v)^2}$ which comes from
the translation zero modes on $\mathbb{R}^8$, one obtains
\begin{equation}
  -\frac{(t+t^3)(u+u^{-1}+v+v^{-1})}{(1-tu)(1-t/u)(1-tv)(1-t/v)}\ .
\end{equation}
This shows that the contribution from 11d supergravity to the index
should be
\begin{equation}\label{gravity-index}
  PE\left[-\alpha\frac{(t+t^3)(u+u^{-1}+v+v^{-1})}{(1-tu)(1-t/u)(1-tv)(1-t/v)}
  \frac{q^2}{1-q^2}\right]\ ,
\end{equation}
with an unknown constant $\alpha$. However,
with the second line of (\ref{Nf=8-neutral}) accounted for by
the 10d $E_8$ SYM, note that the remaining part of the ADHM index takes the form of
\begin{equation}\label{remainder-Nf=8}
  PE\left[\frac{\rm regular}{(1-tu)(1-t/u)}\right]\cdot
  PE\left[-\frac{(t+t^3)(u+u^{-1}+v+v^{-1})}{2(1-tu)(1-t/u)(1-tv)(1-t/v)}
  \frac{q^2}{1-q^2}\right]
\end{equation}
where `regular' numerators do not diverge in the $\epsilon_{1,2},m\rightarrow 0$ limit.
Now, (\ref{remainder-Nf=8}) divided by (\ref{gravity-index}) should give the QFT
index which counts the 4d BPS particles, from the low energy decoupling of the
D0-D4-D8 system. But the ratio would count 4d particles only if $\alpha=\frac{1}{2}$:
otherwise, one would have a left-over 8d particle poles in the single
particle index, with coefficient $\alpha-\frac{1}{2}\neq 0$, contradicting with
the decoupling. So this proves $\alpha=\frac{1}{2}$, based on string theory decoupling.

Collecting all, we have shown that $Z_{\rm extra}=PE[f_{\rm extra}]$ with
\begin{eqnarray}
  f_{\rm extra}&=&
  -\frac{(t+t^3)(u+u^{-1}+v+v^{-1})}{2(1-tu)(1-t/u)(1-tv)(1-t/v)}\frac{q^2}{1-q^2}\\
  &&-\frac{t^2}{(1-tu)(1-t/u)(1-tv)(1-t/v)}\left[\chi(y_i)^{SO(16)}_{\bf 120}
  \frac{q^2}{1-q^2}+\chi(y_i)^{SO(16)}_{\bf 128}\frac{q}{1-q^2}\right]\nonumber
\end{eqnarray}
is the contribution from the string theory or UV sector.
The first term of (\ref{Nf=8-neutral}),
\begin{equation}
  \frac{t(v+v^{-1}-u-u^{-1})}{(1-tu)(1-t/u)}\frac{q^2}{1-q^2}
  =\frac{\sinh\frac{m+\epsilon_-}{2}\sinh\frac{m-\epsilon_-}{2}}
  {\sinh\frac{\epsilon_1}{2}\sinh\frac{\epsilon_2}{2}}\frac{q^2}{1-q^2}\equiv
  I_-(\epsilon_{1,2},m)\frac{q^2}{1-q^2}\ ,
\end{equation}
is not included in $Z_{\rm extra}$. It is part of the 6d QFT spectrum.

So far, we relied on the heterotic M-theory physics to factor out
$Z_{\rm extra}$ from the ADHM quantum mechanics index $Z_{\rm QM}$.
Again, we can directly compute $Z_{\rm extra}$ from the index of D0-D8-O8
quantum mechanics, without relying on unproved properties. See section 3.4.3.

So we provided a clear recipe to compute the index of the circle compactified 6d
$(1,0)$ SCFT on the M5-M9 system, $Z_{\rm QFT}=\frac{Z_{\rm QM}}{Z_{\rm extra}}$.
$Z_{\rm QFT}$ for this system will be studied elsewhere \cite{Kim:2014dza}.

\subsubsection{Direct computations of the D0-D8-O8 indices}\label{D0D8subsection}

The computations reported in this short section supplement the discussions of
sections 3.4.1 and 3.4.2. There we extracted out the neutral part $Z^{(0)}$ of
the D0-D4-D8-O8 index and argued that this contains $Z_{\rm extra}$ which is
deducible from string dualities, etc. Instead, we can simply derive the
$Z_{\rm extra}$ factors of the previous subsections directly from the D0-D8-O8
quantum mechanics. One can start from the gauged quantum mechanics for the open
strings connecting D0-D8-O8 with $O(k)$ gauge group. The field contents can be
easily obtained from the previous D0-D4-D8-O8 fields by dropping all
$N\times k$ bi-fundamental fields. The index is also obvious: one just uses
the index in section 3.3 after dropping all the $Z_{\textrm{1-loop}}$ factors from the
fields charged in $Sp(N)$. So we compute these indices, in all examples up to $q^4$ order,
and obtain
\begin{eqnarray}\label{type-I'}
  Z_{N_f=0}&=&{\rm PE}\left[-\frac{t^2q}{(1-tu)(1-t/u)(1-tv)(1-t/v)}\right]\\
  Z_{1\leq N_f\leq 5}&=&
  {\rm PE}\left[-\frac{t^2}{(1-tu)(1-t/u)(1-tv)(1-t/v)}
  q\chi(y_i)^{SO(2N_f)}_{\bf 2^{N_f-1}}\right]\nonumber\\
  Z_{N_f=6}&=&
  {\rm PE}\left[-\frac{t^2}{(1-tu)(1-t/u)(1-tv)(1-t/v)}
  \left(q\chi(y_i)^{SO(12)}_{\bf 32}+q^2\right)\right]\nonumber\\
  Z_{N_f=7}&=&
  {\rm PE}\left[-\frac{t^2}{(1-tu)(1-t/u)(1-tv)(1-t/v)}
  \left(q\chi(y_i)^{SO(14)}_{\bf 64}+q^2\chi(y_i)^{SO(14)}_{\bf 14}\right)\right]\nonumber
\end{eqnarray}
and
\begin{eqnarray}\label{M9-spectrum}
  Z_{N_f=8}&=&{\rm PE}\left[
  -\frac{(t+t^3)(u+u^{-1}+v+v^{-1})}{2(1-tu)(1-t/u)(1-tv)(1-t/v)}\frac{q^2}{1-q^2}\right.\\
  &&\left.-\frac{t^2}{(1-tu)(1-t/u)(1-tv)(1-t/v)}\left(\chi(y_i)^{SO(16)}_{\bf 120}
  \frac{q^2}{1-q^2}+\chi(y_i)^{SO(16)}_{\bf 128}\frac{q}{1-q^2}\right)\right]\ .\nonumber
\end{eqnarray}
These all directly justify the $Z_{\rm extra}$ factors that we argued using string dualities.
In particular, (\ref{type-I'}) supports the non-perturbative duality between the type I' and
heterotic strings by finding a spectrum which allows $E_{N_f+1}$ enhancement. (\ref{M9-spectrum}) supports that non-perturbative physics of type I' strings reconstructs the physics of M9-plane
compactified on a circle.

\subsubsection{$U(N)$ theories for 5d SCFTs}

Our last example is the $U(N)$ SYM with $N_f$ fundamental hypermultiplets and bare
Chern-Simons term at level $\kappa$, satisfying $N_f+2|\kappa|\leq 2N$. The indices
for the theories saturating the last inequality have $Z_{\rm extra}$
contributions. These partition functions are studied in great detail in
\cite{Bao:2013pwa,Hayashi:2013qwa,Bergman:2013ala,Bergman:2013aca,Taki:2013vka,Taki:2014pba}.

We first discuss the theories with $U(2)$ gauge group, with $N_f\leq 4$ fundamental
matters and CS level $\kappa$ satisfying $N_f+2|\kappa|\leq 2N$. The 5-brane webs
engineering some of these theories are shown in Fig.~\ref{U(2)-web}.
The $SU(2)$ part of the $U(2)$ gauge group is identified with the $Sp(1)$ gauge group,
while the overall $U(1)$ is non-dynamical. The information on the overall $U(1)$,
especially the CS level $\kappa$, should be
irrelevant for $Z_{\rm QFT}$, since the QFT is just the $Sp(1)$
theory coupled to $N_f$ fundamental hypermultiplets. So we expect
\begin{equation}
  \frac{Z^{U(2)}_{\rm QM}(N_f,\kappa)}{Z^{Sp(1)}_{\rm QFT}(N_f)}
  =Z_{\rm extra}^{U(2)}(N_f,\kappa)
\end{equation}
for all $\kappa$, where $Z^{Sp(1)}_{\rm QFT}$ is the QFT index that one obtains
by dividing $Z_{\rm QM}^{U(2)}$ by $Z_{\rm extra}^{U(2)}$. (We suppressed the
$\alpha_i,\epsilon_{1,2},m,y_i,\zeta$ dependence.)
At $N_f+2|\kappa|<2N$, there is no continuum from the string theory which are
attached to the instanton quantum mechanics, and the right hand side is $1$.
At $N_f+2|\kappa|=2N$, the right hand side is not $1$ and further experiences
a wall crossing as the FI parameter $\zeta$ changes.

Before explaining the results, one should realize that the 5d $Sp(N)$ theories can be
classified into two \cite{Douglas:1996xp}, labeled by two discrete theta angles.
Namely, there are two topologically distinct configurations due to
$\pi_4(Sp(N))=\mathbb{Z}_2$. This also descends to the two topologically distinct
configurations in the $O(k)$ ADHM quantum mechanics,
due to $\pi_0(O(k))=\mathbb{Z}_2$ \cite{Bergman:2013ala}. In both 5d/1d cases, the sector
with nontrivial element of $\mathbb{Z}_2$ has a relative $-1$ sign in the path integral.
So the instanton calculus rule $Z^k_{\theta=0}=\frac{Z^k_++Z^k_-}{2}$ changes
to \cite{Bergman:2013ala}
\begin{equation}
  Z^k_{\theta=\pi}=(-1)^{k}\frac{Z^k_+-Z^k_-}{2}\ .
\end{equation}
The overall factor of $(-1)^k$ was argued in \cite{Bergman:2013ala}
at $k=1,2$ in a somewhat indirect way. At $N_f=0$, the two cases with $\theta=0,\pi$
were shown (based on the instanton partition function calculus) in \cite{Bergman:2013ala}
to uplift to the so-called $E_1$ and $\tilde{E}_1$ theories, respectively
\cite{Morrison:1996xf}. With $N_f\geq 1$, the relative minus signs from $\mathbb{Z}_2$
nontrivial sector can be canceled by flipping the sign of a mass parameter. In the
following, we stick to our previous definition of $m_i$ parameters, which implies
that we should insert the relative minus sign for the $Z^k_-$ when we explicitly
write $\theta=\pi$. But this is related to new SCFT only when $N_f=0$. In other cases,
inserting extra minus sign is simply changing our convention for $m_i$.
\cite{Bergman:2013ala} finds that $Z^{U(2)}_{\rm QM}(N_f,\kappa)$ is related to
$Z^{Sp(1)}_{\rm QFT}(N_f,\theta=0)$ when $N-(\kappa+\frac{N_f}{2})$ is even, while
it is related to $Z^{Sp(1)}_{\rm QFT}(N_f,\theta=\pi)$ when $N-(\kappa+\frac{N_f}{2})$
is odd. To make the comparison between the $U(2)$ and $Sp(1)$ observables, we shall
identify $\alpha_1+\alpha_2=0$ in the $U(2)$ results.

One first finds \cite{Bergman:2013ala,Bergman:2013aca}
\begin{equation}
  \frac{Z_{\rm QM}^{U(2)}(N_f,\kappa)}{Z_{\rm QFT}^{Sp(1)}(N_f,e^{i\theta}=\pm 1)}=1
\end{equation}
when $N_f+2|\kappa|<2N$, with $e^{i\theta}=\pm 1$ if $N-(\kappa+\frac{N_f}{2})$ is even/odd,
respectively. We checked this fact for $N=2$, and all possible $N_f,\kappa$ satisfying
$N_f+2|\kappa|<2N$ up to $q^3$ order. Although this was already analyzed in \cite{Bergman:2013ala,Bergman:2013aca}, we checked it for our own sake.
In proving this, it is crucial to insert the factor $(-1)^{k(\kappa+N_f/2)}$ in the
$k$ instanton index of the $U(2)$ theory, as explained in \cite{Bergman:2013ala,Bergman:2013aca}.
Secondly, one finds
\begin{equation}
  \frac{Z_{\rm QM}^{U(2)}(N_f=2N-2|\kappa|,\kappa,\zeta)}
  {Z_{\rm QFT}^{Sp(1)}(N_f=2N-2|\kappa|,e^{i\theta}=\pm 1)}
  =Z_{\rm extra}(\zeta)
\end{equation}
when the 5d SCFT bound $N_f+2|\kappa|\leq 2N$ is saturated. The theta angle is
chosen between $e^{i\theta}=\pm 1$ depending on whether $N-(\kappa+\frac{N_f}{2})$
is even or odd.
Namely, when $\kappa\geq 0$ and saturates 5d SCFT bound $\kappa=N-\frac{N_f}{2}$,
one takes $e^{i\theta}=+1$. On the other hand, when $\kappa<0$, one takes
$e^{i\theta}=(-1)^{N_f}$. At $N_f=0$ when $\theta$ acquires physical meaning,
we find $e^{i\theta}=1$ for both $\kappa=\pm 2$.\footnote{More generally, $U(N)$ theory
with $N_f = 0$ have the same value of $e^{i\theta}$ at Chern-Simons levels $\kappa$, $-\kappa$.}
The extra
$Z_{\rm extra}$ factor is naturally expected, since there always exist D1-branes which
can be separated from the QFT system in this case,
as explained in section 3.2. This is given by
\begin{equation}\label{U(N)-string-1}
  Z_{\rm extra}(\zeta)=\left\{\begin{array}{ll}
  {\rm PE}\left[-\frac{qt}{(1-tu)(1-t/u)}(ty_1\cdots y_{N_f})\right]&{\rm when}\ \zeta<0\\
  {\rm PE}\left[-\frac{qt}{(1-tu)(1-t/u)}(t^{-1}y_1\cdots y_{N_f})\right]
  &{\rm when}\ \zeta>0\end{array}\right.
\end{equation}
for $\kappa>0$,
\begin{equation}\label{U(N)-string-1}
  Z_{\rm extra}(\zeta)=\left\{\begin{array}{ll}
  {\rm PE}\left[-\frac{qt}{(1-tu)(1-t/u)}(ty_1\cdots y_{N_f})^{-1}\right]&{\rm when}\ \zeta<0\\
  {\rm PE}\left[-\frac{qt}{(1-tu)(1-t/u)}(t^{-1}y_1\cdots y_{N_f})^{-1}\right]
  &{\rm when}\ \zeta>0\end{array}\right.
\end{equation}
for $\kappa<0$, and
\begin{equation}\label{U(N)-string-2}
  Z_{\rm extra}(\zeta)=\left\{\begin{array}{ll}
  {\rm PE}\left[-\frac{qt}{(1-tu)(1-t/u)}
  \left(ty_1\cdots y_{N_f}+\frac{1}{ty_1\cdots y_{N_f}}\right)\right]&{\rm when}\ \zeta<0\\
  {\rm PE}\left[-\frac{qt}{(1-tu)(1-t/u)}\left(\frac{y_1\cdots y_{N_f}}{t}+
  \frac{t}{y_1\cdots y_{N_f}}\right)\right]
  &{\rm when}\ \zeta>0\end{array}\right.
\end{equation}
for $\kappa=0$ and $N_f=2N$. Here we defined $y_i\equiv e^{m_i/2}$.
We have checked these results for
$N=2$ and $(N_f,\kappa)=(0,\pm 2)$, $(1,\pm\frac{3}{2})$, $(2,\pm 1)$, $(3,\pm\frac{1}{2})$,
$(4,0)$ up to $q^3$ order. These results are known from
\cite{Hayashi:2013qwa,Bao:2013pwa,Bergman:2013ala,Bergman:2013aca,Taki:2013vka,Taki:2014pba}.
In particular,
\cite{Hayashi:2013qwa} explains that it is consistent with the structure of the index
for M2-branes wrapping 2-cycles in CY$_3$ which can escape from the QFT.
Note that  $Z_{\rm QM}^{U(N)}$ at
$N_f+2|\kappa|=2N$ lacks the $\epsilon_+\rightarrow-\epsilon_+$ (or $t\rightarrow t^{-1}$)
invariance, which is inconsistent either as a half-BPS index of 5d SYM or the index of
5d SCFT with $SU(2)_R$ symmetry \cite{Bergman:2013ala}. This
asymmetry all goes to $Z_{\rm extra}$, leaving $Z_{\rm QFT}^{U(2)}=Z_{\rm QFT}^{Sp(1)}$
invariant under the sign flip of $\epsilon_+$. The bulk contribution is not invariant under
$\epsilon_+\rightarrow-\epsilon_+$. We are not aware of the half-BPS state interpretation
of this part of the index in an $SU(2)_R$ invariant theory, so the asymmetry should be fine.

Note that the ratios $\frac{Z_{\rm QM}^{U(2)}(\zeta<0)}{Z_{\rm QM}^{U(2)}(\zeta>0)}$ are
always given by
\begin{equation}
  {\rm PE}\left[-\!R_0\!-\!R_\infty\right]
  =\left\{\begin{array}{ll}{\rm PE}\left[\frac{\mathrm{sign}(\kappa)qt}{(1-tu)(1-t/u)}
  (t^{-1}-t)\left(w_1w_2y_1\cdots y_{N_f}\right)^{\mathrm{sign}(\kappa)}\right]&{\rm when}\ \kappa\neq 0\\
  {\rm PE}\left[\frac{qt}{(1-tu)(1-t/u)}
  (t^{-1}-t)\left(w_1w_2y_1\cdots y_{N_f}-\frac{1}{w_1w_2y_1\cdots y_{N_f}}\right)\right]&
  {\rm when}\ \kappa=0\end{array}\right.\ ,
\end{equation}
where $w_i\equiv e^{\alpha_i}$, and we have listed the results without taking $w_1w_2=1$.
$R_0,R_\infty$ are the residues
of the holomorphic measure for the rank $1$  integrand. This is consistent with
what we found for the rank $1$ case in section 2.2. At $w_1w_2=1$, it just reduces
to the ratio of two $Z_{\rm extra}$ factors at $\zeta\lessgtr 0$ that we found by comparing
$Z^{U(2)}_{\rm QM}$ with $Z^{Sp(1)}_{\rm QFT}$. For $U(N)$ with $N\geq 3$, we cannot directly
disentangle $Z_{\rm QM}^{U(N)}=Z_{\rm QFT}^{SU(N)}Z_{\rm extra}$. However, from the 5-brane
web diagram, we could naturally expect that the D1-branes escaping  the QFT would behave in
exactly the same way as those in the $U(2)$ theory. For instance, see Fig.~\ref{U(3)-web}
where horizontal D1-branes can escape from the QFT by moving downwards. \cite{Bergman:2013aca}
used this strategy to extract $Z_{\rm QFT}^{SU(3)}$, by dividing out the $Z_{\rm extra}$
that one could get from the $U(2)$ theory at $\kappa=2$. This is
also consistent with the ratio of $Z^{U(3)}_{\rm QM}$ at $\zeta<0$ and $\zeta>0$, which is
\begin{equation}
	\label{eq:u3decoupled}
  \frac{Z^{U(3)}_{\rm QM}(\zeta<0)}{Z^{U(3)}_{\rm QM}(\zeta>ß0)}=
  {\rm PE}\left[\frac{\mathrm{sign}(\kappa)qt}{(1-tu)(1-t/u)}(t^{-1}-t)\left(w_1w_2w_3y_1\cdots y_{N_f}\right)^{\mathrm{sign}(\kappa)}\right]
\end{equation}
for $\kappa \neq 0$, or which is the product of two expressions \eqref{eq:u3decoupled} for positive/negative $\kappa$ if $\kappa = 0$. Setting $U(1)\subset U(3)$ fugacity to $w_1w_2w_3=1$, the right hand side
equals $\frac{Z_{\rm QM}^{U(2)}(\zeta<0)}{Z_{\rm QM}^{U(2)}(\zeta>0)}$
at $w_1w_2=1$.

\section{5d SCFT from D4-D8-O8 and enhanced symmetry}

In this section, we use the QFT instanton partition function
$Z_{\rm QFT}=\frac{Z_{\rm QM}}{Z_{\rm extra}}$ for the $Sp(N)$ theory with $1$
antisymmetric and $N_f\leq 7$ fundamental hypermultiplets to study the 5d SCFT of
\cite{oai:arXiv.org:hep-th/9608111}. The relevant $Z_{\rm extra}$ factors are all identified
in section 3.4.1. In particular, we would like to study the superconformal index
\cite{Kinney:2005ej} for the 5d SCFTs. This index is a supersymmetric partition
function on $S^4\times S^1$. When the 5d SCFT admits a relevant deformation
to a 5d SYM, \cite{Kim:2012gu} studied this quantity in detail. One can define it by
\begin{equation}
  I(t,u,m_i,q)={\rm Tr}\left[(-1)^Fe^{-\beta\{Q,S\}}
  t^{2(J_r+J_R)}u^{2J_l}e^{-F\cdot m}q^k\right]\ .
\end{equation}
$J_r,J_l$ are rotations of $SO(4)\subset SO(5)$ on $S^4$, being
the Cartans of $SU(2)_r\times SU(2)_l\subset SO(4)$. $J_r,J_l$ have two
fixed points at the north and south poles of $S^4$. $J_R$ is the Cartan of
the $SU(2)_R$ symmetry of the $F(4)$ superconformal symmetry. $F$ are the
global symmetries of the SCFT which are visible in the 5d SYM as Noether charges.
$k$ is the instanton number in 5d SYM. This index counts BPS local operators on
$\mathbb{R}^5$, or BPS states on $S^4\times\mathbb{R}$, which saturate the
following bound
\begin{equation}
  \{Q,S\}=E-2J_r-3J_R\geq 0
\end{equation}
for the scale dimension (or energy) $E$.

In 5d SYM, \cite{Kim:2012gu} showed that this index
can be expressed as a unitary matrix integral of group $G$, the gauge
group of 5d SYM. The measure of the integrand is given by a product of two
instanton partition functions of the 5d gauge theory, or more abstractly the
partition function of 5d SCFT on Omega-deformed $\mathbb{R}^4\times S^1$. Especially
in the latter abstract viewpoint, one should be using $Z_{\rm QFT}$ rather than
$Z_{\rm QM}$. The precise form is given by
\begin{equation}\label{SCI}
  I(t,u,m_i,q)=\int[d a]Z_{\rm pert}(ia,t,u,m_i)
  Z_{\rm inst}(ia,t,u,m_i,q)Z_{\rm inst}(-ia,t,u,-m_i,q^{-1})\ .
\end{equation}
$[da]$ is the integral over holonomies of $G$,
including its Haar measure. $Z_{\rm pert}$ is given by \cite{Kim:2012gu}
\begin{equation}
  Z_{\rm pert}={\rm PE}\left[f_{\rm vec}(t,u,e^{ia})+f_{\rm fund}(t,u,e^{ia},e^{m_l})+f_{\rm anti}(t,u,e^{ia},e^{m})\right]\ ,
\end{equation}
where
\begin{eqnarray}
  f_{\rm vec}&=&-\frac{t(u+u^{-1})}{(1-tu)(1-t/u)}\left[\sum_{i<j}^N
  e^{\pm ia_i\pm ia_j}+\sum_{i=1}^Ne^{\pm 2ia_i}+N\right]\nonumber\\
  f_{\rm fund}&=&\frac{t}{(1-tu)(1-t/u)}\sum_{i=1}^N\sum_{l=1}^{N_f}
  e^{\pm ia_i\pm m_l}\nonumber\\
  f_{\rm anti}&=&\frac{t(e^m+e^{-m})}{(1-tu)(1-t/u)}\left[\sum_{i<j}^N
  e^{\pm ia_i\pm ia_j}+N\right]\ .
\end{eqnarray}
Here we use the notation $e^{\pm x} = e^{+x} + e^{-x}$, and so on.
Of course for $Sp(1)$, we do not include $f_{\rm anti}$ in $Z_{\rm pert}$.
Each $Z_{\rm inst}$ is the instanton contribution, which is given by
our $Z_{\rm QFT}$ in section 3.

\subsection{$Sp(1)$ indices}

Since $Z_{\rm QFT}$ from the ADHM quantum mechanics with
$n_A=1$ (our work) and with $n_A=0$ (computed in \cite{Kim:2012gu}) are same
for $N_f\leq 5$, we do not have to compute the superconformal indices again. So we
just review the results of \cite{Kim:2012gu}. For $N_f=0$, one obtains
\begin{align*}
I&=1 + \chi^{E_1}_{\bf 3} t^2 +\chi_2(u)\big[1 + \chi^{E_1}_{\bf 3}\big] t^{3} + \Big(\chi_3(u)\big[1 +\chi^{E_1}_{\bf 3}\big] +1+\chi^{E_1}_{\bf 5}\Big) t^4 \\
& + \Big(\chi_4(u) \big[1+\chi^{E_1}_{\bf 3}\big] +\chi_2(u)\big[1+\chi^{E_1}_{\bf 3}  + \chi^{E_1}_{\bf 5}\big]\Big) t^{5} \\
&+ \Big(\chi_5(u) \big[1+\chi^{E_1}_{\bf 3} \big] + \chi_3(u)\big[1 + \chi^{E_1}_{\bf 3} + \chi^{E_1}_{\bf 5}  + \chi^{E_1}_{\bf 3}\chi^{E_1}_{\bf 3}\big] +\chi^{E_1}_{\bf 3} + \chi^{E_1}_{\bf 7}-1\Big)t^6\\
&+\Big(\chi_6(u) \big[1+\chi^{E_1}_{\bf 3}] +
\chi_4(u) \big[2+4\chi^{E_1}_{\bf 3}+2\chi^{E_1}_{\bf 5}\big]
+ \chi_2(u)\big[ 1+3\chi^{E_1}_{\bf 3}+2\chi^{E_1}_{\bf 5}+\chi^{E_1}_{\bf 7}\big]
\Big) t^7\\
&+\Big(\chi_7(u)\big[1+\chi^{E_1}_{\bf 3}] +\chi_5(u)\big[3\chi^{E_1}_{\bf 5}+5\chi^{E_1}_{\bf 3} +4\big]
+\chi_3(u)\big[2\chi^{E_1}_{\bf 7}+3\chi^{E_1}_{\bf 5}+7 \chi^{E_1}_{\bf 3}+2\big]\\
&+\chi^{E_1}_{\bf 9}+2\chi^{E_1}_{\bf 5}+2\chi^{E_1}_{\bf 3}+3\Big)t^8+{\cal O}(t^{9}),
\end{align*}
where $\chi_n(u)$ is the character of $n$-dimensional representation of $SU(2)$. The enhanced symmetry $E_1=SU(2)$ appears rather trivially, as the superconformal index
is manifestly invariant under the $q\rightarrow q^{-1}$ Weyl symmetry. For $N_f=1$,
one obtains
\begin{align*}
I&=1 + \chi^{E_2}_{\bf 4} t^2 +\chi_2(u)\big[1 + \chi^{E_2}_{\bf4}\big] t^{3} + \Big(\chi_3(u)\big[1 +\chi^{E_2}_{\bf 4}\big] +1+\chi^{SU(2)}_{\bf 5}-\chi_{\bf 4}(f)\Big) t^4 \\
& + \Big(\chi_4(u) \big[1+\chi^{E_2}_{\bf 4}\big]
+\chi_2(u)\big[\chi^{E_2}_{\bf 4}  + \chi^{SU(2)}_{\bf 3}+\chi^{SU(2)}_{\bf 5}-\chi_{\bf 4}(f)\big]\Big) t^{5} \\
&+ \Big(\chi_5(u) \big[1+\chi^{E_2}_{\bf 4} \big] + \chi_3(u)\big[
4 \chi^{E_2}_{\bf 4} + 2\chi^{SU(2)}_{\bf 5}-\chi_{\bf 4}(f) \big]
+\chi^{SU(2)}_{\bf 7} +3 \chi^{SU(2)}_{\bf 3}+1\Big)t^6\\
&+\Big(\chi_6(u) \big[1+\chi^{E_2}_{\bf 4}] +
\chi_4(u) \big[5\chi^{E_2}_{\bf 4}+2\chi^{SU(2)}_{\bf 3}+2\chi^{SU(2)}_{\bf 5}-\chi_{\bf 4}(f)\big]\\
&
+ \chi_2(u)\big[6\chi^{E_2}_{\bf 4}+2\chi^{SU(2)}_{\bf 5}+\chi^{SU(2)}_{\bf 7}-\chi^{SU(2)}_{\bf 3}\chi_{\bf 4}(f)\big]
\Big) t^7\\
&+\Big(\chi_7(u)\big[1+\chi^{E_2}_{\bf 4}]
+\chi_5(u)\big[9\chi^{E_2}_{\bf 4}+3\chi^{SU(2)}_{\bf 5} -\chi_{\bf 4}(f) \big]
+\chi_3(u)\big[9\chi^{E_2}_{\bf 4}+2\chi^{SU(2)}_{\bf 7} +4\chi^{SU(2)}_{\bf 5}\\
&
+2 \chi^{SU(2)}_{\bf 3}-(\chi^{E_2}_{\bf 4}+\chi^{SU(2)}_{\bf 3})\chi_{\bf 4}(f)\big]
+3\chi^{E_2}_{\bf 4}+\chi^{SU(2)}_{\bf 9}+2\chi^{SU(2)}_{\bf 5}+2-\chi^{E_2}_{\bf 4}\chi_{\bf 4}(f)\Big)t^8+{\cal O}(t^{9}),\nn
\end{align*}
with $E_2=SU(2)\times U(1)$. $\chi^{E_2}_{\bf 4}$ is the adjoint character
$1+\chi^{SU(2)}_{\bf 3}$ of $E_2$, while other $SU(2)$ characters with boldfaced subscripts
are for its $SU(2)$ subgroup. $\chi_{\bf 4}(f)$ is given by \cite{Kim:2012gu}
\begin{align}\label{E2-fermion}
\chi_{\bf 4}(f)= \left(e^{i \frac{\rho}{2}}+e^{-i \frac{\rho}{2}}\right) \chi_\mathbf{2}\ ,
\end{align}
where $\chi_\mathbf{2}$ is the $SU(2)$ character and $\rho$ is the $U(1)$ chemical potential
in $E_2 = SU(2) \times U(1)$. The embedding of $SO(2) \times U(1)_I$ into $E_2$ is given by
\begin{align}
E_2 = SU(2)_{\frac{1}{2} (m_1+w)} \times U(1)_{\frac{1}{2} (7 m_1-w)} \supset SO(2)_{m_1} \times {U(1)_I}_w.
\end{align}
Therefore, $\chi_\mathbf{2}$ and $e^{i \frac{\rho}{2}}$ are written in terms of $SO(2) \times U(1)_I$ fugacities $y_1=e^{m_1/2},q=e^{w/2}$ by
\begin{gather}
\chi_\mathbf{2} = y_1^\frac{1}{2} q^\frac{1}{2}+y_1^{-\frac{1}{2}} q^{-\frac{1}{2}}\ \ ,
\ \ e^{i \frac{\rho}{2}} = y_1^{7/2} q^{-1/2}\ .
\end{gather}
For $2\leq N_f\leq 5$, one obtains
\begin{align*}
I&= 1+ \chi_{\bf adj} \,t^2 + \chi_2(u) \big[1+\chi_{\bf adj}\big]t^{3} + \Big(\chi_3(u) \big[1+ \chi_{\bf adj}\big] +1+ \chi_{\bf adj^2}\Big)t^4 \\
&+\Big(\chi_4(u) \big[1+ \chi_{\bf adj}\big] +\chi_2(u)\big[1+ \chi_{\bf adj^2} + \chi_{({\bf adj\otimes adj})_A}\big]\Big)t^{5} \\
&+\Big(\chi_5(u) \big[1+ \chi_{\bf adj}\big] +\chi_3(u)\big[1+\chi_{\bf adj}+ \chi_{\bf adj^2}+ \chi_{\bf adj\otimes adj}\big]
+ \chi_{\bf adj}+ \chi_{\bf adj^3}+ \chi_{({\bf adj\otimes adj })_A}\Big)t^{6}+ {\cal O}(t^{7}),
\end{align*}
where ${\bf adj}$ denotes the adjoint representation of $E_{N_f+1}$, and
$({\bf adj}\otimes{\bf adj})_A$ denotes antisymmetrized tensor product of two adjoint
representations. A brief explanation of $E_n$ characters is provided in Appendix B.

Before proceeding, we comment on the calculations of the superconformal index in
series expansion. Unlike Nekrasov's partition function in which the instanton fugacity
$q$ is the main expansion parameter, the superconformal index is expanded in
$t=e^{-\epsilon_+}$, so comes in both positive and negative powers in $q$.  One should first
fix the order $t^n$ to which one wishes to expand $I$. Then one investigates the $q$ expansion
or $q^{-1}$ expansion of the two $Z_{\rm inst}$'s, and see how many instantons one has to
keep.

Now we explain the $Sp(1)$ index with $N_f=6$ matters, which is a new result. One obtains
\begin{align}\label{Nf=6-SCI}
\begin{aligned}
I &= 1+\chi^{E_7}_\mathbf{133} t^2+\chi_2(u) \left[1+\chi^{E_7}_\mathbf{133}\right] t^3+\left[1+\chi^{E_7}_\mathbf{7371}+\chi_3(u) \left(1+\chi^{E_7}_\mathbf{133}\right)\right] t^4 \\
&\quad +\left[\chi_2(u) \left(1+\chi^{E_7}_\mathbf{133}+\chi^{E_7}_\mathbf{7371}+\chi^{E_7}_\mathbf{8645}\right)+\chi_4(u) \left(1+\chi^{E_7}_\mathbf{133}\right)\right] t^5 \\
&\quad +\left[2 \chi^{E_7}_\mathbf{133}+\chi^{E_7}_\mathbf{8645}+\chi^{E_7}_\mathbf{238602}+\chi_3(u) \left(2+2 \chi^{E_7}_\mathbf{133}+\chi^{E_7}_\mathbf{1539}+2 \chi^{E_7}_\mathbf{7371}+\chi^{E_7}_\mathbf{8645}\right)\right. \\
&\quad \qquad \left.+\chi_5(u) \left(1+\chi^{E_7}_\mathbf{133}\right)\right] t^6+\mathcal O\left(t^7\right)\ ,
\end{aligned}
\end{align}
showing the $E_7$ enhancement. The branching rules for $E_7\rightarrow SO(12)\times U(1)$
are\footnote{The names of representations displayed on the right hand sides, especially the
barred ones, follow the chirality convention in \cite{Feger}. For instance, our (unbarred)
chiral spinors used in (\ref{9d-SYM-inst}) are anti-chiral spinors for $N_f=2,3,6,7$ in
\cite{Feger} and our (\ref{E7-branching}), (\ref{E8-branching}), while they are still chiral
spinors for $N_f=4,5$ in \cite{Feger}.}
\begin{align}\label{E7-branching}
\mathbf{133} &=  \mathbf{1}_2+\mathbf{1}_0+\mathbf{1}_{-2}+\mathbf{\overline{32}}_1+\mathbf{\overline{32}}_{-1}+\mathbf{66}_0, \nonumber\\
\mathbf{1539} &= \mathbf{1}_0+\mathbf{\overline{32}}_1+\mathbf{\overline{32}}_{-1}+\mathbf{66}_2+\mathbf{66}_0+\mathbf{66}_{-2}+\mathbf{77}_0+\mathbf{\overline{352}}_1
+\mathbf{\overline{352}}_{-1} +\mathbf{495}_0, \nonumber\\
\mathbf{7371} &= \mathbf{1}_4+\mathbf{1}_2+2 \times \mathbf{1}_0+\mathbf{1}_{-2}+\mathbf{1}_{-4}+\mathbf{\overline{32}}_3+2 \times \mathbf{\overline{32}}_1+2 \times \mathbf{\overline{32}}_{-1}+\mathbf{\overline{32}}_{-3}\nonumber\\
&\quad +\mathbf{66}_2+\mathbf{66}_0+\mathbf{66}_{-2} +\mathbf{\overline{462}}_2+\mathbf{\overline{462}}_0+\mathbf{\overline{462}}_{-2}+\mathbf{495}_0+\mathbf{1638}_0 +\mathbf{\overline{1728}}_1+\mathbf{\overline{1728}}_{-1}, \nonumber\\
\mathbf{8645} &= \mathbf{1}_2+\mathbf{1}_0+\mathbf{1}_{-2}+\mathbf{\overline{32}}_3+2 \times \mathbf{\overline{32}}_1+2 \times \mathbf{\overline{32}}_{-1}+\mathbf{\overline{32}}_{-3} +\mathbf{66}_2+2 \times \mathbf{66}_0+\mathbf{66}_{-2}\nonumber\\
&\quad +\mathbf{\overline{352}}_1+\mathbf{\overline{352}}_{-1}+\mathbf{\overline{462}}_0 +\mathbf{495}_2+\mathbf{495}_0+\mathbf{495}_{-2}+\mathbf{\overline{1728}}_1+\mathbf{\overline{1728}}_{-1}+\mathbf{2079}_0, \nonumber\\
\mathbf{238602} &= \mathbf{1}_6+\mathbf{1}_4+2 \times \mathbf{1}_2+2 \times \mathbf{1}_0+2 \times \mathbf{1}_{-2}+\mathbf{1}_{-4}+\mathbf{1}_{-6} \nonumber\\
&\quad +\mathbf{\overline{32}}_5+2 \times \mathbf{\overline{32}}_3+3 \times \mathbf{\overline{32}}_1+3 \times \mathbf{\overline{32}}_{-1}+2 \times \mathbf{\overline{32}}_{-3}+\mathbf{\overline{32}}_{-5} \nonumber\\
&\quad +\mathbf{66}_4+\mathbf{66}_2+2 \times \mathbf{66}_0+\mathbf{66}_{-2}+\mathbf{66}_{-4} +\mathbf{\overline{462}}_4+2 \times \mathbf{\overline{462}}_2+3 \times \mathbf{\overline{462}}_0+2 \times \mathbf{\overline{462}}_{-2}+\mathbf{\overline{462}}_{-4} \nonumber\\
&\quad +\mathbf{495}_2+\mathbf{495}_0+\mathbf{495}_{-2}+\mathbf{1638}_2+\mathbf{1638}_0+\mathbf{1638}_{-2} \nonumber\\
&\quad +\mathbf{\overline{1728}}_3+2 \times \mathbf{\overline{1728}}_1+2 \times \mathbf{\overline{1728}}_{-1}+\mathbf{\overline{1728}}_{-3} \nonumber\\
&\quad +\mathbf{\overline{4224}}_3+\mathbf{\overline{4224}}_1+\mathbf{\overline{4224}}_{-1}+\mathbf{\overline{4224}}_{-3}+\mathbf{\overline{8800}}_1+\mathbf{\overline{8800}}_{-1} \nonumber\\
&\quad +\mathbf{21021}_0+\mathbf{\overline{21450}}_2+\mathbf{\overline{21450}}_0+\mathbf{\overline{21450}}_{-2}+\mathbf{23100}_0 +\mathbf{\overline{36960}}_1+\mathbf{\overline{36960}}_{-1}.
\end{align}
To completely obtain all contributions up to $t^6$ order, we count the orders
as follows. Firstly, one can check that $Z_{\rm QM}$ at $4$-instanton order starts from
$t^6$, while $Z_{\rm QM}$ at $5$-instanton order starts from $t^9$. So it may
appear that the result up to $t^6$ will be consistently obtained by making a 4-instanton
expansion in both $Z_{\rm inst}$'s in (\ref{SCI}). However, note that $Z_{\rm inst}$ in
(\ref{SCI}) should be $Z_{\rm QFT}=\frac{Z_{\rm QM}}{Z_{\rm extra}}$, and
$Z_{\rm extra}$ obeys a different upper bound on instanton number with given order in
$t$. Namely, in (\ref{9d-SYM-inst}), the single particle index $f_6$ contains
$t^2q^2$. So in $Z_{\rm extra}=PE[f_6]$, $t^6$ can come with
$t^6q^6=(t^2q^2)^3$, which contain more than $4$-instanton order at $t^6$. Actually this
is the reason why the branching rule of ${\bf 238602}$
contains $5,6$ instanton contributions. However, we know $Z_{\rm extra}$ exactly so
that all contributions at $k>4$ can be easily traced. Thus, we expand $Z_{\rm QM}$ that
appears in (\ref{SCI}) up to $4$-instantons, and $Z_{\rm extra}$
up to $6$-instantons, which consistently yields all contributions till $t^6$ order.

Finally, we consider the $Sp(1)$ index at $N_f=7$. The superconformal index is given by
\begin{align}\label{Nf=7-SCI}
\begin{aligned}
I &= 1+\chi^{E_8}_\mathbf{248} t^2+\chi_2(u) \left[1+\chi^{E_8}_\mathbf{248}\right] t^3+\left[1+\chi^{E_8}_\mathbf{27000}+\chi_3(u) \left(1+\chi^{E_8}_\mathbf{248}\right)\right] t^4 \\
&\quad +\left[\chi_2(u) \left(1+\chi^{E_8}_\mathbf{248}+\chi^{E_8}_\mathbf{27000}+\chi^{E_8}_\mathbf{30380}\right)+\chi_4(u) \left(1+\chi^{E_8}_\mathbf{248}\right)\right] t^5 \\
&\quad +\left[2 \chi^{E_8}_\mathbf{248}+\chi^{E_8}_\mathbf{30380}+\chi^{E_8}_\mathbf{1763125}+\chi_3(u) \left(2+2 \chi^{E_8}_\mathbf{133}+\chi^{E_8}_\mathbf{3875}+2 \chi^{E_8}_\mathbf{27000}+\chi^{E_8}_\mathbf{30380}\right)\right. \\
&\quad \qquad \left.+\chi_5(u) \left(1+\chi^{E_8}_\mathbf{248}\right)\right] t^6+\mathcal O\left(t^7\right)\ ,
\end{aligned}
\end{align}
with $E_8$ enhancement. The relevant $E_8\rightarrow SO(14)\times U(1)$ branching rules are
\begin{align}\label{E8-branching}
\mathbf{248} &= \mathbf{1}_0+\mathbf{14}_2+\mathbf{14}_{-2}+\mathbf{64}_{-1}+\mathbf{\overline{64}}_1+\mathbf{91}_0, \nonumber\\
\mathbf{3875} &= \mathbf{1}_4+\mathbf{1}_0+\mathbf{1}_{-4}+\mathbf{14}_2+\mathbf{14}_{-2}+\mathbf{64}_{3}+\mathbf{64}_{-1}+\mathbf{\overline{64}}_1+\mathbf{\overline{64}}_{-3}+\mathbf{91}_0 \nonumber\\
&\quad +\mathbf{104}_0+\mathbf{364}_2+\mathbf{364}_{-2}+\mathbf{832}_{-1}+\mathbf{\overline{832}}_1+\mathbf{1001}_0, \nonumber\\
\mathbf{27000} &= 2 \times \mathbf{1}_0+\mathbf{14}_2+\mathbf{14}_{-2}+2 \times \mathbf{64}_{-1}+2 \times \mathbf{\overline{64}}_1+2 \times \mathbf{91}_0 \nonumber\\
&\quad +\mathbf{104}_4+\mathbf{104}_0+\mathbf{104}_{-4}+\mathbf{364}_2+\mathbf{364}_{-2} \nonumber\\
&\quad +\mathbf{832}_3+\mathbf{832}_{-1}+\mathbf{\overline{832}}_1+\mathbf{\overline{832}}_{-3}+\mathbf{896}_2+\mathbf{896}_{-2}+\mathbf{1001}_0 \nonumber\\
&\quad +\mathbf{1716}_{-2}+\mathbf{\overline{1716}}_2+\mathbf{3003}_0+\mathbf{3080}_0+\mathbf{4928}_{-1}+\mathbf{\overline{4928}}_1, \nonumber\\
\mathbf{30380} &= \mathbf{1}_0+2 \times \mathbf{14}_2+2 \times \mathbf{14}_{-2}+\mathbf{64}_{3}+2 \times \mathbf{64}_{-1}+2 \times \mathbf{\overline{64}}_1+\mathbf{\overline{64}}_{-3} \nonumber\\
&\quad +\mathbf{91}_4+3 \times \mathbf{91}_0+\mathbf{91}_{-4}+\mathbf{104}_0+\mathbf{364}_2+\mathbf{364}_{-2} \nonumber\\
&\quad +\mathbf{832}_3+2 \times \mathbf{832}_{-1}+2 \times \mathbf{\overline{832}}_1+\mathbf{\overline{832}}_{-3}+\mathbf{896}_2+\mathbf{896}_{-2} \nonumber\\
&\quad +\mathbf{1001}_0+\mathbf{2002}_2+\mathbf{2002}_{-2}+\mathbf{3003}_0+\mathbf{4004}_0+\mathbf{4928}_{-1}+\mathbf{\overline{4928}}_1, \nonumber\\
\mathbf{1763125} &= 2 \times \mathbf{1}_0+2 \times \mathbf{14}_2+2 \times \mathbf{14}_{-2}+3 \times \mathbf{64}_{-1}+3 \times \mathbf{\overline{64}}_1+3 \times \mathbf{91}_0 \nonumber\\
&\quad +\mathbf{104}_4+\mathbf{104}_0+\mathbf{104}_{-4}+\mathbf{364}_2+\mathbf{364}_{-2}+\mathbf{546}_{6}+\mathbf{546}_{2}+\mathbf{546}_{-2}+\mathbf{546}_{-6} \nonumber\\
&\quad +2 \times \mathbf{832}_3+2 \times \mathbf{832}_{-1}+2 \times \mathbf{\overline{832}}_1+2 \times \mathbf{\overline{832}}_{-3}+2 \times \mathbf{896}_2+2 \times \mathbf{896}_{-2} \nonumber\\
&\quad +2 \times \mathbf{1001}_0+2 \times \mathbf{1716}_{-2}+2 \times \mathbf{\overline{1716}}_{2}+\mathbf{2002}_2+\mathbf{2002}_{-2} \nonumber\\
&\quad +3 \times \mathbf{3003}_0+2 \times \mathbf{3080}_0+\mathbf{4004}_4+2 \times \mathbf{4004}_0+\mathbf{4004}_{-4} \nonumber\\
&\quad +3 \times \mathbf{4928}_{-1}+3 \times \mathbf{\overline{4928}}_1+\mathbf{5625}_4+\mathbf{5625}_0+\mathbf{5625}_{-4} \nonumber\\
&\quad +\mathbf{5824}_3+\mathbf{5824}_{-1}+\mathbf{5824}_{-5}+\mathbf{\overline{5824}}_5+\mathbf{\overline{5824}}_1+\mathbf{\overline{5824}}_{-3} \nonumber\\
&\quad +\mathbf{11648}_2+\mathbf{11648}_{-2}+\mathbf{17472}_3+\mathbf{17472}_{-1}+\mathbf{\overline{17472}}_1+\mathbf{\overline{17472}}_{-3} \nonumber\\
&\quad +\mathbf{18200}_2+\mathbf{18200}_{-2}+\mathbf{21021}_0+\mathbf{21021}_{-4}+\mathbf{\overline{21021}}_4+\mathbf{\overline{21021}}_0 \nonumber\\
&\quad +\mathbf{24024'}_2+\mathbf{24024'}_{-2}+\mathbf{27456}_3+\mathbf{\overline{27456}}_{-3}+\mathbf{36608}_2+\mathbf{36608}_{-2} \nonumber\\
&\quad +\mathbf{40768}_{-1}+\mathbf{\overline{40768}}_1+\mathbf{45760}_3+\mathbf{45760}_{-1}+\mathbf{\overline{45760}}_1+\mathbf{\overline{45760}}_{-3} \nonumber\\
&\quad +\mathbf{58344}_0+\mathbf{58968}_0+\mathbf{64064'}_{-1}+\mathbf{\overline{64064'}}_1+\mathbf{115830}_{-2}+\mathbf{\overline{115830}}_2 \nonumber\\
&\quad +\mathbf{146432}_{-1}+\mathbf{\overline{146432}}_1+\mathbf{200200}_0.
\end{align}
The instanton order counting for the $t$ expansion up to $t^6$ goes as follows.
We computed $Z_{\rm QM}$ up to $5$-instantons to get these results. $5$-instanton
results start at $t^6$, so assuming that higher instantons come with higher powers in
$t$, our result should be reliable up to $t^6$ order.\footnote{Here we made a small
assumption that $6$ and higher instantons do not contribute till $t^6$ order. We could
not check this due to large computational time at $k=6$. So the $E_8$ enhancement
at $t^6$ that we find from $5$ instanton calculus is justified with this assumption.} Again
$Z_{\rm extra}$ up to $t^6$ order can come with higher instantons. Since $f_7$ in
(\ref{9d-SYM-inst}) comes with $t^2q^2$, we can maximally have $q^6$ from
$Z_{\rm extra}=PE[f_7]$ at $t^6$. This is the reason why we find contribution at
$k=\pm 6$ in the branching rule of ${\bf 1763125}$. Again, since we know
$Z_{\rm extra}$ exactly, we expand it up to $t^6$ and also expand $Z_{\rm QM}$ up to
$5$-instantons to consistently get all terms up to $t^6$.

This finishes our illustration that the $Sp(1)$ index at $N_f=6,7$ exhibits $E_7$
and $E_8$ enhancement, respectively, complementing the results of \cite{Kim:2012gu}
at $N_f\leq 5$. We close this subsection by a few comments on related works.
The first line of the index (\ref{Nf=6-SCI}) was obtained in \cite{Hayashi:2013qwa},
by computing $Z_{\rm inst}$ from a suitably Higgsed 5d $T_4$ theory \cite{Benini:2009gi}.
The microscopic computation of the index (\ref{Nf=7-SCI}) with
$E_8$ symmetry appears to be new.

\subsection{$Sp(2)$ indices}

By following the same procedures, we can use $Z_{\rm QFT}=Z_{\rm QM}/Z_{\rm extra}$
for the $Sp(2)$ theories as $Z_{\rm inst}$ and compute the superconformal indices.
For $0\leq N_f\leq 7$, we simply note that the superconformal index up to $t^6$ order
takes the following form:
\begin{align}
\begin{aligned}
I&= 1+ \chi_2(e^{m})\,t + \Big(\chi_2(u)\chi_2(e^{m})  +2\chi_3(e^{m})+ \chi_{\bf adj}  \Big) t^2\\
&\quad +\Big(\chi_3(u) \chi_2(e^{m}) + \chi_2(u) \left[2\chi_3(e^{m})+2+\chi_{\bf adj}\right]
+ 2\chi_4(e^{m}) + \chi_2(e^{m}) (1+2\chi_{\bf adj})
\Big) t^3\\
&\quad +\Big( \chi_4(u)\chi_2(e^{m})+\chi_3(u) \left[3\chi_3(e^{m})+2+\chi_{\bf adj}\right]
+ \chi_2(u) \left[3\chi_4(e^{m})+\chi_2(e^{m}) (5+3 \chi_{\bf adj})\right]\\
&\quad \qquad + 3\chi_5(e^{m})+ \chi_3(e^{m})(1+3 \chi_{\bf adj}) + 3+\chi_{\bf adj}
+\chi_{({\bf adj}\otimes {\bf adj})_S}\Big)t^4 \\
&\quad +\Big(\chi_5(u) \chi_2(e^{m})+\chi_4(u) \left[3 \chi_3(e^{ m})+3+\chi_{\bf adj}\right]+\chi_3(u) \left[5 \chi_4(e^{ m})+\chi_2(e^{ m}) (8+4 \chi_{\bf adj})\right] \\
&\qquad +\chi_2(u) \left[4 \chi_5(e^{ m})+\chi_3(e^{ m}) (9+6 \chi_{\bf adj})+5+4 \chi_{\bf adj}
+\chi_{{\bf adj} \otimes {\bf adj}}\right] \\
&\quad \qquad +3 \chi_6(e^{ m})+\chi_4(e^{ m}) (3+4 \chi_{\bf adj})+\chi_2(e^{ m}) (6+3 \chi_{\bf adj}+\chi_{{\bf adj}^2}+\chi_{{\bf adj} \otimes {\bf adj}}-\chi_\text{fer}^{N_f})\Big) t^5 \\
&\quad +\Big(\chi_6(u) \chi_2(e^{ m})+\chi_5(u)\left[4 \chi_3(e^{ m})+3+\chi_{\bf adj}\right]+\chi_4(u)\left[7 \chi_4(e^{ m})+\chi_2(e^{ m}) (11+5 \chi_{\bf adj})\right] \\
&\qquad +\chi_3(u)\left[8 \chi_5(e^{m})+\chi_3(e^{ m}) (16+ 10 \chi_{\bf adj})+13+7 \chi_{\bf adj}+2 \chi_{({\bf adj} \otimes {\bf adj})_S}+\chi_{({\bf adj} \otimes {\bf adj})_A}\right] \\
&\quad \qquad +\chi_2(u)\left[5 \chi_6(e^{ m})+\chi_4(e^{ m}) (14+9 \chi_{\bf adj})+\chi_2(e^{ m}) (16+12 \chi_{\bf adj}+\chi_{{\bf adj}^2}+3 \chi_{{\bf adj} \otimes {\bf adj}}-\chi_\text{fer}^{N_f})\right] \\
&\qquad +4 \chi_7(e^{ m})+\chi_5(e^{ m}) (3+5 \chi_{\bf adj})+\chi_3(e^{ m}) (14+6 \chi_{\bf adj}+2 \chi_{{\bf adj}^2}+2 \chi_{({\bf adj} \otimes {\bf adj})_S}+\chi_{({\bf adj} \otimes {\bf adj})_A}) \\
&\qquad +4+6 \chi_{\bf adj}+\chi_{{\bf adj}^2}+2 \chi_{({\bf adj} \otimes {\bf adj})_A}+\chi_\text{res}^{N_f}-2 \chi_\text{fer}^{N_f}\Big) t^6+\mathcal O\left(t^7\right).
\end{aligned}\nonumber
\end{align}
$m$ is the chemical potential for $SU(2)_F$ global symmetry rotating the anti-symmetric
$Sp(N)$ hypermultiplet. ${\bf adj}$ denotes the adjoint representation of $E_{N_f+1}$
The terms $\chi_{\rm res}^{N_f}$ and $-\chi^{N_f}_{\rm fer}$ are non-universal terms
which depend on $N_f$. $-\chi^{N_f}_{\rm fer}$ is nonzero only for $N_f=1$, given by
\begin{align}
\chi_\text{fer}^{N_f=1}=1+\chi_{\bf 4}(f)
= 1+\left(e^{i \frac{\rho}{2}}+e^{-i \frac{\rho}{2}}\right) \chi_\mathbf{2}\ .
\end{align}
$\chi_{\rm 4}(f)$ and the fugacities in it are explained around (\ref{E2-fermion}).
$\chi_\text{res}^{N_f}$ is given for each $N_f$ by
\begin{align}
\chi_\text{res}^0 &= \chi_\mathbf{3}+\chi_\mathbf{7} = \chi_{(\mathbf{3} \times \mathbf{3} \times \mathbf{3})_S}, \nonumber\\
\chi_\text{res}^1 &= 1+\chi_\mathbf{3}+\chi_\mathbf{5}+\chi_\mathbf{7}, \nonumber\\
\chi_\text{res}^2 &= \chi_3+\chi_7+\chi_\mathbf{8} (1+\chi_5)+\chi_\mathbf{10}+\chi_\mathbf{\overline{10}}+\chi_\mathbf{27} (1+\chi_3)+\chi_\mathbf{64},
\nonumber\\
\chi_\text{res}^3 &= \chi_\mathbf{24}+\chi_\mathbf{126}+\chi_\mathbf{\overline{126}}+\chi_\mathbf{200}+\chi_\mathbf{1000}+\chi_\mathbf{1024}, \nonumber\\
\chi_\text{res}^4 &= \chi_\mathbf{45}+\chi_\mathbf{945}+\chi_\mathbf{1386}+\chi_\mathbf{5940}+\chi_\mathbf{7644}, \nonumber\\
\chi_\text{res}^5 &= \chi_\mathbf{78}+\chi_\mathbf{2925}+\chi_\mathbf{34749}+\chi_\mathbf{43758},
\nonumber\\
\chi_\text{res}^6 &= \chi_\mathbf{133}+\chi_\mathbf{8645}+\chi_\mathbf{152152}+\chi_\mathbf{238602},
\nonumber\\
\chi_\text{res}^7 &= \chi_\mathbf{248}+\chi_\mathbf{30380}+\chi_\mathbf{779247}+\chi_\mathbf{1763125} = \chi_{(\mathbf{248} \otimes \mathbf{248} \otimes \mathbf{248})_S}
\end{align}
where $\chi_\mathbf{n}$ is the character of the ${\bf n}$ dimensional irrep of $E_{N_f+1}$
for $N_f\neq 1,2$.
For $N_f = 1$, $\chi_\mathbf{n}$ is a character of $SU(2)$ in $E_2 = SU(2) \times U(1)$. For
$N_f = 2$, $\chi_\mathbf{n}$ is the character of $SU(3)$ and $\chi_n$ is the character of $SU(2)$
in $E_3 = SU(3) \times SU(2)$. The $Sp(2)$ superconformal indices all show the $E_{N_f+1}$ symmetry enhancements to the $t^6$ order that we checked.

\vskip 0.5cm

\hspace*{-0.8cm} {\bf\large Acknowledgements}
\vskip 0.2cm

\hspace*{-0.75cm} We thank Dongmin Gang, Babak Haghighat, Sung-Soo Kim, Kimyeong Lee,
Guglielmo Lockhart, Cumrun Vafa, Futoshi Yagi, Gabi Zafrir and especially Hee-Cheol Kim 
for helpful discussions. This work is supported by the NRF Grants No. 2012-009117 (JP), 
2012-046278 (CH, JP), 2012R1A1A2042474 (JK,SK),
2012R1A2A2A02046739 (JK, SK), NRF-2015R1A2A2A01003124 (SK). J.P. also appreciates APCTP 
for its stimulating environment for research.

\appendix

\section{ADHM degrees from 5d hypermultiplets}

When there are hypermultiplets in 5d SYM, one only finds fermion zero modes in
the instanton background. However, in the UV ADHM quantum mechanics, there could be
more (bosonic) degrees of freedom. Since the new bosonic degrees appear during
the UV completion of the SUSY sigma model on instanton moduli space, the extra
bosons do not represent the degrees of QFT. In particular, when extra bosons exist,
they will form a hypermultiplet in the mechanics of the form (\ref{hyper-hyper}), which
we call 1d twisted hypermultiplet. While
the bosonic degrees $a_{\alpha\dot\beta}$, $q_{\dot\alpha}$ of the ADHM data
represent the instanton degrees of freedom on spacetime $\mathbb{R}^4$ (instanton positions,
scale), the scalars $\Phi^A$ in (\ref{hyper-hyper}) probe the stringy realization
of instantons moving in the `internal direction,' away from the QFT.
So it is natural for them to have internal $SU(2)_R$ doublet indices rather than
the spacetime $SU(2)_r$ doublet indices as (\ref{ADHM-multiplet}). This indeed is
the case with examples that we explain below.

Consider a 5d hypermultiplet in the fundamental representation of $G$.
The Dirac fermion in this multiplet can be written as a pair of chiral and
anti-chiral fermions in $SO(4)=SU(2)_l\times SU(2)_r\subset SO(4,1)$.
The chiral fermion has $k$ complex zero modes in the background of $k$
self-dual instantons. In the ADHM gauged quantum mechanics, these zero modes
become Fermi multiplets in the fundamental representation of $\hat{G}$. These
multiplets are responsible for the $Z_{\textrm{1-loop}}$ factors (\ref{U(N)-fund-hyper}),
(\ref{Sp(N)-fund-1}), (\ref{Sp(N)-fund-2}), (\ref{Sp(N)-fund-3}).

We next explain the adjoint hypermultiplet. These can be easily motivated
by D-branes. Since adding one adjoint hypermultiplet to the pure $\mathcal{N}=1$
theory makes a maximal SYM, one can engineer this system using D4-branes for $U(N)$,
or D4-branes with an O4-plane for $SO(N)$, $Sp(N)$. Apart from the symmetries
$SU(2)_l\times SU(2)_r\times SU(2)_R$, we have $SU(2)_F$ flavor symmetry which rotates
the adjoint hypermultiplet field in 5d SYM. Placing $k$ D0-branes,
one can deduce the degrees of freedom in the ADHM quantum mechanics by studying
the massless modes of D0-D0 and D0-D4 strings. Including the ADHM data plus
the quantum mechanical gauge fields that we explained in section 2.1, one obtains
\begin{eqnarray}
  \textrm{D0-D0}&:& {\rm adj}(\hat{G})\ \ \ \ \ \ \ (A_t,\varphi,
  \underline{\varphi_{aA}})\ ,\ \
  (\bar\lambda^A_{\dot\alpha},
  \underline{\bar\lambda^a_{\dot\alpha}})\nonumber\\
  &&R(\hat{G})\ {\rm rep.}\ \ \ (a_{\alpha\dot\beta})\ ,\ \ (\lambda^A_\alpha,
  \underline{\lambda^a_\alpha})\nonumber\\
  \textrm{D0-D4}&:&G\times\hat{G}\ \textrm{bi-fundamental}\ \ \
  (q_{\dot\alpha})\ ,\ \ (\psi^A,\underline{\psi^a})\ .
\end{eqnarray}
$R(U(k)),R(Sp(k)),R(O(k))$ are adjoint, antisymmetric, symmetric, respectively.
$a=1,2$ is the doublet index of $SU(2)_F$. The underlined degrees are
coming from the 5d adjoint hypermultiplet.
$\varphi_{aA},\bar\lambda^a_{\dot\alpha}$ combine with
$A_t,\varphi,\bar\lambda^A_{\dot\alpha}$ to form the $\mathcal{N}=(4,4)$ vector
multiplet, by forming a separate $(0,4)$ twisted hypermultiplet of type (\ref{hyper-hyper}).
$\lambda^a_\alpha$ combines with $a_{\alpha\dot\beta},\lambda^A_\alpha$
to form a $(4,4)$ hypermultiplet.
$\psi^a$ combine with $q_{\dot\alpha},\psi^A$ to form a $(4,4)$ hypermultiplet.
These account for $Z_{\rm adj}$
factors of section 3.1. The full action is governed by $\mathcal{N}=(4,4)$ SUSY, and can be found
in \cite{oai:arXiv.org:1110.2175} for instance.

We make a comment on the degrees $\varphi_{aA}$,
$\bar\lambda^a_{\dot\alpha},\lambda^a_\alpha$, $\psi^a$. There is a good sense
in which they all come from 5d hypermultiplet, since they all carry the index
$a$ for $SU(2)_F$ which rotates it. However, some degrees are clearly
extra UV degrees. In particular, zero modes like
$\varphi_{aA}$ never appear in the QFT instanton background. Their eigenvalues actually
represent the location of D0-branes transverse to D4, and together with $\varphi$
form the transverse space $\mathbb{R}^5$. One finds the following bosonic potentials
$\sim|q_{\dot\alpha}\varphi_{aA}|^2$, $\sim[a_{\alpha\dot\beta},\varphi_{aA}]^2$, which
make $\varphi_{aA}$ to decouple with the ADHM fields $(a_m,q_{\dot\alpha})$ in the $g_{QM}\rightarrow\infty$ limit, in the same way as $\varphi$ decouples.

We also explain what degrees are incurred by an $Sp(N)$ antisymmetric hypermultiplet
in the ADHM mechanics. Again one can answer this from the D0-D4-D8-O8 system,
by studying the massless modes of the D0-D0, D0-D4, D0-D8-O8 open strings. One
finds \cite{Douglas:1996uz,Aharony:1997pm}
\begin{eqnarray}
  \textrm{D0-D0}&:&O(k)\ {\rm antisymmetric}\ \ \ (A_t,\varphi)\ ,\ \
  (\bar\lambda^A_{\dot\alpha},\underline{\lambda^a_\alpha})\nonumber\\
  &&O(k)\ {\rm symmetric}\ \ \ (a_{\alpha\dot\beta},\underline{\varphi_{aA}})\ ,\ \
  (\lambda^A_\alpha,\underline{\bar\lambda^a_{\dot\alpha}})\nonumber\\
  \textrm{D0-D4}&:&Sp(N)\times O(k)\ {\rm bif.}\ \ \ (q_{\dot\alpha})\ ,\ \
  (\psi^A,\underline{\psi^a})\nonumber\\
  \textrm{D0-D8}&:&SO(2N_f)\times O(k)\ {\rm bif.}\ \ \ (\underline{\Psi_l})
\end{eqnarray}
where $a=1,2$ is the $SU(2)_F$ doublet which rotates the 5d antisymmetric hypermultiplet,
and $l=1,\cdots,N_f$. The degrees without underlines are the ADHM data or the
vector multiplet fields. Among the underlined degrees, $\Psi_l$ on the last line
comes from the $N_f$ fundamental 5d hypermultiplets. The rest of the underlined degrees
come from the $Sp(N)$ antisymmetric hypermultiplet.
$(\lambda^a_\alpha)$, $(\psi^a)$, $(\Psi_l)$ form Fermi multiplets, while
$(\varphi_{aA},\bar\lambda^a_{\dot\alpha})$ form $(0,4)$ hypermultiplet. $\varphi_{aA}$
represent the motion of D0-branes along D8-O8, transverse to the D4's. Again they
represent extra degrees appearing in the UV ADHM description of QFT instantons.

These degrees form $(0,4)$ quantum mechanics, by which we mean a 1d reduction
of 2d $(0,4)$ gauge theory. Its potential can be understood from the $(0,2)$ SUSY point
of view, as follows \cite{Tong:2014yna}. For each $(0,2)$ Fermi multiplet field $\Psi$,
one can turn on two holomorphic potential functions $E_\Psi(\Phi)$, $J_\Psi(\Phi)$ depending
on the chiral multiplets of the theory. They contribute to be bosonic potential as
$|E_\Psi(\phi)|^2+|J_\Psi(\phi)|^2$. With $(0,4)$ SUSY, the vector multiplet with a gauge
group $G$ is decomposed into $(0,2)$ vector multiplet and an adjoint Fermi multiplet $\Lambda$.
For $(0,4)$ SUSY, each hypermultiplet $\Phi_{\dot\alpha}=(q,\tilde{q}^\dag)$ charged in
$G$ should contribute to $J_\Lambda\sim q\tilde{q}$. On the other hand,
each twisted hypermultiplet $\Phi_A=(\phi,\tilde\phi^\dag)$ charged in $G$ should contribute
to $E_\Lambda\sim\phi\tilde\phi$. However, in all $(0,2)$ systems, one should
have $\sum_\Psi E_\Psi J_\Psi=0$ for SUSY, which is violated with the above contributions
only. So when both hypermultiplets and twisted hypermultiplets are charged in the same
gauge group $G$ in a $(0,4)$ theory, there should be more Fermi multiplet fields $\Psi,\cdots$
in the theory with $E_\Psi\sim q\phi$, $J_\Psi\sim-\tilde{q}\tilde\phi$, etc., so that
$E\cdot J=0$ condition is met. See \cite{Tong:2014yna} for more details. So if a hypermultiplet
and a twisted hypermultiplet is charged in the same gauge group $G$, they necessarily have
a potential $\sim |\Phi_{\dot\alpha}\Phi_A|^2$. This means that in a branch with nonzero
hypermultiplet, twisted hypermultiplets become massive and decouple in the infrared.
Indeed there are such potentials in \cite{Douglas:1996uz,Aharony:1997pm}.

As emphasized in many places in this paper, the ADHM quantum mechanics may contain
extra degrees irrelevant for QFT problems. So determining their couplings in the ADHM mechanics
in principle should depend on how we embed the instanton mechanics into string theory.
However, here we note that \cite{Shadchin:2005mx} more abstractly considered the
ADHM degrees coming from 5d hypermultiplets of the classical groups in tensor product
representations. In fact, \cite{Shadchin:2005mx} just discussed the
equivariant index for the ADHM degrees coming from 5d hypermultiplets, whose Plethystic
exponential provides their $Z_{\textrm{1-loop}}$. The $Z_{\textrm{1-loop}}$
computed from the recipe of \cite{Shadchin:2005mx} agrees with all examples discussed
in this paper (with small corrections on factors, etc. which we believe are simply typos
or minor errors). Since \cite{Shadchin:2005mx} writes down $Z_{\textrm{1-loop}}$ which
only requires our knowledge on the free theory,
it may not be too sensitive to the physical string theory embedding of our QFT problem.
We sometimes blindly used
the rules of \cite{Shadchin:2005mx} to compute the indices, even when we cannot
naturally motivate the system from D-branes as in this appendix. Such cases are:
classification  of the possible poles $R_0,R_\infty$ at the infinities,
provided at the beginning of section 3; the $Sp(1)$ theory with
$N_f\leq 6$ flavors at $n_A=0$.

\section{Characters of $SO(2N_f)$}

$SO(2 N_f)$ characters in this paper can be obtained by the Weyl character formula \cite{Choi:2008za}:
\begin{align}
\chi(h,m) = \frac{\det[\sinh(m_i (h_j+N_f-j))]+\det[\cosh(m_i (h_j+N_f-j))]}{\det[\cosh(m_i (N_f-j))]}
\end{align}
where $h$ denotes the highest weight of the representation with $h_1 \geq h_2 \geq \cdots \geq h_{N_f-1} \geq |h_{N_f}|$ and $m$ denotes chemical potential. For example, two spinor representations of the highest weights $(\frac{1}{2},\cdots,\pm\frac{1}{2})$ have the following characters:
\begin{align}
\chi_\pm^{N_f} = \frac{1}{2} \prod_{i = 1}^{N_f} \left(y_i+y_i^{-1}\right)\pm\frac{1}{2}\prod_{i = 1}^{N_f} \left(y_i-y_i^{-1}\right)
\end{align}
where we use chemical potential $y_i = e^{m_i/2}$. In our paper we use two chirality conventions for such spinor representations. In section 3 we call $(\frac{1}{2},\cdots,\frac{1}{2})$ the chiral spinor and call $(\frac{1}{2},\cdots,-\frac{1}{2})$ the anti-chiral spinor, which has a bar on its name. On the other hand, in section 4 we follow the convention of \cite{Feger} for computational convenience. In that case we call $(\frac{1}{2},\cdots,-\frac{1}{2})$ the chiral spinor for $N_f = 2,3,6,7$ while we still call $(\frac{1}{2},\cdots,\frac{1}{2})$ the chiral spinor for $N_f = 4,5$.

All of the $E_n$ characters can be read off from the branching rules of $E_n$ into
its subgroup specified in the main text. For example, $E_5=S0(10)$ adjoint has the following decomposition under $SO(8) \times U(1)_I$:
\begin{equation}
{\bf 45}\rightarrow{\bf 1}_0+{\bf 28}_0+({\bf 8}_s)_1+({\bf 8}_s)_{-1}.
\end{equation}
The corresponding character is given by
\begin{equation}
\chi^{E_5}_{\bf 45}=\chi_{\bf 1}^{SO(8)}+\chi_{\bf 28}^{SO(8)}+q\chi_{{\bf 8}_s}^{SO(8)}
+q^{-1}\chi_{{\bf 8}_s}^{SO(8)}.
\end{equation}

\end{document}